\pdfoutput=1
\documentclass[12pt]{article}
\usepackage[utf8]{inputenc}%
\usepackage[tracking=true, letterspace=50, expansion=false]{microtype}

\usepackage{amsmath} %
\usepackage{amssymb} 
\usepackage{amsthm} %

\usepackage[margin=1in]{geometry} %
\usepackage{setspace} %
\usepackage[title]{appendix} %
\usepackage{multirow} %
\usepackage{booktabs} %
\usepackage[inline]{enumitem} %
\usepackage[normal, online, flushleft]{threeparttable}
\usepackage{pdflscape}

\newtheorem{theorem}{Theorem}[section] %
\newtheorem{proposition}{Proposition}[section] %
\newtheorem{lemma}{Lemma}[section] %
\newtheorem{corollary}{Corollary}[section] %
\newtheorem{assumption}{Assumption}[section] %

\theoremstyle{definition}%
\newtheorem{example}{Example}[section]%
\newtheorem{remark}{Remark}[section]

\DeclareMathOperator{\var}{var} 
\DeclareMathOperator{\se}{se} 
\DeclareMathOperator{\cva}{cva} %
\newcommand{\WW}{\mathcal{W}} 

\usepackage{mathtools}
\DeclarePairedDelimiter\hor{[}{)}

\DeclarePairedDelimiter\abs{\lvert}{\rvert}
\DeclarePairedDelimiter\norm{\lVert}{\rVert}
\DeclarePairedDelimiter\indicatorfence{\{}{\}}
\newcommand\1{\operatorname{I}\indicatorfence}

\usepackage{xcolor}%
\definecolor{webbrown}{rgb}{.6,0,0}

\usepackage{array}
\newcolumntype{C}[1]{>{\centering\let\newline\\\arraybackslash\hspace{0pt}}m{#1}}

\allowdisplaybreaks

\usepackage[nolist]{acronym}
\begin{acronym}
  \acro{DFM}{dynamic factor model}
  \acro{HPD}{highest posterior density}
  \acro{EBCI}{empirical Bayes confidence interval}
  \acro{EB}{empirical Bayes}
  \acro{CI}{confidence interval}
  \acro{MSE}{mean squared error}
  \acro{CLT}{central limit theorem}
  \acro{OLS}{ordinary least squares}
  \acro{ACI}{average coverage interval}
  \acro{CZ}{commuting zone}
  \acro{FPLIB}{flat prior limited information Bayes}
  \acro{PMT}{posterior mean trimming}
  \acro{DGP}{data generating process}
  \acroplural{DGP}[DGPs]{data generating processes}
  \acro{FDR}{false discovery rate}
  \acro{FDP}{false discovery proportion}
  \acro{NPMLE}{nonparametric maximum likelihood estimator}
  \acroplural{NPMLE}[NPMLEs]{nonparametric maximum likelihood estimators}
\end{acronym}

\usepackage{natbib} %
\usepackage{tikz} %

\usepackage{xr}
\usepackage{hyperref}
\hypersetup{%
  breaklinks = true,%
  colorlinks = true,%
  anchorcolor = webbrown,%
  citecolor = webbrown,%
  filecolor = webbrown,%
  linkcolor = webbrown,%
  menucolor = webbrown,%
  urlcolor= webbrown,%
  citebordercolor= 1 0 0,%
  menubordercolor=1 0 0,%
  urlbordercolor=1 0 0,%
  runbordercolor=1 0 0,%
  pdftitle=Robust Empirical Bayes Confidence Intervals,%
  pdfauthor=Timothy B. Armstrong \& Michal Kolesár \& Mikkel Plagborg-Møller}
\usepackage{cleveref}
\Crefname{equation}{Eq.}{Eqs.}
\crefname{assumption}{assumption}{assumptions}
\crefname{remark}{remark}{remarks}
\crefname{proposition}{proposition}{propositions}
\crefname{sappsec}{Supplemental Appendix}{Supplemental Appendices}
\crefname{sappsubsec}{Supplemental Appendix}{Supplemental Appendices}
\crefname{sappsubsubsec}{Supplemental Appendix}{Supplemental Appendices}
\crefname{appsec}{Appendix}{Appendices}
\crefname{appsubsec}{Appendix}{Appendices}

\title{\texorpdfstring{\vspace{-\baselineskip}}{}Robust Empirical Bayes
  Confidence Intervals\thanks{This paper is dedicated to the memory of Gary
    Chamberlain, who had a profound influence on our thinking about decision
    problems in econometrics, and empirical Bayes methods in particular. Luther
    Yap provided excellent research assistance. We received helpful comments
    from four anonymous referees, Ot\'{a}vio Bartalotti, Toru Kitagawa, Laura Liu, Ulrich Müller, Stefan
    Wager, Mark Watson, Martin Weidner, and numerous seminar participants. We
    are especially indebted to Bruce Hansen and Roger Koenker for inspiring our
    simulation study. Armstrong acknowledges support by the National Science
    Foundation Grant SES-2049765. Kolesár acknowledges support by the Sloan
    Research Fellowship and by the National Science Foundation Grant
    SES-22049356. Plagborg-Møller acknowledges support by the National Science
    Foundation Grant SES-1851665.}}
\author{Timothy B. Armstrong\thanks{email: \texttt{timothy.armstrong@usc.edu}}\\
  University of Southern California \and
  Michal Kolesár\thanks{email: \texttt{mkolesar@princeton.edu}}\\
  Princeton University\and
  Mikkel Plagborg-Møller\thanks{email: \texttt{mikkelpm@princeton.edu}}\\
  Princeton University}%
\date{\today}

\begin{document}
\maketitle

\begin{abstract}
  We construct robust \acp{EBCI} in a normal means problem. The intervals are
  centered at the usual linear empirical Bayes estimator, but use a critical
  value accounting for shrinkage. Parametric \acp{EBCI} that assume a normal
  distribution for the means \citep{morris83} may substantially undercover when
  this assumption is violated. In contrast, our \acp{EBCI} control coverage
  regardless of the means distribution, while remaining close in length to the
  parametric \acp{EBCI} when the means are indeed Gaussian. If the means are
  treated as fixed, our \acp{EBCI} have an average coverage guarantee: the
  coverage probability is at least $1-\alpha$ on average across the $n$
  \acp{EBCI} for each of the means. Our empirical application considers the
  effects of U.S.\ neighborhoods on intergenerational mobility.\\[1ex]
\end{abstract}

\clearpage

\section{Introduction}\label{sec:introduction}

Empirical researchers in economics are often interested in estimating effects
for many individuals or units, such as estimating teacher quality for teachers
in a given geographic area. In such problems, it is common to shrink unbiased
but noisy preliminary estimates of these effects toward baseline values, say the
average effect for teachers with the same experience. In addition to estimating
teacher quality \citep{KaSt08,JaLe08,cfr14i}, shrinkage techniques are used in a
wide range of applications including estimating school quality \citep{ahpw17},
hospital quality \citep{hull20}, the effects of neighborhoods on
intergenerational mobility \citep{chetty_impacts_2018}, and patient risk scores
across regional health care markets \citep{fghw17}.

The shrinkage estimators used in these applications can be motivated by an
\ac{EB} approach. One imposes a working assumption that the individual effects
are drawn from a normal distribution (or, more generally, a known family of
distributions). The \ac{MSE} optimal point estimator then has the form of a
Bayesian posterior mean, treating this distribution as a prior distribution.
Rather than specifying the unknown parameters in the prior distribution ex ante,
the \ac{EB} estimator replaces them with consistent estimates, just as in random
effects models. This approach is attractive because one does not need to assume
that the effects are in fact normally distributed, or even take a ``Bayesian''
or ``random effects'' view: the \ac{EB} estimators have lower
\ac{MSE} (averaged across units) than the unshrunk unbiased estimators, even
when the individual effects are treated as nonrandom \citep{JaSt61}.

In spite of the popularity of \ac{EB} methods, it is currently not known how to
provide uncertainty assessments to accompany the point estimates without
imposing strong parametric assumptions on the effect distribution. Indeed,
\citet[p. 116]{Hansen2016} describes inference in shrinkage settings as an open
problem in econometrics. The natural \ac{EB} version of a \ac{CI} takes the form
of a Bayesian credible interval, again using the postulated effect distribution
as a prior \citep{morris83}. If the distribution is correctly specified, this
\emph{parametric \acf{EBCI}} will cover 95\%, say, of the true effect
parameters, under repeated sampling of the observed data \emph{and} of the
effect parameters. We refer to this notion of coverage as ``\ac{EB} coverage'',
following the terminology in \citet{morris83}. Unfortunately, we show that, in
the context of a normal means model, the parametric \ac{EBCI} with nominal level
95\% can have actual \ac{EB} coverage as low as 74\% for certain non-normal
effect distributions. The potential undercoverage is increasing in the degree of
shrinkage, and we derive a simple ``rule of thumb'' for gauging the potential
coverage distortion.

To allow easy uncertainty assessment in \ac{EB} applications that is reliable
irrespective of the degree of shrinkage, we construct novel \emph{robust
  \acp{EBCI}} that take a simple form and control \ac{EB} coverage
\emph{regardless} of the true effect distribution. Our baseline model is an
(approximate) normal means problem $Y_i \sim N(\theta_{i}, \sigma_i^2)$,
$i=1, \dotsc, n$. In applications, $Y_i$ represents a preliminary estimate of
the effect $\theta_i$ for unit $i$. Like the parametric \ac{EBCI} that assumes a
normal distribution for $\theta_i$, the robust \ac{EBCI} we propose is centered
at the normality-based \ac{EB} point estimate $\hat{\theta}_i$ that shrinks
$Y_{i}$ toward some baseline value, but it uses a larger critical value to
account for bias due to shrinkage.\footnote{\label{fn:software} Our methods are implemented in the Stata package \texttt{ebreg}, R package \texttt{ebci}, and Matlab package \texttt{ebci\_matlab}, which are available at SSC, CRAN, and GitHub, respectively.} \ac{EB} coverage is controlled in the class of all
distributions for $\theta_{i}$ that satisfy certain moment bounds, which we
estimate consistently from the data (similarly to the parametric \ac{EBCI},
which uses the second moment). We show that the baseline implementation of our
robust \ac{EBCI} is ``adaptive'': its length is close to that
of the parametric \ac{EBCI} when the $\theta_{i}$'s are in fact normally
distributed. Thus, little efficiency is lost from using the robust \ac{EBCI} in
place of the non-robust parametric one.\footnote{\label{fn:nonnormal} If the
  $\theta_{i}$'s are not normally distributed, our robust \acp{EBCI} are valid but may leave room for greater efficiency improvement, as we discuss in \Cref{sec:compar_optim}.}

In addition to controlling \ac{EB} coverage, the robust \acp{EBCI} with level
$1-\alpha$ have a frequentist \emph{average coverage} property: If the means
$\theta_{1}, \dotsc, \theta_{n}$ are treated as \emph{fixed}, the coverage
probability, averaged across the $n$ parameters $\theta_i$, is at least
$1-\alpha$. In fact, under mild conditions, at least a fraction $1-\alpha$ of the $n$ \acp{EBCI} will contain their respective parameters (with high probability as $n \to\infty$). This weakening of the usual requirement of coverage for \emph{each}
parameter $\theta_i$ allows our robust \ac{EBCI} to be shorter than the usual
\ac{CI} centered at the unshrunk estimate $Y_{i}$, and often substantially
so.\footnote{\label{fn:impossibility}Relaxing the usual notion of coverage in some way is necessary to
  obtain intervals that reflect the efficiency improvement of the empirical
  Bayes approach. In particular, the results in \citet{pratt61} imply that for
  \acp{CI} with coverage 95\%, one cannot achieve expected length improvements
  greater than 15\% relative to the usual unshrunk \acp{CI}, even if one happens
  to optimize length for the true parameter vector
  $(\theta_{1}, \dotsc, \theta_{n})$. See, for example, Corollary 3.3 in
  \citet{armstrong_optimal_2018} and the discussion following it.} Intuitively,
the average coverage criterion only requires us to guard against the
\emph{average} coverage distortion induced by the biases of the individual
shrinkage estimators $\hat{\theta}_i$, and the data is quite informative about
whether \emph{most} of these biases are large, even though individual biases are
difficult to estimate. To complement the frequentist properties, our \acp{EBCI}
can be viewed as Bayesian credible sets that are robust to the prior on
$\theta_i$, in terms of \emph{ex ante} coverage.

The average coverage criterion has the same motivation as the usual frequentist
justification of the \ac{EB} \emph{point estimator}: the \ac{EB} point estimator
achieves lower \ac{MSE} on average across units at the expense of potentially
worse performance for some individual units \citep[see, for example,][Ch.
1.3]{Efron2010}. Thus, researchers who use \ac{EB} estimators instead of the
unshrunk $Y_{i}$'s prioritize favorable group performance over protecting
individual performance. Our average coverage intervals make an analogous
tradeoff: they guarantee coverage and achieve short length on average across
units at the expense of giving up on a coverage guarantee for every individual
unit. We examine this tradeoff in more detail in \Cref{sec:compar}.

We caution, however, that the average coverage criterion is typically
inappropriate in applications where shrinkage point estimation is unattractive. This includes settings where one is interested in the magnitude or
the identity of the largest $\theta_{i}$, or the true effect for the largest
observed $Y_{i}$ (as in, for example, \citealp{HuFi19}, or
\citealp{Andrews2021}).\footnote{\label{fn:cutoff_diverge}As we show in
  \Cref{sec:cover_selection}, our methods do extend to settings where we keep a
  subset of units $i$ that exceed a given cutoff. However, we do not allow this
  cutoff to diverge with the sample size, such as when one focuses on the unit $i$ with the single largest
  observed $Y_{i}$.} It also includes settings where a particular effect, say
$\theta_{1}$, is of primary interest, or, more generally, settings where the
effects are not exchangeable, and their ordering is relevant \citep{GrRi19}. Our methods are also not applicable if one is
interested in functionals of the random effects \emph{distribution} (as in
\citealp{bonhomme2020}, or \citealp{ignatiadis2019}), rather than in the effects
themselves.
Finally, the justification for our methods is asymptotic in the number of
parameters $n$. In our Monte Carlo simulations, we find that our \acp{EBCI} have
close to nominal coverage over a range of \acp{DGP} once $n$ is greater than
$100$.

We illustrate our results by computing \acp{EBCI} for the causal effects of
growing up in different U.S. neighborhoods (specifically commuting zones) on
intergenerational mobility. We follow \citet{chetty_impacts_2018}, who apply
\ac{EB} shrinkage to initial fixed effects estimates. Depending on the
specification, we find that the robust \acp{EBCI} are on average 12--25\% as
long as the unshrunk \acp{CI}.

Our underlying ideas extend to other linear and non-linear shrinkage settings
with possibly non-Gaussian data. For example, our techniques allow for the
construction of robust \acp{EBCI} that contain (nonlinear) soft thresholding
estimators, as well as average coverage confidence bands for nonparametric
regression functions.

The average coverage criterion was originally introduced in the literature on
nonparametric regression (\citealp{Wahba1983}; \citealp{Nychka1988};
\citealp[Ch. 5.8]{Wasserman2006}). \citet{Cai2015} construct adaptive average
coverage confidence bands. These procedures are challenging to implement in our
\ac{EB} setting, and lack a clear finite-sample justification, unlike our
procedure. \citet{Liu2019} construct forecast intervals in a dynamic panel data
model that guarantee average coverage in a Bayesian sense (for a fixed prior).
We discuss alternative approaches to inference in \ac{EB} settings in
\Cref{sec:compar}.

The rest of this paper is organized as follows. \Cref{sec:simple-example}
illustrates our methods in the context of a simple homoskedastic Gaussian model.
\Cref{sec:pract-impl} presents our recommended baseline procedure and discusses
practical implementation issues. \Cref{sec:general-results} presents our main
results on the coverage and efficiency of the robust \ac{EBCI}, and on the
coverage distortions of the parametric \ac{EBCI}; we also verify the
finite-sample coverage accuracy of the robust \ac{EBCI} through extensive
simulations. \Cref{sec:compar} compares our \ac{EBCI} with other inference
approaches. \Cref{sec:extensions} discusses extensions of the basic framework.
\Cref{sec:empir-appl} contains an empirical application to inference on
neighborhood effects.
\Cref{sec:moment_estimates,sec:comput,coverage_results_sec_append} give details
on finite-sample corrections, computational details, and formal asymptotic
coverage results. The Online Supplement contains proofs as well as further
technical results. Applied readers are encouraged to focus on
\Cref{sec:simple-example,sec:pract-impl,sec:empir-appl}.

\section{Simple example}\label{sec:simple-example}

This section illustrates the construction of the robust \acp{EBCI} that we
propose in a simplified setting with no covariates and with known, homoskedastic
errors. \Cref{sec:pract-impl} relaxes these restrictions, and discusses other
empirically relevant extensions of the basic framework, as well as
implementation issues.

We observe $n$ estimates $Y_{i}$ of elements of the parameter vector
$\theta=(\theta_{1}, \dotsc, \theta_{n})'$. Each estimate is normally
distributed with common, known variance $\sigma^{2}$,
\begin{equation}\label{eq:homoskedastic-normal-means}
  Y_{i}\mid \theta \sim N(\theta_{i}, \sigma^{2}), \qquad i=1, \dotsc, n.
\end{equation}
In many applications, the $Y_{i}$'s arise as preliminary least squares estimates
of the parameters $\theta_{i}$. For instance, they may correspond to fixed
effect estimates of teacher or school value added, neighborhood effects, or firm
and worker effects. In such cases, $Y_{i}$ will only be \emph{approximately}
normal in large samples by the \ac{CLT}; we take this explicitly into account in
the theory in \Cref{coverage_results_sec_append}.

A popular approach to estimation that substantially improves upon the raw
estimator $\hat{\theta}_{i}=Y_{i}$ under the compound \ac{MSE}
$\sum_{i=1}^{n}E[(\hat{\theta}_{i}-\theta_{i})^{2}]$ is based on \acf{EB}
shrinkage. In particular, suppose that the $\theta_{i}$'s are themselves
normally distributed,
\begin{equation}\label{eq:normal-distr}
  \theta_{i}\sim N(0, \mu_{2}).
\end{equation}
Our discussion below applies if \Cref{eq:normal-distr} is viewed as a subjective
Bayesian prior distribution for a single parameter $\theta_i$, but for
concreteness we will think of \Cref{eq:normal-distr} as a ``random effects''
sampling distribution for the $n$ mean parameters $\theta_1, \dotsc, \theta_n$.
Under \Cref{eq:normal-distr}, it is optimal to estimate $\theta_i$ using the
posterior mean $\hat{\theta}_i=w_{EB}Y_i$, where
$w_{EB}=1-\sigma^{2}/(\sigma^{2}+\mu_{2})$. To avoid having to specify the
variance $\mu_{2}$, the \ac{EB} approach treats it as an unknown parameter, and
replaces the marginal precision of $Y_{i}$, $1/(\sigma^{2}+\mu_{2})$, with a
method of moments estimate $n/\sum_{i=1}^{n}Y_{i}^{2}$, or the
degrees-of-freedom adjusted estimate $(n-2)/\sum_{i=1}^{n}Y_{i}^{2}$. The latter
leads to the classic estimator of \citet{JaSt61},
$\hat{w}_{EB}=1-\sigma^2(n-2)/\sum_{i=1}^{n}Y_{i}^{2}$.

One can also use~\Cref{eq:normal-distr} to construct \acp{CI} for the
$\theta_{i}$'s. In particular, since the marginal distribution of
$w_{EB}Y_{i}-\theta_{i}$ is normal with mean zero and variance
$(1-w_{EB})^{2}\mu_{2}+w_{EB}^{2}\sigma^{2}=w_{EB}\sigma^{2}$, this leads to the
$1-\alpha$ \ac{CI}
\begin{equation}\label{eq:parametric-ebci}
  w_{EB}Y_{i}\pm z_{1-\alpha/2}w_{EB}^{1/2}\sigma,
\end{equation}
where $z_{\alpha}$ is the $\alpha$ quantile of the standard normal distribution.
Since the form of the interval is motivated by the parametric
assumption~\eqref{eq:normal-distr}, we refer to it as a parametric \ac{EBCI}.
With $\mu_{2}$ unknown, one can replace $w_{EB}$ by
$\hat{w}_{EB}$.\footnote{Alternatively, to account for estimation error in
  $\hat{w}_{EB}$, \citet{morris83} suggests adjusting the variance estimate
  $\hat{w}_{EB}\sigma^{2}$ to
  $\hat{w}_{EB}\sigma^{2}+2Y_{i}^{2}(1-\hat{w}_{EB})^{2}/(n-2)$. The adjustment
  does not matter asymptotically.} This is asymptotically equivalent
to~\eqref{eq:parametric-ebci} as $n\to\infty$.

The coverage of the parametric \ac{EBCI} in~\eqref{eq:parametric-ebci} is
$1-\alpha$ under repeated sampling of $(Y_{i}, \theta_{i})$ according to
\Cref{eq:homoskedastic-normal-means,eq:normal-distr}. To distinguish this notion
of coverage from the case with fixed $\theta$, we refer to coverage under
repeated sampling of $(Y_{i}, \theta_{i})$ as ``empirical Bayes coverage''. This
follows the definition of an \acf{EBCI} in \citet[Eq.\ 3.6]{morris83} and
\citet[Ch. 3.5]{CaLo00}. Unfortunately, this coverage property relies heavily on
the parametric assumption~\eqref{eq:normal-distr}. We show in
\Cref{sec:param-ebci-cover} that the actual \ac{EB} coverage of the nominal
$1-\alpha$ parametric \ac{EBCI} can be as low as $1-1/\max\{z_{1-\alpha/2},1\}$
for certain non-normal distributions of $\theta_{i}$ with variance $\mu_{2}$;
for 95\% \acp{EBCI}, this evaluates to 74\%. This contrasts with existing
results on estimation: although the \ac{EB} estimator is motivated by the
parametric assumption~\eqref{eq:normal-distr}, it performs well even if this
assumption is dropped, with low \ac{MSE} even if we treat $\theta$ as fixed.

This paper constructs an \ac{EBCI} with a similar robustness property: the
interval will be close in length to the parametric \ac{EBCI} when
\Cref{eq:normal-distr} holds, but its \ac{EB} coverage is at least $1-\alpha$
without any parametric assumptions on the distribution of $\theta_{i}$. To
describe the construction, suppose that all that is known is that $\theta_{i}$
is sampled from a distribution with second moment given by $\mu_{2}$ (in
practice, we can replace $\mu_2$ by the consistent estimate
$n^{-1}\sum_{i=1}^{n}Y^{2}_{i}-\sigma^{2}$). Conditional on $\theta_{i}$, the
estimator $w_{EB}Y_{i}$ has bias $(w_{EB}-1)\theta_{i}$ and variance
$w_{EB}^{2}\sigma^{2}$, so that the $t$-statistic
$(w_{EB}Y_{i}-\theta_{i})/w_{EB}\sigma$ is normally distributed with mean
$b_{i}=(1-1/w_{EB})\theta_{i}/\sigma$ and variance $1$. Therefore, if we use a
critical value $\chi$, the non-coverage of the \ac{CI}
$w_{EB}Y_{i}\pm \chi w_{EB}\sigma$, conditional on $\theta_{i}$, will be given
by the probability
\begin{equation}\label{eq:rb}
  r(b_{i}, \chi)=P(\abs{Z-b_{i}}\geq \chi\mid
  \theta_{i})=\Phi(-\chi-b_{i})+\Phi(-\chi+b_{i}),
\end{equation}
where $Z$ denotes a standard normal random variable, and $\Phi$ denotes its cdf. Thus, by iterated expectations, under repeated sampling of
$\theta_{i}$, the non-coverage is bounded by
\begin{equation}\label{eq:non-coverage-bound}
  \rho(\sigma^{2}/\mu_{2}, \chi)=\sup_{F}E_{F}[r(b, \chi)]\quad
  \text{s.t.}\quad E_{F}[b^{2}]=\frac{(1-1/w_{EB})^{2}}{\sigma^{2}}\mu_{2}
  =\frac{\sigma^{2}}{\mu_{2}},
\end{equation}
where $E_{F}$ denotes expectation under $b\sim F$. Although this is an
infinite-dimensional optimization problem over the space of distributions, it
turns out that it admits a simple closed-form solution, which we give in
\Cref{theorem:bound-second-moment} in \Cref{sec:comput}. Moreover, because the
optimization is a linear program, it can be solved even in the more general
settings of applied relevance that we consider in \Cref{sec:pract-impl}.

Set $\chi=\cva_{\alpha}(\sigma^{2}/\mu_{2})$, where
$\cva_{\alpha}(t)=\rho^{-1}(t, \alpha)$, and the inverse is with respect to
the second argument. Then the resulting interval
\begin{equation}\label{eq:robust-ebci}
  w_{EB}Y_{i}\pm \cva_{\alpha}(\sigma^{2}/\mu_{2})w_{EB}\sigma
\end{equation}
will maintain coverage $1-\alpha$ among all distributions of $\theta_{i}$ with
$E[\theta_{i}^{2}]=\mu_{2}$ (recall that we estimate $\mu_{2}$ consistently from
the data). For this reason, we refer to it as a robust \ac{EBCI}.
\Cref{fig:cva05} in \Cref{sec:baseline-model} gives a plot of the critical values for $\alpha=0.05$. We show
in \Cref{sec:efficiency} below that by also imposing a constraint on the fourth
moment of $\theta_{i}$, in addition to the second moment constraint, one can
construct a robust \ac{EBCI} that ``adapts'' to the Gaussian case in the sense
that its
length will be close to that of the parametric \ac{EBCI} in \Cref{eq:parametric-ebci} if these moment constraints are compatible with a normal distribution.

Instead of considering \ac{EB} coverage, one may alternatively wish to assess
uncertainty associated with the estimates $\hat{\theta}_i=w_{EB}Y_{i}$ when $\theta$ is treated
as fixed. In this case, the \ac{EBCI} in \Cref{eq:robust-ebci} has an average
coverage guarantee that
\begin{equation} \label{eq:simple-aci} \frac{1}{n}\sum_{i=1}^{n} P\big(\theta_i
  \in [w_{EB}Y_{i}\pm \cva_{\alpha}(\sigma^{2}/\mu_{2})w_{EB}\sigma] \;\big|\;
  \theta \big)\geq 1-\alpha,
\end{equation}
provided that the moment constraint can be interpreted as a constraint on the
empirical second moment on the $\theta_{i}$'s,
$n^{-1}\sum_{i=1}^{n}\theta_{i}^{2}=\mu_{2}$. In other words, if we condition on
$\theta$, then the coverage is at least $1-\alpha$ on average across the $n$
\acp{EBCI} for $\theta_{1}, \dotsc, \theta_{n}$. To see this, note that the
average non-coverage of the intervals is bounded
by~\eqref{eq:non-coverage-bound}, except that the supremum is only taken over
possible empirical distributions for $\theta_{1}, \dotsc, \theta_{n}$ satisfying
the moment constraint. Since this supremum is necessarily smaller than
$\rho(\sigma^{2}/\mu_{2}, \chi)$, it follows that the average coverage is at
least $1-\alpha$. In fact, if the $Y_i$'s exhibit limited dependence across $i$, a stronger property holds: the probability that at least a fraction $1-\alpha$ of the $n$ \acp{EBCI} contain their respective true parameters converges to 1 as $n \to \infty$, cf.\ \Cref{rem:ac_alt_def_remark} below.

The usual \acp{CI} $Y_{i}\pm z_{1-\alpha/2}\sigma$ also of course achieve
average coverage $1-\alpha$. The robust \ac{EBCI} in \Cref{eq:robust-ebci} will,
however, be shorter, especially when $\mu_{2}$ is small relative to
$\sigma^{2}$---see \Cref{fig:efficiency_unshrunk} below. The reduction in length
is achieved by weakening the requirement that each \ac{CI} covers its true
parameter $1-\alpha$ percent of the time to the requirement that the coverage probability equal
$1-\alpha$ on average across the \acp{CI}.
It may seem surprising that we can construct a narrower \ac{CI} by centering it
at the shrinkage estimates $w_{EB}Y_{i}$. The intuition for this is that the
shrinkage reduces the variability of the estimates, at the expense of
introducing bias in the estimates. The bias necessitates the use of a larger
critical value $\cva_{\alpha}(\sigma^{2}/\mu_{2})$. Because under the average
coverage criterion we only need to control the bias \emph{on average} across
$i$, rather than for each individual $\theta_{i}$, this increase in the critical
value is smaller than the reduction in the standard error.

\section{Practical implementation}\label{sec:pract-impl}

We now describe how to compute a robust \ac{EBCI} that allows for
heteroskedasticity, shrinks towards more general regression estimates rather
than towards zero, and exploits higher moments of the bias to yield a narrower
interval. In \Cref{sec:baseline-model}, we describe the empirical Bayes model
that motivates our baseline approach. \Cref{sec:baseline-implementation}
describes the practical implementation of our baseline approach.

\subsection{Motivating model and robust EBCI}\label{sec:baseline-model}

In applied settings, the unshrunk estimates $Y_i$ will typically have
heteroskedastic variances. Furthermore, rather than shrinking towards zero, it
is common to shrink toward an estimate of $\theta_i$ based on some covariates
$X_i$, such as a regression estimate $X_i'\hat{\delta}$. We now describe how to
adapt the ideas in \Cref{sec:simple-example} to such settings.

Consider a generalization of the model in \Cref{eq:homoskedastic-normal-means}
that allows for heteroskedasticity and covariates,
\begin{equation} \label{eq:hierarch_y}
  Y_{i}\mid \theta_{i}, X_{i}, \sigma_{i} \sim N(\theta_{i},
  \sigma_{i}^2), \qquad i=1,\dotsc, n.
\end{equation}
The covariate vector $X_i$ may contain just the intercept, and it may also
contain (functions of) $\sigma_{i}$. To construct an \ac{EB} estimator of
$\theta_{i}$, consider the working assumption that the sampling distribution of
the $\theta_i$'s is conditionally normal:
\begin{equation} \label{eq:hierarch_theta} \theta_i\mid X_{i}, \sigma_{i} \sim
  N(\mu_{1,i}, \mu_{2}), \quad \text{where} \quad \mu_{1,i}=X_i'\delta.
\end{equation}
The hierarchical model~\eqref{eq:hierarch_y}--\eqref{eq:hierarch_theta} leads to
the Bayes estimate $\hat{\theta}_{i}=\mu_{1, i}+w_{EB, i}(Y_i-\mu_{1, i})$,
where $w_{EB, i}=\frac{\mu_{2}}{\mu_{2}+\sigma_{i}^{2}}$. This estimate shrinks
the unrestricted estimate $Y_{i}$ of $\theta_{i}$ toward
$\mu_{1,i}=X_{i}'\delta$. In contrast to~\eqref{eq:hierarch_y}, the normality
assumption~\eqref{eq:hierarch_theta} typically cannot be justified simply by
appealing to the \ac{CLT}; the linearity of the conditional mean
$\mu_{1,i}=X_i'\delta$ may also be suspect. Our robust \ac{EBCI} will therefore
be constructed so that it achieves valid \ac{EB} coverage even if
assumption~\eqref{eq:hierarch_theta} fails. To obtain a narrow robust \ac{EBCI},
we augment the second moment restriction used to compute the critical value in
\Cref{eq:non-coverage-bound} with restrictions on higher moments of the bias of
$\hat{\theta}_{i}$. In our baseline specification, we add a restriction on the
fourth moment.

In particular, we replace assumption~\eqref{eq:hierarch_theta} with the much
weaker requirement that the conditional second moment and kurtosis of
$\varepsilon_i=\theta_{i}-X_{i}'\delta$ do not depend on $(X_{i}, \sigma_{i})$:
\begin{align}\label{eq:moment_independence}
  E[(\theta_i-X_i'\delta)^{2} \mid X_{i}, \sigma_i]&=\mu_{2}, &
  E[(\theta_i-X_i'\delta)^{4} \mid X_{i}, \sigma_i]/\mu^{2}_{2}&=\kappa,
\end{align}
where $\delta$ is defined as the probability limit of the regression estimate
$\hat{\delta}$.\footnote{Our framework can be modified to let $(X_i, \sigma_i)$
  be fixed, in which case $\delta$ depends on $n$. See the discussion following
  \Cref{thm:coverage_baseline} below.} We discuss this requirement further in
\Cref{rem:conditional_coverage} below, and we relax it in \Cref{rem:nonparam}
below.

We now apply analysis analogous to that in \Cref{sec:simple-example}. Let us
suppose for simplicity that $\delta$, $\mu_{2}$, $\kappa$, and $\sigma_{i}$ are
known; we relax this assumption in \Cref{sec:baseline-implementation} below, and
in the theory in \Cref{sec:general-results}. Denote the conditional bias of
$\hat{\theta}_i$ normalized by the standard error by
$b_{i} = (w_{EB, i}-1)\varepsilon_i/(w_{EB, i}\sigma_i) =
-\sigma_{i}\varepsilon_{i}/\mu_{2}$. Under repeated sampling of $\theta_i$, the
non-coverage of the \ac{CI} $\hat{\theta}_i \pm \chi w_{EB, i}\sigma$,
conditional on $(X_{i}, \sigma_{i})$, depends on the distribution of the
normalized bias $b_{i}$, as in \Cref{sec:simple-example}. Given the moments
$\mu_{2}$ and $\kappa$, the \emph{maximal} non-coverage is given by
\begin{equation}\label{eq:fourth_moment_bound}
  \rho(m_{2,i}, \kappa, \chi)=\sup_{F}E_{F}[r(b, \chi)]\quad
  \text{s.t.}\quad E_{F}[b^{2}]=m_{2,i}, \, E_{F}[b^{4}]=\kappa m_{2,i}^2,
\end{equation}
where $b$ is distributed according to the distribution $F$. Here
$m_{2,i} = E[b_{i}^{2} \mid X_{i}, \sigma_{i}] =\sigma^{2}_{i}/\mu_{2}$. Observe
that the kurtosis of $b_{i}$ matches that of $\varepsilon_{i}$.
\Cref{sec:comput} shows that the infinite-dimensional linear
program~\eqref{eq:fourth_moment_bound} can be reduced to two nested
\emph{univariate} optimizations. We also show that the least favorable
distribution---the distribution $F$
maximizing~\eqref{eq:fourth_moment_bound}---is a discrete distribution with up
to 4 support points (see \Cref{remark:lf-distro}).

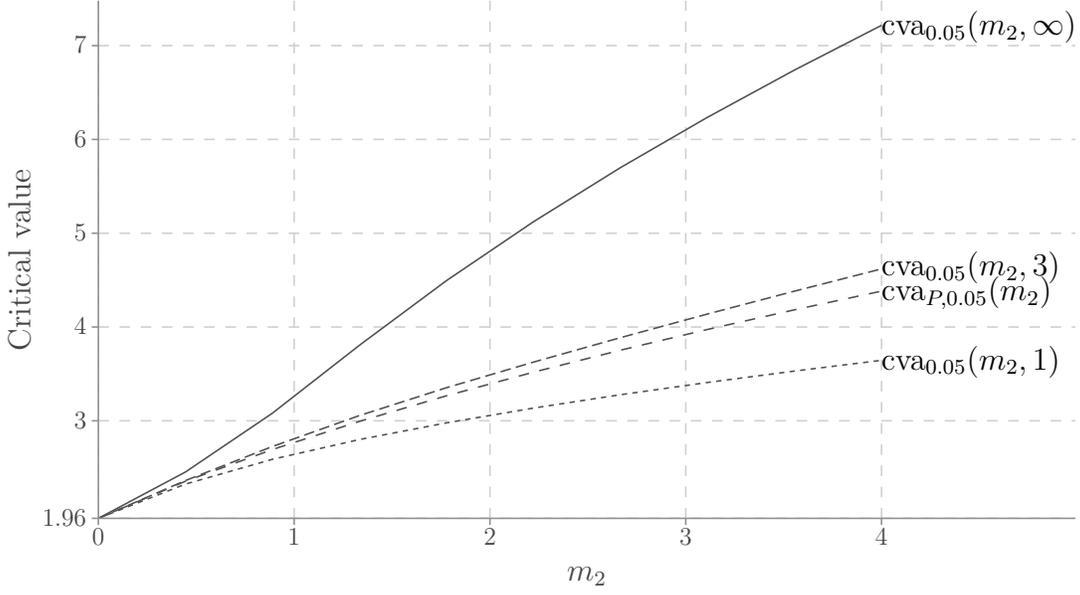
\begin{figure}[t]
  \centering{
\begin{tikzpicture}[x=1pt,y=1pt]
\definecolor{fillColor}{RGB}{255,255,255}
\path[use as bounding box,fill=fillColor,fill opacity=0.00] (0,0) rectangle (411.94,231.26);
\begin{scope}
\path[clip] (  0.00,  0.00) rectangle (411.94,231.26);
\definecolor{drawColor}{RGB}{255,255,255}
\definecolor{fillColor}{RGB}{255,255,255}

\path[draw=drawColor,line width= 0.6pt,line join=round,line cap=round,fill=fillColor] (  0.00,  0.00) rectangle (411.94,231.26);
\end{scope}
\begin{scope}
\path[clip] ( 37.55, 30.40) rectangle (406.94,226.26);
\definecolor{fillColor}{RGB}{255,255,255}

\path[fill=fillColor] ( 37.55, 30.40) rectangle (406.94,226.26);
\definecolor{drawColor}{RGB}{208,208,208}

\path[draw=drawColor,line width= 0.6pt,dash pattern=on 4pt off 4pt ,line join=round] ( 37.55, 30.40) --
	(406.94, 30.40);

\path[draw=drawColor,line width= 0.6pt,dash pattern=on 4pt off 4pt ,line join=round] ( 37.55, 67.31) --
	(406.94, 67.31);

\path[draw=drawColor,line width= 0.6pt,dash pattern=on 4pt off 4pt ,line join=round] ( 37.55,102.79) --
	(406.94,102.79);

\path[draw=drawColor,line width= 0.6pt,dash pattern=on 4pt off 4pt ,line join=round] ( 37.55,138.28) --
	(406.94,138.28);

\path[draw=drawColor,line width= 0.6pt,dash pattern=on 4pt off 4pt ,line join=round] ( 37.55,173.77) --
	(406.94,173.77);

\path[draw=drawColor,line width= 0.6pt,dash pattern=on 4pt off 4pt ,line join=round] ( 37.55,209.26) --
	(406.94,209.26);

\path[draw=drawColor,line width= 0.6pt,dash pattern=on 4pt off 4pt ,line join=round] ( 37.55, 30.40) --
	( 37.55,226.26);

\path[draw=drawColor,line width= 0.6pt,dash pattern=on 4pt off 4pt ,line join=round] (111.58, 30.40) --
	(111.58,226.26);

\path[draw=drawColor,line width= 0.6pt,dash pattern=on 4pt off 4pt ,line join=round] (185.60, 30.40) --
	(185.60,226.26);

\path[draw=drawColor,line width= 0.6pt,dash pattern=on 4pt off 4pt ,line join=round] (259.63, 30.40) --
	(259.63,226.26);

\path[draw=drawColor,line width= 0.6pt,dash pattern=on 4pt off 4pt ,line join=round] (333.65, 30.40) --
	(333.65,226.26);
\definecolor{drawColor}{RGB}{81,81,81}

\path[draw=drawColor,line width= 0.6pt,line join=round] ( 37.55, 30.40) --
	( 70.45, 47.92) --
	(103.35, 70.18) --
	(136.25, 95.88) --
	(169.15,120.34) --
	(202.05,142.62) --
	(234.95,163.10) --
	(267.85,182.15) --
	(300.75,200.03) --
	(333.65,216.94);

\path[draw=drawColor,line width= 0.6pt,dash pattern=on 2pt off 2pt ,line join=round] ( 37.55, 30.40) --
	( 70.45, 43.35) --
	(103.35, 52.74) --
	(136.25, 60.21) --
	(169.15, 66.53) --
	(202.05, 72.12) --
	(234.95, 77.17) --
	(267.85, 81.81) --
	(300.75, 86.13) --
	(333.65, 90.19);

\path[draw=drawColor,line width= 0.6pt,dash pattern=on 4pt off 2pt ,line join=round] ( 37.55, 30.40) --
	( 70.45, 44.69) --
	(103.35, 57.58) --
	(136.25, 69.29) --
	(169.15, 79.82) --
	(202.05, 89.51) --
	(234.95, 98.77) --
	(267.85,107.71) --
	(300.75,116.37) --
	(333.65,124.78);

\path[draw=drawColor,line width= 0.6pt,dash pattern=on 4pt off 4pt ,line join=round] ( 37.55, 30.40) --
	( 70.45, 44.44) --
	(103.35, 56.44) --
	(136.25, 67.09) --
	(169.15, 76.77) --
	(202.05, 85.70) --
	(234.95, 94.03) --
	(267.85,101.87) --
	(300.75,109.30) --
	(333.65,116.37);
\end{scope}
\begin{scope}
\path[clip] ( 37.55, 30.40) rectangle (406.94,226.26);
\definecolor{drawColor}{RGB}{0,0,0}

\node[text=drawColor,anchor=base west,inner sep=0pt, outer sep=0pt, scale=  1.00] at (333.65, 86.75) {$\mathrm{cva}_{0.05}(m_{2}, 1)$};

\node[text=drawColor,anchor=base west,inner sep=0pt, outer sep=0pt, scale=  1.00] at (333.65,112.93) {$\mathrm{cva}_{P, 0.05}(m_{2})$};

\node[text=drawColor,anchor=base west,inner sep=0pt, outer sep=0pt, scale=  1.00] at (333.65,122.32) {$\mathrm{cva}_{0.05}(m_{2}, 3)$};

\node[text=drawColor,anchor=base west,inner sep=0pt, outer sep=0pt, scale=  1.00] at (333.65,213.49) {$\mathrm{cva}_{0.05}(m_{2}, \infty)$};
\end{scope}
\begin{scope}
\path[clip] (  0.00,  0.00) rectangle (411.94,231.26);
\definecolor{drawColor}{RGB}{144,144,144}

\path[draw=drawColor,line width= 0.6pt,line join=round] ( 37.55, 30.40) --
	( 37.55,226.26);
\end{scope}
\begin{scope}
\path[clip] (  0.00,  0.00) rectangle (411.94,231.26);
\definecolor{drawColor}{RGB}{68,68,68}

\node[text=drawColor,anchor=base east,inner sep=0pt, outer sep=0pt, scale=  0.80] at ( 33.05, 27.64) {1.96};

\node[text=drawColor,anchor=base east,inner sep=0pt, outer sep=0pt, scale=  0.80] at ( 33.05, 64.55) {3};

\node[text=drawColor,anchor=base east,inner sep=0pt, outer sep=0pt, scale=  0.80] at ( 33.05,100.04) {4};

\node[text=drawColor,anchor=base east,inner sep=0pt, outer sep=0pt, scale=  0.80] at ( 33.05,135.53) {5};

\node[text=drawColor,anchor=base east,inner sep=0pt, outer sep=0pt, scale=  0.80] at ( 33.05,171.02) {6};

\node[text=drawColor,anchor=base east,inner sep=0pt, outer sep=0pt, scale=  0.80] at ( 33.05,206.50) {7};
\end{scope}
\begin{scope}
\path[clip] (  0.00,  0.00) rectangle (411.94,231.26);
\definecolor{drawColor}{RGB}{144,144,144}

\path[draw=drawColor,line width= 0.6pt,line join=round] ( 35.05, 30.40) --
	( 37.55, 30.40);

\path[draw=drawColor,line width= 0.6pt,line join=round] ( 35.05, 67.31) --
	( 37.55, 67.31);

\path[draw=drawColor,line width= 0.6pt,line join=round] ( 35.05,102.79) --
	( 37.55,102.79);

\path[draw=drawColor,line width= 0.6pt,line join=round] ( 35.05,138.28) --
	( 37.55,138.28);

\path[draw=drawColor,line width= 0.6pt,line join=round] ( 35.05,173.77) --
	( 37.55,173.77);

\path[draw=drawColor,line width= 0.6pt,line join=round] ( 35.05,209.26) --
	( 37.55,209.26);
\end{scope}
\begin{scope}
\path[clip] (  0.00,  0.00) rectangle (411.94,231.26);
\definecolor{drawColor}{RGB}{144,144,144}

\path[draw=drawColor,line width= 0.6pt,line join=round] ( 37.55, 30.40) --
	(406.94, 30.40);
\end{scope}
\begin{scope}
\path[clip] (  0.00,  0.00) rectangle (411.94,231.26);
\definecolor{drawColor}{RGB}{144,144,144}

\path[draw=drawColor,line width= 0.6pt,line join=round] ( 37.55, 27.90) --
	( 37.55, 30.40);

\path[draw=drawColor,line width= 0.6pt,line join=round] (111.58, 27.90) --
	(111.58, 30.40);

\path[draw=drawColor,line width= 0.6pt,line join=round] (185.60, 27.90) --
	(185.60, 30.40);

\path[draw=drawColor,line width= 0.6pt,line join=round] (259.63, 27.90) --
	(259.63, 30.40);

\path[draw=drawColor,line width= 0.6pt,line join=round] (333.65, 27.90) --
	(333.65, 30.40);
\end{scope}
\begin{scope}
\path[clip] (  0.00,  0.00) rectangle (411.94,231.26);
\definecolor{drawColor}{RGB}{68,68,68}

\node[text=drawColor,anchor=base,inner sep=0pt, outer sep=0pt, scale=  0.80] at ( 37.55, 20.39) {0};

\node[text=drawColor,anchor=base,inner sep=0pt, outer sep=0pt, scale=  0.80] at (111.58, 20.39) {1};

\node[text=drawColor,anchor=base,inner sep=0pt, outer sep=0pt, scale=  0.80] at (185.60, 20.39) {2};

\node[text=drawColor,anchor=base,inner sep=0pt, outer sep=0pt, scale=  0.80] at (259.63, 20.39) {3};

\node[text=drawColor,anchor=base,inner sep=0pt, outer sep=0pt, scale=  0.80] at (333.65, 20.39) {4};
\end{scope}
\begin{scope}
\path[clip] (  0.00,  0.00) rectangle (411.94,231.26);
\definecolor{drawColor}{RGB}{68,68,68}

\node[text=drawColor,anchor=base,inner sep=0pt, outer sep=0pt, scale=  1.00] at (222.24,  6.94) {$m_{2}$};
\end{scope}
\begin{scope}
\path[clip] (  0.00,  0.00) rectangle (411.94,231.26);
\definecolor{drawColor}{RGB}{68,68,68}

\node[text=drawColor,rotate= 90.00,anchor=base,inner sep=0pt, outer sep=0pt, scale=  1.00] at ( 11.89,128.33) {Critical value};
\end{scope}
\end{tikzpicture}}
  \caption{Function $\cva_{\alpha}(m_{2}, \kappa)$ for $\alpha=0.05$ and
    selected values of $\kappa$. The function $\cva_{\alpha}(m_{2})$, defined in
    \Cref{sec:simple-example}, that only imposes a constraint on the second
    moment, corresponds to $\cva_{\alpha}(m_{2}, \infty)$. The function
    $\cva_{P, \alpha}(m_{2})=z_{1-\alpha/2}\sqrt{1+m_{2}}$ corresponds to the
    critical value under the assumption that $\theta_{i}$ is normally
    distributed.}\label{fig:cva05}
\end{figure}

Define the critical value
$\cva_{\alpha}(m_{2,i}, \kappa)=\rho^{-1}(m_{2,i}, \kappa, \alpha)$, where the
inverse is in the last argument. \Cref{fig:cva05} plots this function for
$\alpha=0.05$ and selected values of $\kappa$. This leads to the robust
\ac{EBCI}
\begin{equation}\label{eq:conditional_ci}
  \hat{\theta}_{i} \pm \cva_{\alpha}(m_{2, i}, \kappa)w_{EB, i}\sigma_{i},
\end{equation}
which, by construction, has coverage at least $1-\alpha$ under repeated sampling
of $(Y_i, \theta_i)$, conditional on $(X_{i}, \sigma_{i})$, so long as
\Cref{eq:moment_independence} holds; it is not required
that~\eqref{eq:hierarch_theta} holds. Note that both the critical value and the
\ac{CI} length are increasing in $\sigma_{i}$.

\subsection{Baseline implementation}\label{sec:baseline-implementation}

Our baseline implementation of the robust \ac{EBCI} plugs in consistent
estimates of the unknown quantities in \Cref{eq:conditional_ci}, based on the
data $\{Y_{i}, X_{i}, \hat{\sigma}_{i}\}_{i=1}^{n}$, where $\hat{\sigma}_{i}$ is a consistent estimate of $\sigma_{i}$ (such as the standard error of the
preliminary estimate $Y_{i}$), and $X_{i}$ is a vector of covariates that are
thought to help predict $\theta_{i}$.

\begin{enumerate}
\item\label{item:param-estimates} Regress $Y_i$ on $X_i$ to obtain the fitted
  values $X_i'\hat\delta$, with
  $\hat\delta=(\sum_{i=1}^{n}\omega_i X_{i}X_{i}')^{-1}\sum_{i=1}^n\omega_{i}
  X_{i}Y_{i}$ denoting the weighted least squares estimate with precision
  weights $\omega_i$. Two natural choices are setting
  $\omega_{i}=\hat\sigma_i^{-2}$, or setting $\omega_{i}=1/n$ for unweighted
  estimates; see \Cref{sec:weighting} for further discussion. Let
  $\hat\mu_{2}=\max\left\{\frac{\sum_{i=1}^n
      \omega_i(\hat{\varepsilon}_i^2-\hat\sigma_i^2)}{\sum_{i=1}^n\omega_i},
    \frac{2\sum_{i=1}^n\omega_i^2\hat\sigma_i^4}{\sum_{i=1}^n\omega_i \cdot
      \sum_{i=1}^n \omega_i\hat\sigma_i^{2}} \right\}$, and
  $\hat{\kappa}=\max\left\{\frac{\sum_{i=1}^{n}
      \omega_i(\hat{\varepsilon}_{i}^{4}-6\hat{\sigma}_i^2\hat{\varepsilon}_{i}^{2}
      +3\hat{\sigma}_i^4)}{\hat{\mu}_{2}^{2}\sum_{i=1}^n\omega_i}, 1 + \frac{32
      \sum_{i=1}^n
      \omega_i^2\hat\sigma_i^{8}}{\hat\mu_2^2\sum_{i=1}^n\omega_i\cdot
      \sum_{i=1}^n\omega_i\hat\sigma_i^4} \right\}$, where
  $\hat\varepsilon_i=Y_i-X_i'\hat\delta$.
\item\label{item:1} Form the \ac{EB} estimate
  \begin{equation*}
    \hat\theta_i= X_i'\hat\delta + \hat{w}_{EB, i}(Y_i-X_i'\hat\delta),
    \quad \text{where} \quad
    \hat{w}_{EB, i} = \frac{\hat\mu_{2}}{\hat{\mu}_{2} + \hat{\sigma}_{i}^{2}}.
  \end{equation*}
\item Compute the critical value
  $\cva_{\alpha}(\hat\sigma_i^{2}/\hat\mu_{2}, \hat{\kappa})$ defined
  below~\Cref{eq:fourth_moment_bound}.
\item Report the robust \ac{EBCI}
  \begin{equation}\label{eq:ebci_under_indep}
    \hat\theta_i \pm \cva_{\alpha}(\hat\sigma_i^{2}/\hat{\mu}_{2},
    \hat{\kappa})\hat{w}_{EB, i}\hat\sigma_{i}.
  \end{equation}
\end{enumerate}
We provide fast and stable software packages that automate these
steps (see \cref{fn:software}). We now discuss the assumptions needed for
validity of the robust \ac{EBCI}.

\begin{remark}[Conditional \ac{EB} coverage and moment
  independence]\label{rem:conditional_coverage}
  A potential concern about \ac{EB} coverage in a heteroskedastic setting is
  that in order to reduce the length of the \ac{CI} on average, one could choose to overcover
  parameters $\theta_{i}$ with small $\sigma_{i}$ and undercover parameters
  $\theta_{i}$ with large $\sigma_{i}$. Our robust \ac{EBCI} ensures that this
  does not happen by requiring \ac{EB} coverage to hold conditional on
  $(X_{i}, \sigma_{i})$. This also avoids analogous coverage concerns as a
  result of the value of $X_{i}$.

  The key to ensuring this property is assumption~\eqref{eq:moment_independence}
  that the conditional second moment and kurtosis of
  $\varepsilon_{i}=\theta_{i}-X_{i}'\delta$ do not depend on
  $(X_{i}, \sigma_{i})$. Conditional moment independence assumptions of this
  form are common in the literature. For instance, it is imposed in the analysis
  of neighborhood effects in \citet{chetty_impacts_2018} (their approach
  requires independence of the second moment), which is the basis for our
  empirical application in \Cref{sec:empir-appl}. Nonetheless, such conditions
  may be strong in some settings, as argued by \citet{xie_sure_2012} in the
  context of \ac{EB} point estimation. In \Cref{rem:nonparam} below, we drop
  condition~\eqref{eq:moment_independence} entirely by replacing $\hat\mu_{2}$
  and $\hat\kappa$ with nonparametric estimates of these conditional moments;
  alternatively, one could relax it by using a flexible parametric
  specification.\footnote{\label{fn:t-stat_shrinkage}Another way to drop
    condition~\eqref{eq:moment_independence} is to base shrinkage on the
    $t$-statistics $Y_i/\sigma_i$, applying the baseline implementation above
    with $Y_i/\hat\sigma_i$ in place of $Y_i$ and 1 in place of
    $\hat{\sigma}_i$. Then the homoskedastic analysis in
    \Cref{sec:simple-example} applies, leading to valid \acp{EBCI} without any
    assumptions about independence of the moments. See Remark 3.8 and Appendix
    D.1 in \citet{akp20v2} for further discussion.}
\end{remark}

\begin{remark}[Nonparametric moment estimates]\label{rem:nonparam}
  As a robustness check to guard against failure of the moment independence
  assumption~\eqref{eq:moment_independence}, one may replace the critical value
  in~\Cref{eq:ebci_under_indep} with
  $\cva_{\alpha}((1-1/\hat{w}_{EB, i})^{2}\hat{\mu}_{2i}/\hat{\sigma}^{2}_{i},
  \hat{\kappa}_{i})$, where $\hat{\mu}_{2i}$ and $\hat{\kappa}_{i}$ are
  consistent nonparametric estimates of
  $\mu_{2i}=E[(\theta_{i}-X_{i}'\delta)^{2}\mid X_{i}, \sigma_{i}]$ and
  $\kappa_{i}=E[(\theta_{i}-X_{i}'\delta)^{4}\mid X_{i},
  \sigma_{i}]/\mu_{2i}^{2}$. The resulting \ac{CI} will be asymptotically
  equivalent to the \ac{CI} in the baseline implementation if
  \Cref{eq:moment_independence} holds, but it will achieve valid \ac{EB} coverage
  even if this assumption fails. In our empirical application, we use
  nearest-neighbor estimates, as described in \Cref{sec:finite_n}. As a simple
  diagnostic to gauge how much the second moment of $\theta_i-X_i'\delta$ varies with
  $(X_{i}, \sigma_{i})$, one can report the $R^{2}$ gain in predicting
  $\hat{\varepsilon}_{i}^{2}-\hat{\sigma}^{2}_{i}$ using $\hat{\mu}_{2i}$ rather
  than the baseline estimate $\hat{\mu}_{2}$, as we illustrate in our empirical
  application.
\end{remark}

\begin{remark}[Average coverage and non-independent
  sampling]\label{rem:average_coverage_independence}
  We show in \Cref{sec:general-results} that the robust \ac{EBCI} satisfies an
  average coverage criterion of the form~\eqref{eq:simple-aci} when the
  parameters $\theta=(\theta_{1}, \dotsc, \theta_{n}$) are considered fixed, in
  addition to achieving valid \ac{EB} coverage when the $\theta_i$'s are viewed
  as random draws from some underlying distribution. To guarantee average
  coverage or \ac{EB} coverage, we do not need to assume that the $Y_i$'s and
  $\theta_i$'s are drawn independently across $i$. This is because the average
  coverage and \ac{EB} coverage criteria only depend on the marginal
  distribution of $(Y_{i}, \theta_{i})$, not the joint distribution. Indeed, in
  deriving the infeasible \ac{CI} in \Cref{eq:conditional_ci}, we made no
  assumptions about the dependence structure of $(Y_{i}, \theta_{i})$ across
  $i$. Consequently, to guarantee asymptotic coverage of the feasible interval
  in \Cref{eq:ebci_under_indep} as $n\to\infty$, we only need to ensure that the
  estimates $\hat{\mu}_2, \hat{\kappa}, \hat{\delta}, \hat{\sigma}_i$
  are consistent for $\mu_{2}, \kappa, \delta, \sigma_i$, which is the case
  under many forms of weak dependence or clustering. Furthermore, our baseline
  implementation above does not require the researcher to take an explicit stand
  on the dependence of the data; for example, in the case of clustering, the
  researcher does not need to take an explicit stand on how the clusters are
  defined.
\end{remark}

\begin{remark}[Estimating moments of the distribution of
  $\theta_{i}$]\label{rem:estimating_moments}
  The estimators $\hat{\mu}_{2}$ and $\hat{\kappa}$ in
  step~\ref{item:param-estimates} of our baseline implementation above are based
  on the moment conditions
  $E[(Y_{i}-X_{i}'\delta)^{2}-\sigma^{2}_{i}\mid X_{i}, \sigma_{i}]=\mu_{2}$ and
  $E[(Y_{i}-X_{i}'\delta)^{4}+3\sigma^{4}_{i}-6\sigma^{2}_{i}(Y_{i}-X_{i}'\delta)^{2}
  \mid X_{i}, \sigma_{i}]=\kappa \mu_{2}^{2}$, replacing population expectations
  by weighted sample averages. In addition, to avoid small-sample coverage
  issues when $\mu_2$ and $\kappa$ are near their theoretical lower bounds of 0
  and 1, respectively, these estimates incorporate truncation on $\hat\mu_2$ and
  $\hat\kappa$. These truncated estimates approximate the Bayesian posterior means under a flat prior on $\mu_2$ and $\kappa$, as in
  \citet{morris83pebci,morris83}. Although the resulting \acp{EBCI} do not
  directly account for estimation uncertainty in $\mu_{2}$ and $\kappa$, we
  verify their small-sample coverage accuracy via extensive simulations in
  \Cref{sec:sim}. \Cref{sec:finite_n} discusses the choice of the moment
  estimates, as well as other ways of performing truncation.
\end{remark}

\begin{remark}[Using higher moments and other forms of shrinkage]\label{rem:higher_moments}
  In addition to using the second and fourth moment of bias, one may
  augment~\eqref{eq:fourth_moment_bound} with restrictions on higher moments of
  the bias in order to further tighten the critical value. In
  \Cref{sec:efficiency}, we show that using other moments in addition to the
  second and fourth moment does not substantially decrease the critical value in
  the case where $\theta_i$ is normally distributed. Thus, the CI in our
  baseline implementation is robust to failure of the normality
  assumption~\eqref{eq:hierarch_theta}, while being near-optimal when this
  assumption does hold. \Cref{sec:efficiency} also shows that further efficiency
  gains are possible if one uses the linear estimator
  $\tilde{\theta}_{i}=\mu_{1,i}+w_{i}(Y_i-\mu_{1,i})$ with the shrinkage
  coefficient $w_{i}$ chosen to optimize \ac{CI} length, instead of using the
  \ac{MSE}-optimal shrinkage $w_{EB, i}$. For efficiency under a non-normal
  distribution of $\theta_{i}$, one needs to consider non-linear shrinkage; we
  discuss this extension in \Cref{sec:general_shrinkage}.
\end{remark}

\section{Main results}\label{sec:general-results}

This section provides formal statements of the coverage properties of the
\acp{CI} presented in \Cref{sec:simple-example,sec:pract-impl}. Furthermore, we
show that the \acp{CI} presented in \Cref{sec:simple-example,sec:pract-impl} are
highly efficient when the mean parameters are in fact normally distributed.
Next, we calculate the maximal coverage distortion of the parametric \ac{EBCI},
and derive a rule of thumb for gauging the potential coverage distortion.
Finally, we present a comprehensive simulation study of the finite-sample
performance of the robust \ac{EBCI}. Applied readers interested primarily in
implementation issues may skip ahead to the empirical application in
\Cref{sec:empir-appl}.

\subsection{Coverage under baseline implementation}\label{sec:cover-under-basel}

In order to state the formal result, let us first carefully define the notions
of coverage that we consider. Consider intervals $CI_{1}, \dotsc, CI_{n}$ for
elements of the parameter vector $\theta=(\theta_1, \dotsc, \theta_n)'$. The
probability measure $P$ denotes the joint distribution of $\theta$ and
$CI_1, \dotsc, CI_n$. Following \citet[Eq.\ 3.6]{morris83} and \citet[Ch.
3.5]{CaLo00}, we say that the interval $CI_i$ is an (asymptotic) $1-\alpha$
\acf{EBCI} if
\begin{equation}\label{eq:ebci_def}
  \liminf_{n\to\infty} P(\theta_i\in CI_i) \ge 1-\alpha.
\end{equation}
We say that the intervals $CI_i$ are (asymptotic) $1-\alpha$ \acp{ACI} under the
parameter sequence $\theta_1, \dotsc, \theta_n$ if
\begin{equation}\label{eq:aci_def}
  \liminf_{n\to\infty} \frac{1}{n}\sum_{i=1}^n P(\theta_i\in CI_i\mid \theta) \ge 1-\alpha.
\end{equation}
The average coverage property~\eqref{eq:aci_def} is a property of the
distribution of the data conditional on $\theta$ and therefore does not require
that we view the $\theta_i$'s as random (as in a Bayesian or ``random effects''
analysis). To maintain consistent notation, we nonetheless use the
conditional notation $P(\cdot\mid \theta)$ when considering average coverage.
See \Cref{coverage_results_sec_append} for a formulation with
$\theta$ treated as nonrandom.

Observe that under the exchangeability condition that
$P(\theta_i\in CI_i)=P(\theta_j\in CI_j)$ for all $i, j$, if the \ac{ACI}
property~\eqref{eq:aci_def} holds almost surely, then the \ac{EBCI}
property~\eqref{eq:ebci_def} holds, since then
\begin{equation*}
  P(\theta_{i}\in CI_{i})= \frac{1}{n}\sum_{j=1}^n P(\theta_{j}\in CI_{j})
  \ge 1-\alpha+o(1)\quad\text{for all $i$.}
\end{equation*}

We now provide coverage results for the baseline implementation described in
\Cref{sec:baseline-implementation}. To keep the statements in the main text as
simple as possible, we (i) maintain the assumption that the unshrunk estimates
$Y_i$ follow an exact normal distribution conditional on the parameter
$\theta_i$, (ii) state the results only for the homoskedastic case where the
variance $\sigma_i$ of the unshrunk estimate $Y_i$ does not vary across $i$, and
(iii) consider only unconditional coverage statements of the
form~\eqref{eq:ebci_def} and~\eqref{eq:aci_def}. In
\Cref{coverage_results_sec_append}, we allow the estimates $Y_i$ to be only
approximately normally distributed and allow $\sigma_i$ to vary, and we verify
that our assumptions hold in a linear fixed effects panel data model. We also
formalize the statements about conditional coverage made in
\Cref{rem:conditional_coverage}.

\begin{theorem}\label{thm:coverage_baseline}
  Suppose $Y_i\mid \theta\sim N(\theta_i, \sigma^2)$. Let
  $\mu_{j, n}=\frac{1}{n}\sum_{i=1}^n (\theta_i-X_i'\delta)^j$ and let
  $\kappa_n=\mu_{4,n}/\mu_{2,n}^2$. Suppose the sequence
  $\theta=\theta_1, \dotsc, \theta_n$ and the conditional distribution
  $P(\cdot\mid \theta)$ satisfy the following conditions with probability one:
  \begin{enumerate}
  \item $\mu_{2,n}\to \mu_{2}$ and $\mu_{4,n}/\mu_{2,n}^2\to \kappa$ for some
    $\mu_{2}\in (0,\infty)$ and $\kappa\in (1,\infty)$.
  \item Conditional on $\theta$,
    $(\hat\delta, \hat\sigma, \hat\mu_{2}, \hat\kappa)$ converges in probability
    to $(\delta, \sigma, \mu_{2}, \kappa)$.
  \end{enumerate}
  Then the \acp{CI} in \Cref{eq:ebci_under_indep} with $\hat\sigma_i=\hat\sigma$
  satisfy the \ac{ACI} property~\eqref{eq:aci_def} with probability one.
  Furthermore, if $\theta_1, \dotsc, \theta_n$ follow an exchangeable
  distribution and the estimators $\hat\delta$, $\hat\sigma$, $\hat\mu_{2}$ and
  $\hat\kappa$ are exchangeable functions of the data
  $(X_1', Y_1)', \dotsc, (X_n', Y_n)'$, then these \acp{CI} satisfy the \ac{EB}
  coverage property~\eqref{eq:ebci_def}.
\end{theorem}

\Cref{thm:coverage_baseline} follows immediately from
\Cref{indep_shrinkage_coverage_thm} in \Cref{coverage_results_sec_append}. In
order to cover both the \ac{EB} coverage condition~\eqref{eq:ebci_def} and the
average coverage condition~\eqref{eq:aci_def}, \Cref{thm:coverage_baseline}
considers a random sequence of parameters $\theta_1, \dotsc, \theta_n$, and
shows average coverage conditional on these parameters. See
\Cref{coverage_results_sec_append} for a formulation with $\theta$ treated as
nonrandom.

The condition on the moments $\mu_2$ and $\kappa$ avoids degenerate cases such
as when $\mu_{2}=0$, in which case the \ac{EB} point estimator $\hat{\theta}_i$
shrinks each preliminary estimate $Y_i$ all the way to $X_i'\hat{\delta}$. Note
also that the theorem does not require that $\hat{\delta}$ be the \ac{OLS}
estimate in a regression of $Y_{i}$ onto $X_{i}$, and that $\delta$ be the
population analog; one can define $\delta$ in other ways, the theorem only
requires that $\hat{\delta}$ be a consistent estimate of it. The definition of
$\delta$ does, however, affect the plausibility of the moment independence
assumption in \Cref{eq:moment_independence} needed for conditional coverage
results stated in
\Cref{coverage_results_sec_append}.\footnote{\label{fn:effect_on_width}The
  specification of $\mu_{1i}=X_{i}'\delta$ also affects the \ac{EBCI} width
  through its effect on $\mu_{2}$ and $\kappa$.}
\begin{remark}\label{rem:ac_alt_def_remark}
  As shown in \Cref{coverage_results_sec_append}, if \acp{CI} satisfy the average coverage condition~\eqref{eq:aci_def}
  given $\theta_{1}, \dotsc, \theta_{n}$, they will typically also satisfy the stronger
  condition
  \begin{equation}\label{eq:alt_aci_def}
    \frac{1}{n}\sum_{i=1}^n \1{\theta_i\in CI_i} \ge 1-\alpha + o_{P(\cdot\mid \theta)}(1),
  \end{equation}
  where $o_{P(\cdot\mid \theta)}(1)$ denotes a sequence that converges in
  probability to zero conditional on $\theta$ (\Cref{eq:alt_aci_def} implies
  \Cref{eq:aci_def} since the left-hand side is uniformly bounded). That is, at
  least a fraction $1-\alpha$ of the $n$ \acp{CI} contain their respective true
  parameters, asymptotically. This is analogous to the result that for
  estimation, the difference between the squared error
  $\frac{1}{n}\sum_{i=1}^n(\hat\theta_i-\theta_i)^2$ and the \ac{MSE}
  $\frac{1}{n}\sum_{i=1}^{n} E[(\hat\theta_i-\theta_i)^2\mid \theta]$ typically
  converges to zero.
\end{remark}

\subsection{Relative efficiency}\label{sec:efficiency}

The robust \ac{EBCI} in \Cref{eq:conditional_ci}, unlike the parametric
\ac{EBCI} $\hat{\theta}_{i}\pm z_{1-\alpha/2}\sigma_{i}\sqrt{w_{EB, i}}$, does
not rely on the normality assumption in \Cref{eq:hierarch_theta} for its
validity. We now show that this robustness does not come at a high cost in terms
of efficiency: if the normality assumption~\eqref{eq:hierarch_theta} in fact
holds, the efficiency loss is limited unless the signal-to-noise ratio
$\mu_{2}/\sigma^{2}_{i}$ is very small.

There are two reasons for the inefficiency of the robust \ac{EBCI}. First, the
robust \ac{EBCI} only makes use of the second and fourth moment of the
conditional distribution of $\theta_{i}-X_{i}'\delta$, rather than its full
distribution. Second, if we only have knowledge of these two moments, it is no
longer optimal to center the \ac{EBCI} at the estimator $\hat{\theta}_{i}$: one
may need to consider other, perhaps non-linear, shrinkage estimators, as we do
below in \Cref{sec:general_shrinkage}.

We decompose the sources of inefficiency by studying the relative length of the
robust \ac{EBCI} relative to the \ac{EBCI} that picks the amount of shrinkage
optimally. For the latter, we maintain
assumption~\eqref{eq:moment_independence}, and consider a more general class of
estimators $\tilde{\theta}(w_{i})=\mu_{1, i}+w_{i}(Y_{i}-\mu_{1, i})$. For
tractability, we focus on fixed-length \acp{CI} based on linear
shrinkage estimators, but allow the amount of shrinkage $w_{i}$ to be optimally
determined. The normalized bias of $\tilde{\theta}(w_{i})$ is given by
$b_{i}=(1/w_{i}-1)\varepsilon_{i}/\sigma_{i}$, which leads to the \ac{EBCI}
\begin{equation*}
  \mu_{1, i}+w_{i}(Y_{i}-\mu_{1, i})\pm
  \cva_{\alpha}((1-1/w_{i})^{2}\mu_{2}/\sigma^{2}_{i}, \kappa)w_{i}\sigma_{i}.
\end{equation*}
The half-length of this \ac{EBCI},
$\cva_{\alpha}((1-1/w_{i})^{2}\mu_{2}/\sigma^{2}_{i}, \kappa)w_{i}\sigma_{i}$,
can be numerically minimized as a function of $w_{i}$ to find the \ac{EBCI}
length-optimal shrinkage. Denote the minimizer by
$w_{opt}(\mu_{2}/\sigma^{2}_{i}, \kappa, \alpha)$. Like $w_{EB, i}$, the optimal
shrinkage depends on $\mu_{2}$ and $\sigma_{i}^{2}$ only through the
signal-to-noise ratio $\mu_{2}/\sigma^{2}_{i}$. Numerically evaluating the
minimizer shows that $w_{opt}(\cdot, \kappa, \alpha)\geq w_{EB, i}$ for
$\kappa\geq 3$ and $\alpha\in\{0.05,0.1\}$. The resulting \ac{EBCI} is optimal
among all fixed-length \acp{EBCI} centered at linear estimators
under~\eqref{eq:moment_independence}, and we call it the optimal robust
\ac{EBCI}.\footnote{\label{fn:optimal_robust}Since the optimal robust \ac{EBCI}
  is always shorter than the robust \ac{EBCI} in~\Cref{eq:conditional_ci}, the former is
  preferable on efficiency grounds. It may not contain the \ac{MSE}-optimal point estimator $\hat{\theta}_i$, however.}

\begin{figure}[t]
  \centering{\input{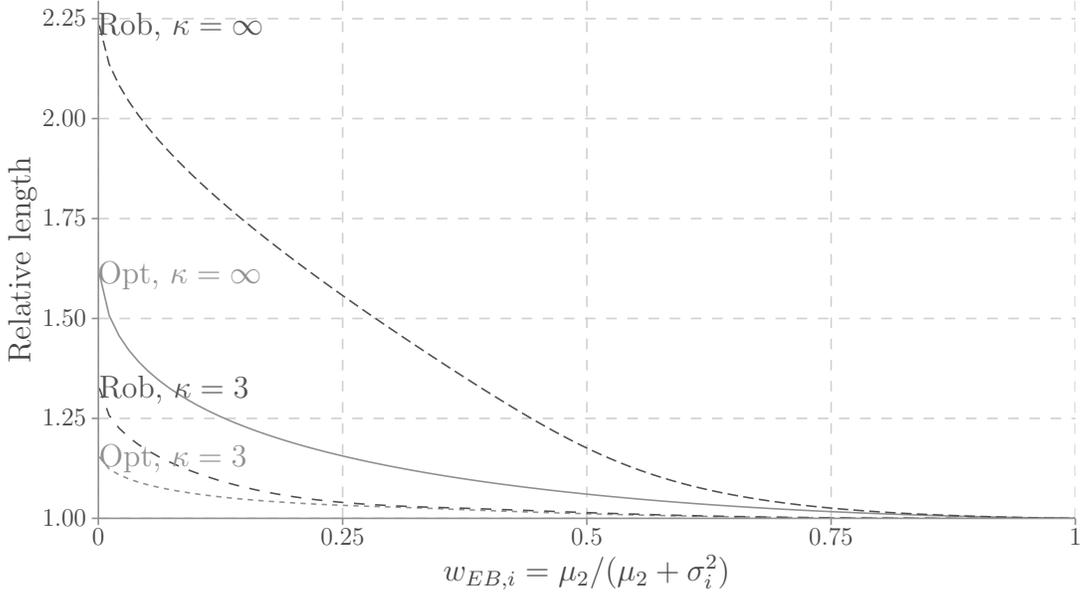}}
  \caption{Relative efficiency of robust \ac{EBCI} (Rob) and optimal robust
    \ac{EBCI} (Opt) relative to the normal benchmark, for $\alpha=0.05$. The
    figure plots ratios of Rob length,
    $2\cva_{\alpha}(\sigma_{i}^{2}/\mu_{2}, \kappa) \cdot \sigma_{i}
    {\mu_{2}/(\mu_{2}+\sigma_{i}^{2})}$, and Opt length,
    $2\cva_{\alpha}((1-1/w_{opt}(\mu_{2}/\sigma_{i}^{2}, \kappa,
    \alpha))^{2}\mu_{2}/\sigma_{i}^{2}, \kappa)\cdot \sigma_{i}
    w_{opt}(\mu_{2}/\sigma_{i}^{2}, \kappa, \alpha)$, relative to the parametric
    \ac{EBCI} length
    $2z_{1-\alpha/2}\sqrt{\mu_{2}/(\mu_{2}+\sigma_{i}^{2})}\sigma_{i}$ as a
    function of the shrinkage factor
    $w_{EB, i}=\mu_{2}/(\mu_{2}+\sigma_{i}^{2})$, which maps the signal-to-noise
    ratio $\mu_{2}/\sigma_{i}^{2}$ to the interval
    $[0,1]$.}\label{fig:ebci_efficiency}
\end{figure}

\Cref{fig:ebci_efficiency} plots the ratio of lengths of the optimal robust
\ac{EBCI} and robust \ac{EBCI} relative to the parametric \ac{EBCI}, for
$\alpha=0.05$. The figure shows that to maintain efficiency relative to the
normal benchmark, it is important to impose the fourth moment constraint. If
this constraint is imposed, the efficiency loss of the robust \ac{EBCI} is
modest unless the signal-to-noise ratio is very small: if $w_{EB, i}\geq 0.1$
(which is equivalent to $\mu_{2}/\sigma^{2}_{i}\geq 1/9$), the efficiency loss
is at most $11.4\%$ for $\alpha=0.05$; up to half of the efficiency loss is due
to not using the optimal shrinkage. For $\alpha=0.1$ (not plotted), the results
are very similar; in particular, if $w_{EB, i}\geq 0.1$, the efficiency loss is
at most $12.9\%$.

\begin{figure}[t]
  \centering{\input{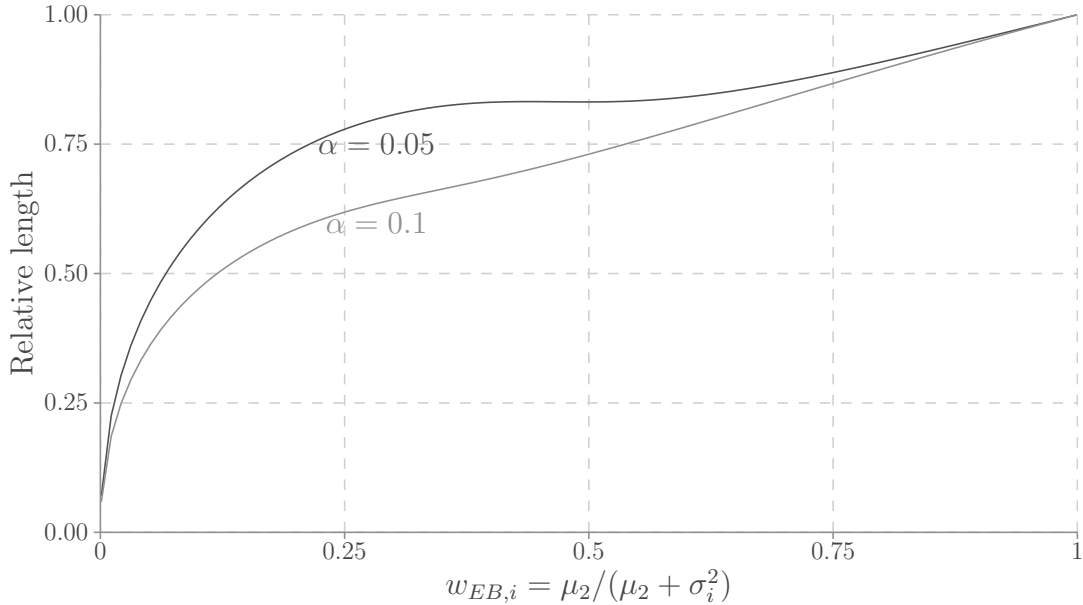}}
  \caption{Efficiency of robust \ac{EBCI}
    $\hat{\theta}_{i}\pm \cva_{\alpha}(\sigma_i^{2}/\mu_{2}, \kappa=\infty)
    \cdot \sigma {\mu_{2}/(\mu_{2}+\sigma_i^{2})}$ relative to the unshrunk
    \ac{CI} $Y_{i}\pm z_{1-\alpha/2}\sigma_i$. The figure plots the ratio of the
    length of the robust \ac{EBCI} relative to the unshrunk \ac{CI} as a
    function of the shrinkage factor
    $w_{EB, i} =
    \mu_{2}/(\mu_{2}+\sigma_{i}^{2})$.}\label{fig:efficiency_unshrunk}
\end{figure}

When the signal-to-noise ratio is very small, so that $w_{EB, i}<0.1$, the
efficiency loss of the robust \ac{EBCI} is higher (up to 39\% for $\alpha=0.05$
or $0.1$). Using the optimal robust \ac{EBCI} ensures that the efficiency loss
is below 20\%, irrespective of the signal-to-noise ratio. On the other hand,
when the signal-to-noise ratio is small, any of these \acp{CI} will be
significantly tighter than the unshrunk \ac{CI}
$Y_{i}\pm z_{1-\alpha/2}\sigma_{i}$. To illustrate this point,
\Cref{fig:efficiency_unshrunk} plots the efficiency of the robust \ac{EBCI} that
imposes the second moment constraint only, relative to this unshrunk \ac{CI}. It
can be seen from the figure that shrinkage methods allow us to tighten the
\ac{CI} by 44\% or more when $\mu_{2}/\sigma^{2}_{i}\leq 0.1$.

\subsection{Undercoverage of parametric EBCI}\label{sec:param-ebci-cover}

The parametric \ac{EBCI}
$\hat\theta_i \pm z_{1-\alpha/2}w_{EB, i}^{1/2}\sigma_{i}$ is an
\ac{EB} version of a Bayesian credible interval that
treats~\eqref{eq:hierarch_theta} as a prior. We now assess its potential
undercoverage when \Cref{eq:hierarch_theta} is violated.

\begin{figure}[tp]
\centering{\input{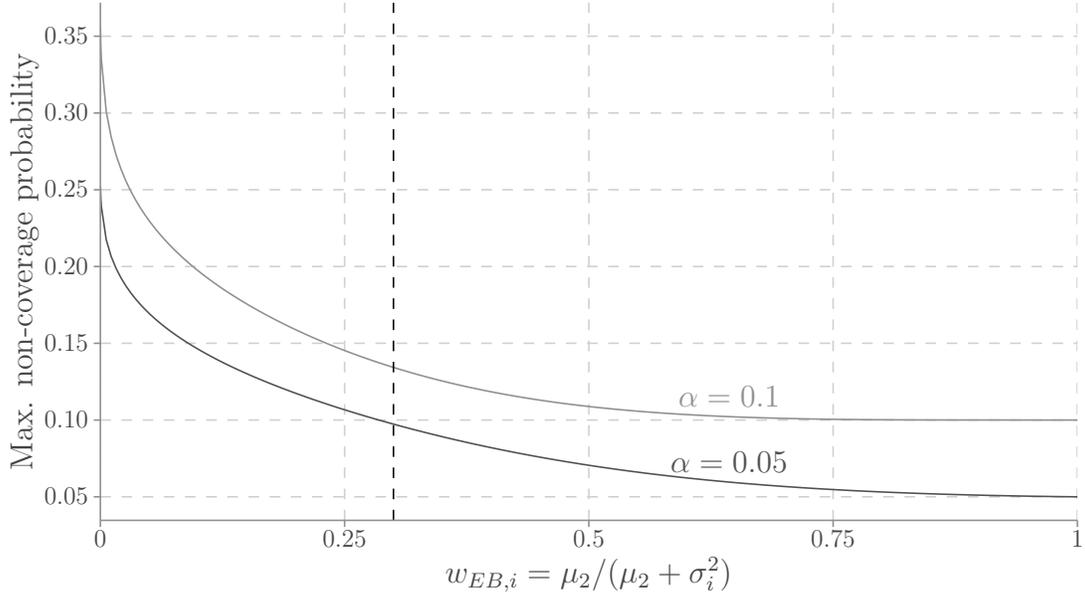}}
\caption{Maximal non-coverage probability of parametric \ac{EBCI},
  $\alpha \in \lbrace 0.05, 0.10 \rbrace$. The vertical line marks the ``rule of
  thumb'' value $w_{EB, i}=0.3$, above which the maximal coverage distortion is
  less than 5 percentage points for these two values of
  $\alpha$.}\label{fig:noncov_param}
\end{figure}

Given knowledge of only the second moment $\mu_{2}$ of
$\varepsilon_i=Y_i-X_i'\delta$, the maximal undercoverage of this interval is
given by
\begin{equation} \label{eq:eb_param_noncov}
\rho(1/w_{EB, i}-1, z_{1-\alpha/2}/\sqrt{w_{EB, i}}),
\end{equation}
since $w_{EB, i}= \mu_{2}/(\mu_{2}+\sigma_i^2)$. Here $\rho$ is the non-coverage
function defined in \Cref{eq:non-coverage-bound}. \Cref{fig:noncov_param} plots
the maximal non-coverage probability as a function of $w_{EB, i}$, for
significance levels $\alpha=0.05$ and $\alpha=0.10$. The figure suggests a
simple ``rule of thumb'': if $w_{EB, i} \geq 0.3$, the maximal coverage
distortion is less than 5 percentage points for these values of $\alpha$.

The following lemma confirms that the maximal non-coverage is decreasing in
$w_{EB, i}$, as suggested by the figure. It also gives an expression for the
maximal non-coverage across all values of $w_{EB, i}$ (which is achieved in the
limit $w_{EB, i} \to 0$).
\begin{lemma}\label{thm:eb_param}
  The non-coverage probability~\eqref{eq:eb_param_noncov} of the parametric \ac{EBCI}
is weakly decreasing as a
  function of $w_{EB, i}$, with the supremum given by $1/\max\{z_{1-\alpha}^2,1\}$.
\end{lemma}
The maximal non-coverage probability $1/\max\{z_{1-\alpha/2}^2,1\}$ equals
$0.260$ for $\alpha=0.05$ and $0.370$ for $\alpha=0.10$. For
$\alpha>2\Phi(-1)\approx 0.317$, the maximal non-coverage probability is 1.

If we additionally impose knowledge of the kurtosis of $\varepsilon_i$, the
maximal non-coverage of the parametric \ac{EBCI} can be similarly computed using
\Cref{eq:fourth_moment_bound}, as illustrated in the application in
\Cref{sec:empir-appl}.

\subsection{Monte Carlo simulations}\label{sec:sim}
Here we show through simulations that the robust \ac{EBCI} achieves accurate average coverage in finite samples.

\subsubsection{Design}
The \ac{DGP} is a simple linear fixed effects panel data model. We first draw
$\theta_i$, $i=1,\dotsc, n$, i.i.d.\ from a random effects distribution specified
below. Then we simulate panel data from the model
\begin{equation*}
  W_{it} = \theta_i + U_{it}, \quad i=1, \dotsc, n, \quad t=1, \dotsc, T,
\end{equation*}
where the errors $U_{it}$ are mean zero and i.i.d.\ across $(i, t)$ and
independent of the $\theta_i$'s. The unshrunk estimator of $\theta_i$ is the
sample average of $W_{it}$ for unit $i$, with standard error obtained from the
usual unbiased variance estimator:
\begin{equation*}
  Y_i = \frac{1}{T}\sum_{t=1}^T W_{it},
  \quad \hat{\sigma}_i = \sqrt{\frac{1}{T(T-1)}\sum_{t=1}^T (W_{it}-Y_i)^2}.
\end{equation*}

We draw $U_{it}$ from one of two distributions: (1) a normal distribution and
(2) a (shifted) chi-squared distribution with 3 degrees of freedom. In case (1),
$Y_i$ is exactly normal conditional on $\theta_i$, but $\hat{\sigma}_i^2$ does
not exactly equal $\var(Y_i \mid \theta_i)$ for finite $T$. In case (2), $Y_i$
is non-normal and positively skewed (conditional on $\theta_i$) for finite $T$.

We consider six random
effects distributions for $\theta_i$ (see \Cref*{sec:sim-details} for detailed
definitions):
\begin{enumerate*}[label=(\roman*)]%
\item normal (kurtosis $\kappa=3$);
\item scaled chi-squared with 1 degree of freedom ($\kappa=15$);
\item two-point distribution ($\kappa \approx 8.11$);
\item three-point distribution ($\kappa=2$);
\item the least favorable distribution for the robust \ac{EBCI} that exploits
  only second moments ($\kappa$ depends on $\mu_{2}$, see \Cref{sec:comput});
  and
\item the least favorable distribution for the parametric \ac{EBCI}.
\end{enumerate*}

Given $T$, we scale the $\theta_i$ distribution to match one of four signal-to-noise ratios
$\mu_2/\var(Y_i \mid \theta_i) \in \lbrace 0.1,0.5,1,2 \rbrace$, for a total of
$6 \times 4 = 24$ \acp{DGP} for each distribution of $U_{it}$. We shrink towards
the grand mean ($X_i=1$ for all $i$). We construct the robust \acp{EBCI}
following the baseline implementation in~\Cref{sec:baseline-implementation}
(with $\omega_{i}=1/n$), as well as a version that does not impose constraints
on the kurtosis.

As $T \to \infty$, we recover the idealized setting in
\Cref{sec:simple-example}, with $(Y_{i}-\theta_{i})/\sqrt{\var(Y_i \mid \theta_i)}$
converging in distribution to a standard normal (conditional on $\theta_i$), and
$\hat{\sigma}_i^2/\var(Y_i \mid \theta_i)$ converging in probability to 1, for
each $i$.

\subsubsection{Results}

\Cref{tab:sim_panel_normal} shows that the 95\% robust \acp{EBCI} achieve good
average coverage when the panel errors $U_{it}$ are normally distributed. This
is true for all \acp{DGP}, panel dimensions $n$ and $T$, and whether we exploit
one or both of the (estimated) moments $\mu_2$ and $\kappa$. When the time
dimension $T$ equals 10, the maximal coverage distortion across all \acp{DGP} and all
cross-sectional dimensions $n \in \lbrace 100,200,500\rbrace$ is 3.2 percentage
points. For $T \geq 20$, the coverage distortion of the robust \acp{EBCI} is
always below 2.1 percentage points.

\begin{table}[tp]
  \centering
  \begin{threeparttable}
    \caption{Monte Carlo simulation results, panel data with normal
      errors.}\label{tab:sim_panel_normal}
    \begin{tabular*}{0.95\linewidth}{@{\extracolsep{\fill}}@{}l cccc cccc cccc@{}}
      & \multicolumn{4}{@{}c}{Robust, $\mu_2$ only}
      & \multicolumn{4}{@{}c}{Robust, $\mu_2$ \& $\kappa$}
      & \multicolumn{4}{@{}c}{Parametric} \\
      \cmidrule(rl){2-5}\cmidrule(rl){6-9}\cmidrule(rl){10-13}
      $T$&10&20&$\infty$&ora&10&20&$\infty$&ora&10&20&$\infty$&ora\\
      \midrule
      \multicolumn{13}{@{}l}{Panel A\@{}: Average coverage (\%), minimum across 24 \acsp{DGP}}\\
      \cmidrule(r){1-9}
      $n=100$ & 92.1 & 93.7 & 94.0 & 95.0 & 91.8 & 93.2 & 93.2 & 94.6 & 79.2 & 79.7 & 79.3 & 86.9 \\
$n=200$ & 91.9 & 93.4 & 92.9 & 95.0 & 91.8 & 93.3 & 92.9 & 94.8 & 80.7 & 80.3 & 81.0 & 86.3 \\
$n=500$ & 91.9 & 93.6 & 94.8 & 95.0 & 91.9 & 93.5 & 94.3 & 94.9 & 84.2 & 85.1 & 85.1 & 85.6
\\
      \multicolumn{13}{@{}l}{Panel B\@{}: Relative average length, average across 24 \acsp{DGP}}\\
      \cmidrule(r){1-9}
      $n=100$ & 1.09 & 1.10 & 1.11 & 1.16 & 1.03 & 1.02 & 1.02 & 1.00 & 0.81 & 0.82 & 0.83 & 0.86 \\
$n=200$ & 1.09 & 1.10 & 1.12 & 1.16 & 1.02 & 1.02 & 1.01 & 1.00 & 0.81 & 0.82 & 0.84 & 0.86 \\
$n=500$ & 1.10 & 1.11 & 1.13 & 1.16 & 1.04 & 1.03 & 1.01 & 1.00 & 0.82 & 0.83 & 0.84 & 0.86

    \end{tabular*}
    \begin{tablenotes}
    \item \emph{Notes:} Normally distributed errors. Nominal average confidence
      level $1-\alpha=95\%$. All \ac{EBCI} procedures use baseline estimate of
      $\hat{\mu}_2$ and (if applicable) $\hat{\kappa}$, except columns labeled
      ``ora'', which use oracle values of $\mu_2$ and $\kappa$. Columns
      $T=\infty$ and ``ora'' use oracle standard errors $\sigma_i$. For each
      \acs{DGP}, ``average coverage'' and ``average length'' refer to averages
      across units $i=1, \dotsc, n$ and across 2,000 Monte Carlo repetitions.
      Average \acs{CI} length is measured relative to the robust \ac{EBCI} that
      exploits the oracle values of $\mu_2$, $\kappa$, and $\sigma_i$ (but not
      of the grand mean $\delta=E[\theta]$).
    \end{tablenotes}
  \end{threeparttable} \\
\bigskip\bigskip
  \begin{threeparttable}
    \caption{Monte Carlo simulation results, panel data with chi-squared errors.}\label{tab:sim_panel_chi2}
    \begin{tabular*}{0.95\linewidth}{@{\extracolsep{\fill}}@{}l cccc cccc cccc@{}}
 & \multicolumn{4}{@{}c}{Robust, $\mu_2$ only} & \multicolumn{4}{@{}c}{Robust, $\mu_2$ \& $\kappa$} & \multicolumn{4}{@{}c}{Parametric} \\
\cmidrule(rl){2-5}\cmidrule(rl){6-9}\cmidrule(rl){10-13}
$T$&10&20&50&ora&10&20&50&ora&10&20&50&ora\\
\midrule
\multicolumn{13}{@{}l}{Panel A\@{}: Average coverage (\%), minimum across 24 \acsp{DGP}}\\
\cmidrule(r){1-9}
$n=100$ & 87.9 & 90.9 & 93.1 & 95.0 & 87.8 & 90.8 & 92.6 & 94.7 & 79.9 & 79.3 & 79.3 & 87.0 \\
$n=200$ & 87.9 & 90.8 & 93.0 & 94.9 & 87.8 & 90.8 & 92.8 & 94.8 & 77.8 & 79.8 & 80.3 & 86.2 \\
$n=500$ & 87.8 & 90.8 & 93.0 & 95.0 & 87.8 & 90.7 & 92.9 & 94.9 & 82.0 & 84.1 & 84.8 & 85.6
\\
\multicolumn{13}{@{}l}{Panel B\@{}: Relative average length, average across 24 \acsp{DGP}}\\
\cmidrule(r){1-9}
$n=100$ & 1.05 & 1.08 & 1.10 & 1.16 & 1.01 & 1.02 & 1.02 & 1.00 & 0.79 & 0.81 & 0.82 & 0.86 \\
$n=200$ & 1.04 & 1.08 & 1.10 & 1.16 & 0.99 & 1.00 & 1.00 & 1.00 & 0.78 & 0.81 & 0.82 & 0.86 \\
$n=500$ & 1.05 & 1.09 & 1.11 & 1.16 & 0.99 & 1.00 & 1.00 & 1.00 & 0.79 & 0.82 & 0.83 & 0.86

    \end{tabular*}
    \begin{tablenotes}
    \item \emph{Notes:} Chi-squared distributed errors. See caption for \Cref{tab:sim_panel_normal}. Results for $T=\infty$ are by definition the same as in \Cref{tab:sim_panel_normal}.
    \end{tablenotes}
  \end{threeparttable}
\end{table}

\Cref{tab:sim_panel_chi2} shows that coverage distortions are somewhat larger
when the panel errors $U_{it}$ are chi-squared distributed and $T$ is small. The
robust \acp{EBCI} undercover by up to 7.2 percentage points when $T=10$ due to
the pronounced non-normality of $Y_i$ given $\theta_i$. However, the distortion
is at most 4.3 percentage points when $T=20$, and at most 2.4 percentage points
when $T \geq 50$. The coverage distortion due to non-normality when $T$ is small
is similar to the coverage distortion of the usual unshrunk \ac{CI} (not
reported).

Importantly, in all cases considered in
\Cref{tab:sim_panel_normal,tab:sim_panel_chi2}, the worst-case coverage
distortion of the parametric \ac{EBCI} substantially exceeds that of the
corresponding robust \acp{EBCI}, sometimes by more than 10 percentage points.
Nevertheless, the cost of robustness in terms of extra \ac{CI} length is modest
and consistent with the theoretical results in \Cref{sec:efficiency}.

Both the estimation of the standard errors $\sigma_i$ and the estimation of the
moments $\mu_2$ and $\kappa$ contribute to the finite-sample coverage
distortions. The ``ora'' columns in \Cref{tab:sim_panel_normal} exploit the
oracle (true) values of $\mu_2$, $\kappa$, and $\sigma_i=\sqrt{\var(Y_i \mid \theta_i)}$, while the
$T=\infty$ columns use oracle standard errors but not oracle moments. By
comparing these columns, we see that estimation of $\mu_2$ and $\kappa$ is
responsible for modest coverage distortions when $n=100$ or $200$. However,
estimation of the standard errors $\sigma_i$ also contributes to the
distortions, as can be seen by comparing the $T=10$ and $T=\infty$ columns.

In \Cref*{sec:sim-heterosk} we show that the robust
\ac{EBCI} also has good coverage in a heteroskedastic design calibrated to the
empirical application in \Cref{sec:empir-appl} below.

\section{Comparison with other approaches}\label{sec:compar}

Here we compare our \ac{EBCI} procedure with other approaches to confidence
interval construction in the normal means model. We also discuss other related
inference problems.

\subsection{Average coverage vs.\ alternative coverage concepts}\label{sec:compar_avg_cov}

The average coverage requirement in \Cref{eq:aci_def} is less stringent than the
usual (pointwise) notion of frequentist coverage that
$P(\theta_i\in CI_i\mid \theta) \geq 1-\alpha$ for all $i$. An even stronger
coverage requirement is that of simultaneous coverage:
$P(\forall i\colon \theta_i\in CI_{i}\mid \theta) \geq 1-\alpha$. As outlined
in~\cref{fn:impossibility}, under the pointwise coverage criterion, one cannot
achieve substantial reductions in length relative to the unshrunk \ac{CI}. Under
the simultaneous coverage criterion, it is likewise impossible to substantially
improve upon the usual sup-$t$ confidence band based on the unshrunk estimates
\citep{Cai2015}. Thus, undercoverage for some $\theta_{i}$'s must be tolerated
if one wants to use shrinkage to improve \ac{CI} length.

The fact that our \acp{EBCI} achieve improvements in average length at the
expense of undercovering for certain units $i$ is analogous to well-known
properties of \ac{EB} point estimators. We now show that the units $i$ for which
our \ac{EBCI} undercovers are quantitatively similar to the units for which the
shrinkage estimator $\hat{\theta}_{i}$ has higher \ac{MSE} than the unshrunk
estimator $Y_i$. Let $\varepsilon_{i}=\theta_{i}-X_{i}'\delta$ be the ``shrinkage error'' defined in
\Cref{sec:baseline-model}. The pointwise coverage of our \ac{EBCI} is decreasing in the normalized shrinkage error $\abs{\varepsilon_{i}}/\sqrt{\mu_2}$, for a
fixed signal-to-noise ratio $\mu_2/\sigma_i^2$.\footnote{The pointwise coverage (conditional on
$X_{i}$) equals
$1-r(\sqrt{1/w_{EB, i}-1}\cdot \abs{\varepsilon_{i}}/\sqrt{\mu_{2}},
\cva_{\alpha}(1/w_{EB, i}-1,\kappa))$, with $r$ defined in \Cref{eq:rb} and $w_{EB, i}=\mu_{2}/(\mu_{2}+\sigma_{i}^{2})$.}  Hence, the units $i$ for which our
\ac{EBCI} undercovers are those whose covariate-predicted value $X_i'\delta$
fails to approximate their true effect $\theta_i$ well. The \ac{MSE} of the
shrinkage estimator (for an individual unit $i$), normalized by the \ac{MSE}
of the unshrunk estimator, is
similarly increasing in $\abs{\varepsilon_{i}}/\sqrt{\mu_2}$.\footnote{The ratio of \acp{MSE} equals
$E[(\hat{\theta}_i-\theta_i)^2 \mid \theta_i, X_i]/\sigma_{i}^{2}=w_{EB,
  i}^{2}+(1-w_{EB, i}) w_{EB, i}\cdot \abs{\varepsilon_{i}}/\sqrt{\mu_2}$.}

\begin{figure}[t]
  \centering%
  \tikzset{font=\small} %
  \input{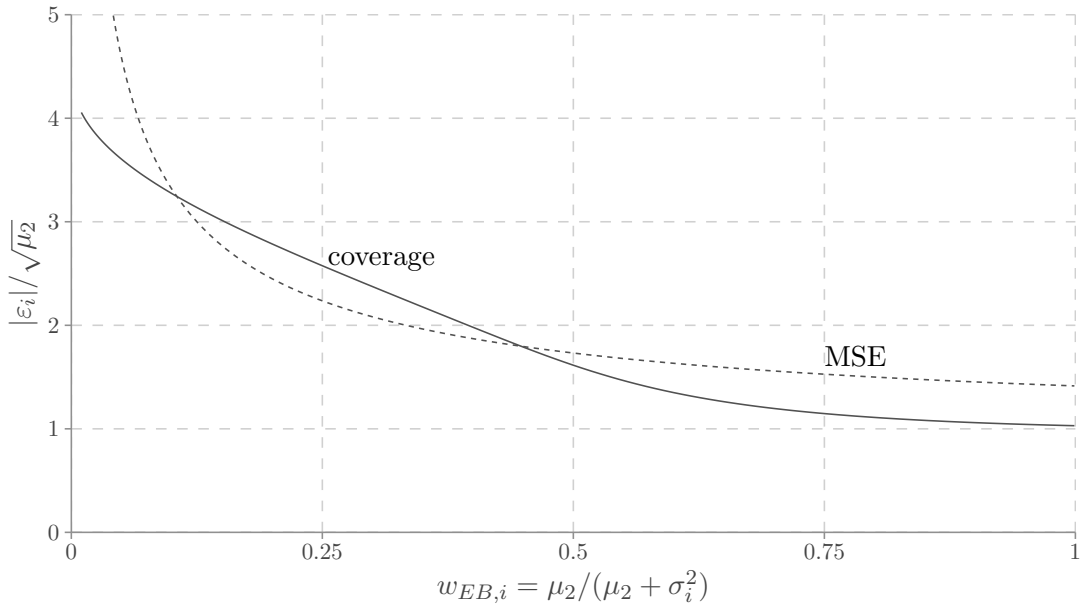}
  \caption{Value of $\abs{\varepsilon_{i}}/\sqrt{\mu_{2}}$, as a function of
    $w_{EB, i}$, such that the \ac{MSE} of the shrinkage point estimator equals
    that of the unshrunk estimator (MSE), and such that the coverage of the
    robust \ac{EBCI} with $\kappa=\infty$ equals the nominal average coverage
    $1-\alpha$ (coverage), for $\alpha=0.05$.}\label{fig:cond_coverage}
\end{figure}

\Cref{fig:cond_coverage} shows that the knife-edge value of
$\abs{\varepsilon_{i}}/\sqrt{\mu_2}$ for which the pointwise coverage of our
\ac{EBCI} equals $1-\alpha$ is quantitatively close to the value of
$\abs{\varepsilon_i}/\sqrt{\mu_2}$ for which the MSE of the shrinkage estimator
equals that of the unshrunk estimator. In other words, to the extent that one
worries about undercoverage for certain types of $\theta_i$ values, one should
simultaneously worry about the relative performance of the shrinkage point
estimator for those same values.

We stress that the pointwise coverage depends on the unobservable shrinkage
error $\varepsilon_{i}$, which cannot be gauged directly from the observables
$(Y_{i}, X_{i})$. If one wishes to avoid systematic differences in coverage
across units $i$ with different genders, say (i.e.,\ one is worried that
$\varepsilon_{i}$ correlates with gender) one can simply add gender to the set
of covariates $X_{i}$: the baseline procedure in
\Cref{sec:baseline-implementation} ensures control of average coverage
conditional on the covariates $X_i$. In \Cref{sec:cover_selection}, we show how
to adapt our \acp{EBCI} to settings where one focuses the analysis on a subset
of units $i$ based on the values of their unshrunk estimates $Y_i$ (e.g.,
keeping only the estimates that exceed a given threshold).

From a Bayesian point of view, our robust \ac{EBCI} can be viewed as an
uncertainty interval that is robust to the choice of prior distribution in the
\emph{unconditional} gamma-minimax sense: the coverage probability of this
\ac{CI} is at least $1-\alpha$ when averaged over the distribution of the data
and over the prior distribution for $\theta_{i}$, for any prior distribution
that satisfies the moment bounds. This follows directly from the derivations in
\Cref{sec:simple-example}, reinterpreting the random effects distribution for
$\theta_i$ as a prior distribution. In contrast, \emph{conditional}
gamma-minimax credible intervals, discussed recently by \citet[p.
6]{Kitagawa2019}, are too stringent in our setting. This notion requires that
the posterior credibility of the interval be at least $1-\alpha$ regardless of
the choice of prior, in any data sample, which would require reporting the
entire parameter space (up to the moment bounds).

\subsection{Finite-sample vs.\ asymptotic coverage}\label{sec:compar_finite}

Our procedures are asymptotically valid as $n\to\infty$, as proved in
\Cref{sec:cover-under-basel}. These asymptotics do not capture the impact of
estimation error in the ``hyper-parameters'' $\hat{\sigma}_i$, $\hat{\delta}$,
$\hat{\mu}_2$, and $\hat{\kappa}$, or the impact of lack of exact normality of
the $Y_{i}$'s, on the finite-sample performance of the \acp{EBCI}. As detailed
in \Cref{sec:baseline-implementation} and \Cref{sec:moment_estimates}, we do
apply a finite-sample adjustment to the moments $\hat{\mu}_2$ and
$\hat{\kappa}$, which is motivated by the same heuristic arguments that
\citet{morris83pebci,morris83} uses to motivate finite-sample adjustments to the
parametric \ac{EBCI}.\footnote{\label{fn:bootstrap}An alternative approach would be to adapt the
  bootstrap adjustment proposed by \citet[Ch. 3.5.3]{CaLo00} in the context of
  parametric \ac{EBCI} construction \citep[see also][]{efron2019}. As with the
  \citet{morris83pebci,morris83} adjustment, we are not aware of a formal result
  justifying it.} The promising simulation results in \Cref{sec:sim}
notwithstanding, these adjustments do not ensure exact average coverage control
in finite samples.\footnote{\label{fn:bonferroni}One could account for hyperparameter uncertainty by
  computing the critical value
  $\sup_{\tilde{\sigma}_i, \tilde{\mu}_2,\tilde{\kappa} \in \hat{\mathcal{C}}_i}
  \cva_{\alpha}(\tilde{\sigma}_i^{2}/\tilde{\mu}_{2}, \tilde{\kappa})$ over an
  initial confidence set $\hat{\mathcal{C}}_i$ for the hyper-parameters, coupled
  with a Bonferroni adjustment of the confidence level $1-\alpha$. This approach
  appears to be highly conservative in practice.}

Our results are thus analogous to standard results on coverage of Eicker-Huber-White
\acp{CI} in cross-sectional \ac{OLS}: asymptotic
validity follows by consistency of the \ac{OLS} variance estimate and asymptotic normality of the outcomes, while adjustments to account for finite-sample issues (such as the HC2 or HC3 variance estimators studied in \citealp{MaWh85}) are justified
heuristically. Deriving \acp{EBCI} with finite-sample coverage guarantees is an
interesting problem that we leave for future research; the problem appears to be
challenging even in the context of constructing parametric \acp{EBCI}.

\subsection{Local vs.\ global optimality}\label{sec:compar_optim}

Our \acp{EBCI} are designed to provide uncertainty assessments to accompany
linear shrinkage estimates that, as the Introduction argues, have been popular
in applied work. Our procedure's global validity, as well as local near-optimality when the $\theta_{i}$'s are normal (cf. \Cref{sec:efficiency}), is analogous to
Eicker-Huber-White \acp{CI} for \ac{OLS} estimators: these \acp{CI} are optimal
under normal homoskedastic regression errors, but remain valid when this
assumption is dropped.

Similar to the Eicker-Huber-White \acp{CI}, our \acp{EBCI} are not globally
efficient: when the $\theta_{i}$'s are not Gaussian, it is generally inefficient to restrict attention to \acp{CI} that are centered at a linear point estimator and have fixed width. While we expect our \acp{EBCI} to remain
near-efficient under mild departures from normality, substantial efficiency
gains may be possible if the effect size distribution is, for example, heavy-tailed or bimodal.\footnote{\label{fn:conservative}Indeed, if the true effect
  distribution puts mass $1/2$ on $\theta_{i}=K$ and $\theta_{i}=-K$, then, as
  $K$ gets large, our \acp{EBCI} become arbitrarily conservative relative to an
  oracle that reports the highest posterior density set under this prior.}
\Cref{sec:general_shrinkage} shows how our method can be adapted to construct
\acp{EBCI} that are locally near-optimal under non-normal baseline priors using
non-linear shrinkage, such as soft thresholding. Since the distribution of
$\theta_{i}$ is nonparametrically identified under the normal
model~\eqref{eq:hierarch_y}, it is in principle possible to construct \acp{EBCI}
that are globally efficient using nonparametric methods. In the context of the
homoskedastic model with no covariates in \Cref{eq:homoskedastic-normal-means},
various approaches to nonparametric point estimation of the $\theta_{i}$'s have
been proposed, including kernels \citep{BrGr09}, splines \citep{efron2019}, or
nonparametric maximum likelihood \citep{KiWo56,JiZh09,KoMi14}. An interesting
problem for future research is to adapt these methods to \ac{EBCI} construction,
while ensuring asymptotic validity, good finite-sample performance, and allowing
for covariates, heteroskedasticity, and possible dependence across $i$.

\subsection{Other inference problems}\label{sec:compar_other_inf}

A number of alternative inference procedures have been proposed in the context
of the normal means model. \citet{Efron2015} develops a formula for the
frequentist standard error of \ac{EB} estimators, but this cannot be used to
construct \acp{CI} without a corresponding estimate of the bias. There is a
substantial literature on shrinkage confidence balls, i.e., confidence sets of
the form $\{\theta\colon\sum_{i=1}^{n} (\theta_i-\hat\theta_i)^2\le \hat c\}$
(see \citealp{casella_shrinkage_2012}, for a review). While theoretically
interesting, these sets can be difficult to visualize and report in
practice.\footnote{Confidence balls can be translated into average coverage
  intervals using Chebyshev's inequality \citep[see][Ch. 5.8]{Wasserman2006}.
  However, such intervals are very conservative compared to the ones we
  construct.}

Finally, while we focus on \ac{CI} length in our relative efficiency
comparisons, our approach can be fruitfully applied when the goal of \ac{CI}
construction is to discern non-null effects, rather than to construct short
\acp{CI}. In particular, suppose one forms a test of the null hypothesis
$H_{0,i}:\theta_i=\theta_0$ for some null value $\theta_0$ by rejecting when
$\theta_0\notin CI_i$, where $CI_i$ is our robust \ac{EBCI} given
in~\eqref{eq:conditional_ci}. In \Cref*{sec:power-details}, we show that the
test based on our \ac{EBCI} has higher average power than the usual $z$-test
based on the unshrunk estimate when $X_i'\delta$ (the regression line towards
which we shrink) is far enough from the null value $\theta_0$, and that these
power gains can be substantial. Furthermore, such tests can be combined with
corrections from the multiple testing literature to form procedures that
asymptotically control the \ac{FDR}, a commonly used criterion for multiple
testing.\footnote{In particular, \citet{storey_direct_2002} shows that the
  \citet{benjamini_controlling_1995} procedure asymptotically controls the
  \ac{FDR} so long as the $p$-values do not exhibit too much statistical
  dependence and the proportion of rejected null hypotheses does not converge
  too quickly to zero. While \citet{storey_direct_2002} assumes that the
  uncorrected tests control size in the classical sense, the argument goes
  through essentially unchanged so long as the tests invert \acp{CI} that
  satisfy (\ref{eq:alt_aci_def}), which holds so long as the \acp{CI} do not
  exhibit too much statistical dependence, as discussed in
  \Cref{rem:ac_alt_def_remark}. We note, however, that this does not hold for
  modifications of the \citet{benjamini_controlling_1995} procedure that use
  initial estimates of the proportion of true null hypotheses.}

\section{Extensions}\label{sec:extensions}

We now discuss two extensions of our method: adapting our intervals to general,
possibly non-linear shrinkage, and constructing intervals that achieve coverage
conditional on $Y_{i}$ falling into a pre-specified interval.

\subsection{General shrinkage}\label{sec:general_shrinkage}

Our method can be generalized to cover general, possibly non-linear shrinkage
based on possibly non-Gaussian data. Let
$\mathcal{S}(y; \chi, \tilde{X}_{i})\subseteq \mathbb{R}$ be a family of
candidate confidence sets for a parameter $\theta_{i}$, which depends on the
data $Y_{i}=y$, a tuning parameter $\chi\in\mathbb{R}$ to be selected below, and
covariates $\tilde{X}_{i}$ (that include any known nuisance parameters) that we
treat as fixed. We assume that $\mathcal{S}$ is increasing in $\chi$, in the
sense of set containment, and that the non-coverage probability conditional on
$\theta$ satisfies
\begin{equation}\label{eq:conditional_coverage_general}
  P(\theta_{i}\not\in
  \mathcal{S}(Y_{i}; \chi, \tilde{X}_{i}) \mid \theta, \tilde{X}^{(n)})=
  \tilde{r}(a_{i}, \chi),
\end{equation}
where $a_{i}$ is some function of $\theta_i$,
$\tilde{X}^{(n)}=(\tilde{X}_{1}, \dotsc, \tilde{X}_{n})$, and $\tilde r$ is a
known function (perhaps computed numerically or through simulation). Similarly
to linear shrinkage in the normal means model,
\Cref{eq:conditional_coverage_general} may only hold approximately if the set
$\mathcal{S}$ depends on estimated parameters (such as standard error estimates
or tuning parameters), or if we use a large-sample approximation to the
distribution of $Y_{i}$. We assume that $a_{i}$ satisfies the moment constraints
$E_{F}[g(a_{i})\mid \tilde{X}^{(n)}]=m$, where $g$ is a $p$-vector of moment
functions, and the expectation is over the conditional distribution $F$ of
$a_{i}$ conditional on $\tilde{X}^{(n)}$.\footnote{The moment functions $g$ need
  not be simple moments, and could incorporate constraints used for selection of
  hyper-parameters, such as constraints on the marginal data distribution or, if
  an unbiased risk criterion is used, the constraint that the derivative of the
  risk equals zero at the selected prior hyper-parameters.} To guarantee \ac{EB}
coverage, we compute the maximal non-coverage
\begin{equation}\label{eq:rho_general}
  \rho_{g}(m, \chi)=\sup_{F}E_{F}[\tilde r(a, \chi)], \qquad E_{F}[g(a)]=m,
\end{equation}
analogously to \Cref{eq:fourth_moment_bound}. This is a linear program, which
can be computed numerically to a high degree of precision even with several
constraints; see \Cref{sec:comput} for details. Given an estimate $\hat m$ of
the moment vector $m$, we form a robust \ac{EBCI} as
\begin{equation}\label{eq:ebci_general}
  \mathcal{S}(Y_{i};\hat \chi, \tilde{X}_{i}),
  \quad\text{where}\quad
  \hat\chi = \inf\{\chi\colon \rho_{g}(\hat{m}, \chi)\leq \alpha\}.
\end{equation}

\begin{example}[Linear shrinkage in the normal model]\label{example:main_case}
  The setting in \Cref{sec:baseline-model} obtains if we set
  $\tilde{X}_{i}=(X_{i}, \sigma_{i})$ and
  $\mathcal{S}(y;\chi, \tilde{X}_{i})=\{(1-w_{EB, i})X_{i}'\delta+w_{EB,
    i}Y_{i}\pm \chi w_{EB, i}\sigma_{i}\}$. Here $a_{i}$ is given by the
  normalized bias $b_{i}=(1/w_{EB, i}-1)(\theta_{i}-X_{i}'\delta)/\sigma_{i}$,
  and the function $\tilde r$ is given by the function $r(b, \chi)$ defined
  in~\eqref{eq:rb}. Our baseline implementation uses constraints on the second
  and fourth moments, $g(a_{i})=(a_{i}^{2}, a_{i}^{4})$.
\end{example}

\begin{example}[Nonlinear soft thresholding]\label{example:soft_thresholding}
  Consider for simplicity the homoskedastic normal model
  $Y_i \mid \theta_i \sim N(\theta_i, \sigma^2)$ without covariates. A popular
  alternative to linear estimators is the soft thresholding estimator
  $\hat{\theta}_{ST, i} = \text{sign}(Y_i)\max\lbrace
  \abs{Y_i}-\sqrt{2\sigma^2/\mu_2},0\rbrace$ \citep[e.g.][]{Abadie2019}. It equals the
  posterior mode corresponding to a baseline Laplace prior with second moment
  $\mu_2$, which has density
  $\pi_0(\theta)=\frac{1}{\sqrt{2\mu_{2}}} \exp(-\abs{\theta}\sqrt{2/\mu_{2}})$
  \citep[Example 2.5]{johnstone19}. To construct a robust \ac{EBCI} that always
  contains the soft thresholding estimator, we calibrate the corresponding
  highest posterior density set:
  \begin{equation}\label{eq:hpd_st}
    \mathcal{S}(Y_i; \chi) = \left\lbrace t \in \mathbb{R} \colon \log
      \frac{\sigma^{-1}\phi((Y_i- t)/\sigma)\pi_0(t)
      }{\int_{-\infty}^\infty \sigma^{-1}\phi((Y_i-
        \tilde{\theta})/\sigma)\pi_0(\tilde{\theta})\, d\tilde{\theta}} + \chi \geq
      0 \right\rbrace,
  \end{equation}
  where $\phi$ is the standard normal density. This set is available in closed
  form and takes the form of an interval (see \Cref*{sec:appendix_softthresh}).
  Here $a_{i}=\theta_{i}$, and the function $\tilde r(a, \chi)$
  in~\eqref{eq:conditional_coverage_general} can be computed via numerical
  integration.

  In contrast to the \acp{EBCI} in \Cref{example:main_case} (which
  may be viewed as calibrating the highest posterior density set under a
  \emph{normal} prior), the Laplace prior $\pi_{0}$ leads to nonlinear
  shrinkage and an \ac{EBCI} whose length depends on the data $Y_i$. This
  reflects the suboptimality of linear shrinkage and fixed-length intervals under
  the Laplace prior.

  In \Cref*{sec:appendix_softthresh}, we show that the resulting robust
  \ac{EBCI} that imposes the constraint $E[\theta_{i}^{2}]=\mu_{2}$ not only has
  robust \ac{EB} coverage (by definition), it also achieves substantial expected
  length improvements when the $\theta_i$'s are in fact Laplace distributed. For
  $\alpha=0.05$ and $\mu_2/\sigma^2 \leq 0.2$, the expected length under the
  Laplace distribution of the soft thresholding \ac{EBCI} is at least 49\%
  smaller than the length of the unshrunk CI\@. This exceeds the length
  reduction achieved by the linear robust \ac{EBCI} shown in
  \Cref{fig:efficiency_unshrunk}.
\end{example}

\begin{example}[Poisson shrinkage]\label{example:poisson}
  \Cref*{sec:appendix_poisson} constructs a robust \ac{EBCI} for the rate
  parameter $\theta_i$ in a Poisson model
  $Y_i \mid \theta_i \sim \text{Poisson}(\theta_i)$. This example demonstrates
  that our general approach does not require normality of the data.
\end{example}

\begin{example}[Linear estimators in other settings]\label{example:general_linear}

  While our focus has been on \ac{EB} shrinkage, our approach applies to other
  settings in which an estimator $\hat\theta_i$ is approximately normally
  distributed with non-negligible bias. In particular, suppose
  $(\hat\theta_i-\theta_i)/\se_i$ is distributed $N(a_i,1)$, where $\se_i$ is
  the standard deviation of the estimate $\hat\theta_i$, which for simplicity we
  take to be known. This holds whenever $\hat\theta_i$ is a linear function of
  jointly normal observations $W_{1}, \dotsc, W_{N}$, i.e.,
  $ \hat{\theta}_{i}=\sum_{j=1}^{N} k_{ij}W_{j}$ for some deterministic weights $k_{ij}$.
  Examples include series, kernel, or local polynomial estimators in a
  nonparametric regression with fixed covariates and normal errors. We can
  construct a confidence interval for $\theta_{i}$ as
  $\hat\theta_i\pm \chi \cdot \se_i$, in which case
  \Cref{eq:conditional_coverage_general} holds with $\tilde{r}=r$ given in
  \Cref{eq:rb}. It follows from \Cref{high_level_conditional_coverage_thm} in
  \Cref{coverage_results_sec_append} that if the moment constraints $m$ on the
  normalized bias in~\Cref{eq:rho_general} are replaced by consistent estimates,
  the resulting robust \ac{EBCI} will satisfy the average coverage
  property~\eqref{eq:aci_def} in large samples. We leave a full treatment of
  these applications for future research.
\end{example}

\subsection{Coverage after selection}\label{sec:cover_selection}

In some applications, researchers may be primarily interested in parameters
corresponding to those units $i$ whose initial estimates $Y_{i}$ fall in a given
interval $[\iota_1,\iota_2]$, where
$-\infty \leq \iota_1 < \iota_2 \leq \infty$. For example, in a teacher value
added application, we may only be interested in the ability $\theta_i$ of those
teachers $i$ whose fixed effect estimates $Y_i$ are positive, corresponding to
setting $\iota_1=0$ and $\iota_2=\infty$. Because of the selection on outcomes,
na\"{i}vely applying our baseline \ac{EBCI} procedure to the selected sample
$\lbrace i\colon Y_i \in [\iota_1,\iota_2]\rbrace$ does not yield the desired
average coverage across the selected units $i$. We now show how to correct for
the selection bias in the simple homoskedastic model
$Y_i \mid \theta_i \sim N(\theta_i, \sigma^2)$ without covariates from
\Cref{sec:simple-example} (reintroducing the extra model features in
\Cref{sec:baseline-model} only complicates notation).

We seek a critical value $\chi$ such that the average coverage of the \ac{CI}
$[\hat{\theta}_i \pm \chi w_{EB} \sigma]$ is at least $1-\alpha$
\emph{conditional} on the sample selection, i.e.,
\begin{equation}\label{eq:conditional_ebci_def}
  P(\theta_i \in \hat{\theta}_i \pm \chi w_{EB} \sigma \mid Y_i \in [\iota_1,\iota_2]) \geq 1-\alpha
\end{equation}
under repeated sampling of $(Y_i, \theta_i)$, regardless of the distribution for
$\theta_i$ (we maintain focus on linear shrinkage for simplicity, but our
approach extends to nonlinear shrinkage using the ideas in
\Cref{sec:general_shrinkage}). Straightforward calculations show that the
non-coverage, conditional on $\theta_{i}$ and on selection, equals
\begin{multline*}
  \tilde{r}_{\iota_1,\iota_2}(\theta_i, \chi) = P(\theta_i \notin \hat{\theta}_i \pm \chi w_{EB} \sigma \mid Y_i \in [\iota_1,\iota_2], \theta_i) \\
  = \min\left\lbrace 1-\frac{\Phi(\min\lbrace
      \chi-b_{i}, (\iota_2-\theta_i)/\sigma\rbrace)-\Phi(\max\lbrace
      -\chi-b_{i}, (\iota_1-\theta_i)/\sigma\rbrace)}{\Phi((\iota_2-\theta_i)/\sigma)-\Phi((\iota_1-\theta_i)/\sigma)},
    1\right\rbrace,
\end{multline*}
where $b_{i}=(1-1/w_{EB})\theta_{i}/\sigma$ as in \Cref{sec:simple-example}.
Among all distributions for $\theta_i$ consistent with the conditional moment
$\tilde{\mu}_{2,\iota_1,\iota_2}=E\left[\theta_i^2 \mid Y_i \in
  [\iota_1,\iota_2]\right]$, the worst-case non-coverage probability,
conditional on selection, is given by
\begin{equation*}
  \tilde{\rho}_{\iota_1,\iota_2}(\tilde{\mu}_{2,\iota_1,\iota_2}, \chi) \equiv
  \sup_{F} E_{F}[\tilde{r}_{\iota_1,\iota_2}(\theta_i, \chi)] \quad \text{s.t.} \quad
  E_{F}[\theta_i^2]=\tilde{\mu}_{2,\iota_1,\iota_2},
\end{equation*}
where $E_{F}$ denotes expectation under $\theta_{i}\sim F$. This is an
infinite-dimensional linear program that can be solved numerically to a high
degree of accuracy, cf.\ \Cref{sec:comput}. To achieve robust conditional
coverage, we solve numerically for the $\chi$ such that
$\tilde{\rho}_{\iota_1,\iota_2}(\tilde{\mu}_{2,\iota_1,\iota_2}, \chi) =
\alpha$.

We can estimate the conditional second moment $\tilde{\mu}_{2,\iota_1,\iota_2}$
as follows. Denote the log marginal density of $Y_i$ by
$\ell(y) \equiv \log \int \phi(y-\theta)\, d\Gamma_0(\theta)$, where $\Gamma_0$
is the true distribution of $\theta_i$. Tweedie's formulas
\citep[e.g.][Eq.~(26)]{efron2019} imply
\begin{equation}\label{eq:cond_cov_moment}
  \tilde{\mu}_{2,\iota_1,\iota_2} = E\left[\theta_i^2 \mid Y_i
    \in [\iota_1,\iota_2]\right] = 1 +
  E\left[(Y_i + \ell'(Y_i))^2 + \ell''(Y_i) \;\big|\; Y_i \in [\iota_1,\iota_2] \right].
\end{equation}
Let $\hat{\ell}(y)$ be a kernel estimate of the log marginal density function of
the data $Y_1, \dotsc, Y_n$. Then the estimate
\begin{equation*}
  \widehat{\tilde{\mu}}_{2,\iota_1,\iota_2} \equiv 1 +
  \frac{\sum_{i \colon Y_i \in [\iota_1,\iota_2]}
    \lbrace (Y_i + \hat{\ell}'(Y_i))^2 + \hat{\ell}''(Y_i)\rbrace}{{\#}\lbrace i \colon Y_i \in [\iota_1,\iota_2] \rbrace}
\end{equation*}
will be consistent as $n\to\infty$ for $\tilde{\mu}_{2,\iota_1,\iota_2}$
in~\eqref{eq:cond_cov_moment} under mild regularity conditions.

The criterion~\eqref{eq:conditional_ebci_def} can be viewed as the \ac{EB}
analogue of the criterion
$P(\theta_i \in CI_i \mid Y_i \in [\iota_1,\iota_2], \theta) \geq 1-\alpha$,
which requires frequentist coverage conditional on the event
$\{Y_i\in [\iota_1,\iota_2]\}$. The latter criterion has been considered in the recent ``selective inference'' literature
\citep{benjamini_false_2005,lee_exact_2013,HuFi19,Andrews2021}. In contrast to this literature, we cannot allow $\iota_1$ to be given by the maximum of the
initial estimates \citep[as in][]{Andrews2021}, as we require $\iota_{1}$ and $\iota_{2}$ to converge in probability to distinct nonrandom limits. On the other hand, weakening the
notion of frequentist coverage to \ac{EB} (or average) coverage allows for
improvements in the length of the intervals, similar to the analysis in \Cref{sec:efficiency} in the absence of selection.

\section{Empirical application}\label{sec:empir-appl}

We illustrate our methods using the data and model in
\citet{chetty_impacts_2018}, who are interested in the effect of neighborhoods
on intergenerational mobility.

\subsection{Framework}
We follow \citet{chetty_impacts_2018} in using two definitions of a
``neighborhood effect'' $\theta_{i}$. The first focuses on effects for children
growing up in low-income families, and defines $\theta_{i}$ as the effect of
spending an additional year of childhood in \ac{CZ} $i$ on children's rank in
the income distribution at age 26, for children with parents at the 25th
percentile of the national income distribution. The second definition is
analogous, except it focuses on children growing up in high-income families, and
consequently conditions on children with parents at the 75th percentile.
\citet{chetty_impacts_2018} argue that these definitions approximately capture
the mean rank effects for children in below-median and above-median income
families. Using de-identified tax returns for all children born between 1980 and
1986 who move across \acp{CZ} exactly once as children,
\citet{chetty_impacts_2018} exploit variation in the age at which children move
between \acp{CZ} to obtain preliminary fixed effect estimates $Y_{i}$ of
$\theta_{i}$.

Since these preliminary estimates are measured with noise, to predict $\theta_{i}$, \citet{chetty_impacts_2018} shrink $Y_{i}$ towards
average outcomes of permanent residents of \ac{CZ} $i$ (children with parents at
the same percentile of the income distribution who spent all of their childhood
in the \ac{CZ}). To give a sense of the accuracy of these forecasts,
\citet{chetty_impacts_2018} report estimates of their unconditional \ac{MSE}
(i.e.,\ treating $\theta_{i}$ as random), under
the implicit assumption that the moment independence assumption in
\Cref{eq:moment_independence} holds. Here we complement their analysis by
constructing robust \acp{EBCI} associated with these forecasts.

Our sample consists of 595 U.S. \acp{CZ}, with population over 25,000 in the
2000 census: this is the sample for which \citet{chetty_impacts_2018} report
baseline estimates $Y_{i}$ of the effects $\theta_{i}$.
These baseline estimates are normalized so that their population-weighted mean
is zero. We may therefore interpret $\theta_{i}$ as the effect relative to an
``average'' \ac{CZ}. We follow the baseline implementation from
\Cref{sec:baseline-implementation} with standard errors $\hat{\sigma}_{i}$
reported by \citet{chetty_impacts_2018}, and covariates $X_{i}$ corresponding to
a constant and the average outcomes for permanent residents. In line with the
original analysis, we use precision weights $\omega_i=1/\hat{\sigma}^{2}_{i}$
when constructing the estimates $\hat{\delta}$, $\hat{\mu}_{2}$ and
$\hat{\kappa}$.

\subsection{Results}

\begin{table}[p]
  \centering
  \begin{threeparttable}
    \caption{Statistics for 90\% \acp{EBCI} for neighborhood
      effects.}\label{tab:ch}
    \begin{tabular*}{0.9\linewidth}{@{\extracolsep{\fill}}@{}l@{}lrrrr@{}}
      && \multicolumn{2}{c}{Baseline} & \multicolumn{2}{c}{Nonparametric} \\
      \cmidrule(rl){3-4}  \cmidrule(rl){5-6}
      && \multicolumn{1}{r}{(1)} & \multicolumn{1}{r}{(2)}
      & \multicolumn{1}{r}{(3)} & \multicolumn{1}{r}{(4)}\\
      \phantom{a}&Percentile &  25th     &       75th  &      25th &   75th \\
      \midrule
      \multicolumn{2}{@{}l}{Panel A\@{}: Summary statistics}\\
      \cmidrule{1-2}
      & $E[\sqrt{\mu_{2,i}}]$                   &   0.079&    0.044&    0.076&     0.042\\
      & $E[\kappa_{i}]$                         & 778.5& 5948.6& 1624.9& 43009.9\\
      & $E[\mu_{2i}/\sigma_i^2]$                &   0.142&    0.040&    0.139&     0.072\\
      & $\hat{\delta}_{\text{intercept}}$             &  $-1.441$&   $-2.162$&   $-1.441$&    $-2.162$\\
      & $\hat{\delta}_{\text{perm.\ resident}}$       &   0.032&    0.038&    0.032&     0.038\\
      & $E[w_{EB, i}]$                          &   0.093&    0.033&    0.093&     0.033\\
      & $E[w_{opt, i}]$                         &   0.191&    0.100&    0.191&     0.100\\
      & $E[\text{non-cov of parametric EBCI}_i]$&   0.227&    0.278&    0.210&     0.292\\
      \multicolumn{2}{@{}l}{Panel B\@: $E[\text{half-length}_i]$}\\
      \cmidrule{1-2}
      & Robust EBCI          & 0.195&    0.122&    0.186&     0.116   \\
      & Optimal robust EBCI  & 0.149&    0.090&    0.145&     0.094   \\
      & Parametric EBCI      & 0.123&    0.070&    0.123&     0.070   \\
      & Unshrunk CI          & 0.786&    0.993&    0.786&     0.993   \\[1ex]
      \multicolumn{2}{@{}l}{Panel C\@: Efficiency relative to robust EBCI}\\
      \cmidrule{1-2}
      & Optimal robust EBCI  & 1.312&    1.352&    1.289&     1.238 \\
      & Parametric EBCI      & 1.582&    1.731&    1.509&     1.648 \\
      & Unshrunk CI          & 0.248&    0.123&    0.237&     0.117 \\
    \end{tabular*}
    \begin{tablenotes}
    \item \emph{Notes:} Columns (1) and (2) correspond to shrinking $Y_{i}$ as
      in the baseline implementation that imposes \Cref{eq:moment_independence},
      so that $\mu_{2i}=E[(\theta_{i}-X_{i}'\delta)^{2}\mid X_{i}, \sigma_{i}]$
      and
      $\kappa_{i}=E[(\theta_{i}-X_{i}'\delta)^{4}\mid X_{i},
      \sigma_{i}]/\mu_{2i}^{2}$ do not vary with $i$. Columns (3) and (4) use
      nonparametric estimates of $\mu_{2i}$ and $\kappa_{i}$, using the nearest
      neighbor estimator described in \Cref{sec:finite_n}. The number of nearest
      neighbors $J=422$ (column (3)) and $J=525$ (column (4)) is selected using
      cross-validation. For all columns,
      $\hat{\delta}=(\hat{\delta}_{\text{intercept}}, \hat{\delta}_{\text{perm.\
          resident}})$ is computed by regressing $Y_{i}$ onto a constant and
      outcomes for permanent residents. ``Optimal Robust EBCI'' refers to a
      robust EBCI based on length-optimal shrinkage $w_{opt, i}$, described in
      \Cref{sec:efficiency}. ``$E[\text{non-cov of parametric EBCI}_i]$'':
      average of maximal non-coverage probability of parametric EBCI, given the
      estimated moments.
    \end{tablenotes}
  \end{threeparttable}
\end{table}

Columns (1) and (2) in \Cref{tab:ch} summarize the main estimation and
efficiency results. The shrinkage magnitude and relative efficiency results are
similar for children with parents at the 25th and 75th percentiles of the income
distribution. In both columns, the estimate of the kurtosis $\kappa$ is large
enough so that it does not affect the critical values or the form of the optimal
shrinkage: specifications that only impose constraints on the second moment
yield identical results.\footnote{The truncation in the $\hat{\kappa}$ formula
  in our baseline algorithm in \Cref{sec:baseline-implementation} binds in
  columns (1) and (2), although the non-truncated estimates 345.3 and 5024.9 are
  similarly large; using these non-truncated estimates yields identical
  results.} In line with this finding, a density plot of the $t$-statistics
(reported as Figure S2 in \citet{akp20v2}) exhibits a fat lower tail. As a
robustness check, columns (3) and (4) show that results based on nonparametric
moment estimates (see \Cref{rem:nonparam,sec:finite_n}) are very similar to our baseline specification. Indeed, the $R^{2}$ gain in predicting
$\hat{\varepsilon}_{i}^{2}-\hat{\sigma}^{2}_{i}$ using $\hat{\mu}_{2i}$ is less
than $0.001$ in both specifications, indicating that there is little evidence in
the data against the moment independence assumption.

The baseline robust 90\% \acp{EBCI} are 75.2--87.7\% shorter than the usual
unshrunk \acp{CI} $Y_{i}\pm z_{1-\alpha/2}\hat{\sigma}_{i}$. To interpret these
gains in dollar terms, for children with parents at the 25th percentile of the
income distribution, a percentile gain corresponds to an annual income gain of
\$818 \citep[p.~1183]{chetty_impacts_2018}. Thus, the average half-length of the
baseline robust \acp{EBCI} in column (1) implies \acp{CI} of the form
$\pm \$160$ on average, while the unshrunk \acp{CI} are of the form $\pm \$643$
on average. These large gains are a consequence of a low signal-to-noise ratio
 $\mu_{2}/\sigma_{i}^{2}$ in this application. Because the shrinkage
magnitude is so large on average, the tail behavior of the bias matters, and
since the kurtosis estimates suggests these tails are fat, it is important to
use the robust critical value: the parametric \ac{EBCI} exhibits average
potential size distortions of 12.7--17.8 percentage points. Indeed, for over 90\% of the \acp{CI} in
the specifications in columns (1) and (2), the shrinkage coefficient $w_{EB, i}$
falls below the ``rule of thumb'' threshold of 0.3 derived in
\Cref{sec:param-ebci-cover}.

\begin{figure}[t]
  \centering%
  \tikzset{font=\small} %
  \input{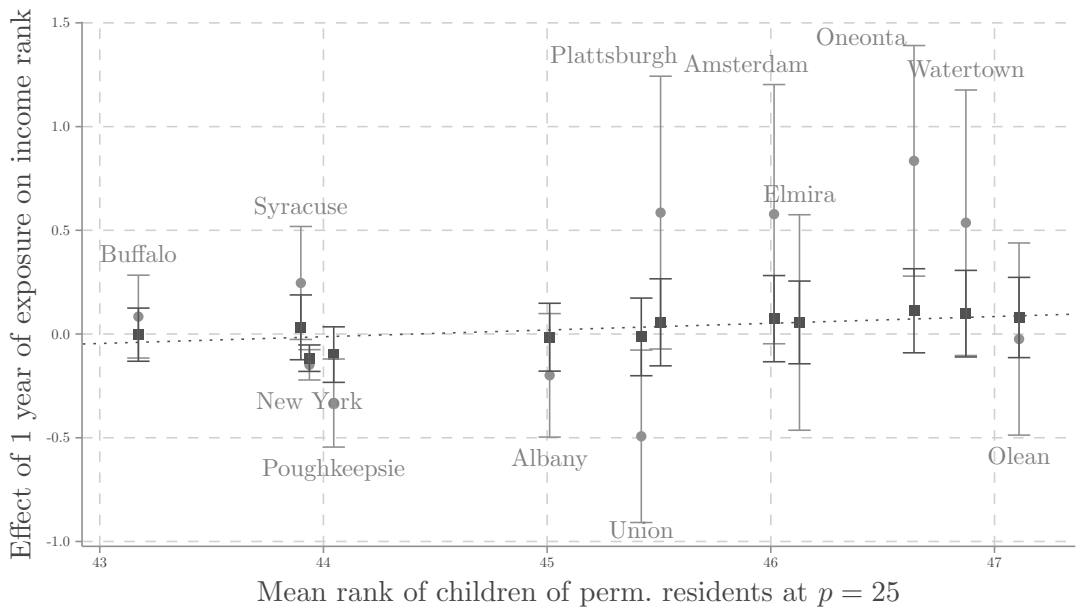}
  \caption{Neighborhood effects for New York and 90\% robust \acp{EBCI} for
    children with parents at the $p=25$ percentile of the national income
    distribution, plotted against mean outcomes of permanent residents. Gray
    lines correspond to \acp{CI} based on unshrunk estimates represented by
    circles, and black lines correspond to robust \acp{EBCI} based on \ac{EB}
    estimates represented by squares that shrink towards a dotted regression
    line based on permanent residents' outcomes. Baseline implementation as in
    \Cref{sec:baseline-implementation}.}\label{fig:ch_cz25NY}
\end{figure}

To visualize these results, \Cref{fig:ch_cz25NY} plots the unshrunk 90\%
\acp{CI} based on the preliminary estimates, as well as robust \acp{EBCI} based
on \ac{EB} estimates for cities in the state of New York for children with
parents at the 25th percentile. While the \acp{EBCI} for large \acp{CZ} like New
York City or Buffalo are similar to the unshrunk \acp{CI}, they are much
tighter for smaller \acp{CZ} like Plattsburgh or Watertown, with point estimates
that shrink the preliminary estimates $Y_{i}$ most of the way toward the regression
line $X_{i}'\hat{\delta}$.

In summary, shrinkage allows us to considerably tighten the \acp{CI} based on
preliminary estimates. This is true even though the \acp{CI} effectively only use second moment constraints---imposing kurtosis constraints does not affect
the critical values in this application.

\begin{appendices}
\crefalias{section}{appsec}
\crefalias{subsection}{appsubsec}

\section{Moment estimates}\label{sec:moment_estimates}

The \ac{EBCI} in our baseline implementation has valid \ac{EB} coverage asymptotically
as $n\to\infty$, so long as the estimates $\hat\mu_2$ and $\hat\kappa$ are
consistent. While the particular choice of the estimates $\hat\mu_2$ and
$\hat\kappa$ does not affect the \ac{CI} asymptotically, finite sample considerations
can be important for small to moderate values of $n$. In particular, unrestricted moment-based estimates of $\mu_{2}$ and $\kappa$ may fall
below their theoretical lower bounds of $0$ and $1$, in which case it is not
clear how to define the \ac{EBCI}.\footnote{Formally, our results are asymptotic
  and require $\mu_2>0$ and $\kappa>1$, so that these issues do not occur when
  $n$ is large enough. We discuss the difficulty of providing finite-sample coverage guarantees in \Cref{sec:compar}.} To address this issue, in analogy to finite-sample
corrections to parametric \acp{EBCI} proposed in \citet{morris83pebci,morris83},
\Cref{sec:finite_n} derives two finite-sample corrections to the unrestricted
estimates that approximate a Bayesian estimate under a flat hyperprior on
$(\mu_{2}, \kappa)$. We verify that these corrections give good coverage in an
extensive set of Monte Carlo designs in \Cref{sec:sim}. We also discuss
implementation of nonparametric moment estimates. \Cref{sec:weighting} discusses
the choice of weights $\omega_{i}$.

\subsection{Finite \texorpdfstring{$n$}{n} corrections and nonparametric moment
estimates}\label{sec:finite_n}

To derive our estimates of $\mu_2$ and $\kappa$, we first consider unrestricted
estimation under the moment independence
condition~\eqref{eq:moment_independence}. For $\mu_2$, this condition implies
the moment condition
$E[(Y_{i}-X_{i}'\delta)^{2}-\sigma^{2}_{i}\mid X_{i}, \sigma_{i}]=\mu_{2}$.
Replacing $Y_{i}-X_{i}'\delta$ with the residual
$\hat\varepsilon_i=Y_i-X_i'\hat\delta$ yields the estimate
\begin{equation}\label{eq:mu2_UC}
  \hat{\mu}_{2, \text{UC}}
  = \frac{\sum_{i=1}^n \omega_{i}\WW_{2i}}{\sum_{i=1}^n\omega_i},
  \quad
  \WW_{2i}=\hat{\varepsilon}_i^2-\hat\sigma_i^2,
\end{equation}
for any weights $\omega_i=\omega_i(X_i, \hat\sigma_i)$. Here, UC stands for
``unconstrained,'' since the estimate $ \hat{\mu}_{2, \text{UC}}$ can be
negative. To incorporate the constraint $\mu_2>0$, we use an approximation to a
Bayesian approach with a flat prior on the set $\hor{0,\infty}$. A full Bayesian
approach to estimating $\mu_2$ would place a hyperprior on possible joint
distributions of $X_i, \sigma_i, \theta_i$, which could potentially lead to using
complicated functions of the data to estimate $\mu_2$. For simplicity, we
compute the posterior mean given $\hat\mu_{2,\text{UC}}$, and we use a normal
approximation to the likelihood. Since the posterior distribution only uses
knowledge of $\hat\mu_{2,\text{UC}}$, we refer to this as a \acf{FPLIB} approach.

To derive this formula, first note that, if $\hat m$ is an estimate of a
parameter $m$ with $\hat m\mid m\sim N(m, V)$, then under a flat prior for $m$
on $\hor{0, \infty}$, the posterior mean of $m$ is given by
\begin{equation*}
  b(\hat m, V) = \hat m + \sqrt{V}\phi(\hat m/\sqrt{V})/\Phi(\hat m/\sqrt{V}),
\end{equation*}
where $\phi$ and $\Phi$ are the standard normal pdf and cdf respectively.
Furthermore, if $\hat{m}=\sum_{i=1}^n\omega_{i} Z_{i}/\sum_{i=1}^{n}\omega_{i}$
where the $Z_i$'s are independent with mean $m$ conditional on the weights
$\omega=(\omega_1, \dotsc, \omega_n)'$, then an unbiased estimate of the
variance of $\hat{m}$ given $\omega$ is given by
\begin{equation*}
  V(Z, \omega) = \frac{\sum_{i=1}^n
    \omega_i^2(Z_i^2-\hat{m}^2)}{\left(\sum_{i=1}^n\omega_i \right)^2 - \sum_{i=1}^n\omega_i^2}.
\end{equation*}
Conditioning on the $X_i$'s and $\sigma_i$'s (and ignoring sampling variation in
$\hat{\delta}$ and the $\hat{\sigma}_i$'s), we can then apply this formula to
$\hat\mu_{2,\text{UC}}$, with $Z_i=\WW_{2i}$, where $\WW_{2i}$ is given
in~\eqref{eq:mu2_UC}. This gives the \ac{FPLIB} estimate for $\mu_2$:
\begin{equation*}
  \hat\mu_{2,\text{FPLIB}} = b(\hat\mu_{2,\text{UC}}, V(\WW_2,\omega)).
\end{equation*}
To derive the \ac{FPLIB} estimate for $\kappa$, we begin with an unconstrained
estimate of $\mu_{4}=E[(\theta_i-X_{i}'\delta)^{4}]$. The moment independence
condition~\eqref{eq:moment_independence} delivers the moment condition
$\mu_{4}=E[(Y_{i}-X_{i}'\delta)^{4}+
3\sigma^{4}_{i}-6\sigma^{2}_{i}(Y_{i}-X_{i}'\delta)^{2} \mid X_{i},
\sigma_{i}]$, which leads to the unconstrained estimate
\begin{equation*}
  \hat\mu_{4,\text{UC}}=\frac{\sum_{i=1}^n \omega_i \WW_{4i}}{\sum_{i=1}^n \omega_i},
  \quad \WW_{4i} = \hat\varepsilon_i^4 - 6\hat\sigma_i^2\hat\varepsilon_i^2 + 3\hat \sigma_i^4.
\end{equation*}
To avoid issues with small values of estimates of $\mu_2$ in the denominator, we
apply the \ac{FPLIB} approach to an estimate of $\mu_{4}-\mu_{2}^{2}$, using a
flat prior on the parameter space $\hor{0,\infty}$. Using the delta method leads
to approximating the variance of $\hat\mu_{4,\text{UC}}-\hat\mu_{2,\text{UC}}^2$
with the variance of
$\sum_{i=1}^n \omega_i(\WW_{4i}-2\mu_2\WW_{2i})/\sum_{i=1}^n\omega_i$, so that
the \ac{FPLIB} estimate of $\mu_{4}-\mu_2^2$ is
$b(\hat\mu_{4,\text{UC}}-\hat\mu_{2,\text{UC}}^2,
V(\WW_{4}-2\hat\mu_{2,\text{FPLIB}}\WW_{2}, \omega))$, and the \ac{FPLIB}
estimate of $\kappa$ is
\begin{equation*}
  \hat\kappa_{\text{FPLIB}} = 1 + \frac{b(\hat\mu_{4,\text{UC}}-\hat\mu_{2,\text{UC}}^2, V(\WW_{4}-2\hat\mu_{2,\text{FPLIB}}\WW_{2}, \omega))}{\hat\mu_{2,\text{FPLIB}}^2}.
\end{equation*}

As a further simplification, we derive approximations in which the posterior
mean formula $b(\hat m, V)$ is replaced by a simple truncation formula. We refer
to this approach as \acf{PMT}. In particular, suppose we apply the formula
$b(\hat m, V)$ to an estimator $\hat{m}$ such that $\hat m\ge m_0$ and
$V\ge V_0$ by construction, where $m_0<0$. Then the posterior mean satisfies
$b(\hat m, V) \ge b(m_0, V_0)$ \citep[][Proposition
1.2]{pinelis_monotonicity_2002}. Thus, a simple approximation to the \ac{FPLIB}
estimator is to truncate $\hat m$ from below at $b(m_0, V_0)$. To obtain an even
simpler formula, we use the approximation
$b(m_0, V_0) = -V_0/m_0 + O(V_0^{3/2})$ \citep[][Proposition
1.3]{pinelis_monotonicity_2002}, which holds as $V_{0}\to 0$ (or, equivalently,
as $n\to\infty$, provided the estimator $\hat{m}$ is consistent). The variance
of $\hat{\mu}_{2,\text{UC}}$ conditional on $(X_{i}, \sigma_{i})$ is bounded
below by
$2\sum_{i=1}^n \omega_i^2 \sigma_i^4/\left(\sum_{i=1}^n \omega_i \right)^2$, and
$\hat\mu_{2,\text{UC}}\ge -\sum_{i=1}^n
\omega_i\sigma_i^2/\sum_{i=1}^n\omega_i$, so we can use
$V_{0}/m_{0}=-\frac{2\sum_{i=1}^n \omega_i^2 \sigma_i^4}{\sum_{i=1}^n
  \omega_i\sigma_i^2\cdot \sum_{i=1}^n \omega_i}$, which gives the \ac{PMT}
estimator
\begin{equation*}
  \hat\mu_{2,\text{PMT}} =
  \max\left\{\hat\mu_{2,\text{UC}}, \frac{2\sum_{i=1}^n \omega_i^2 \sigma_i^4}{\sum_{i=1}^n
      \omega_i\sigma_i^2\cdot \sum_{i=1}^n \omega_i}\right\}.
\end{equation*}
For $\kappa$, we simplify our approach to deriving a trimming rule by treating
$\mu_{2}$ as known, and considering the variance of the infeasible estimate
$\hat\kappa^*_{\text{UC}}=\frac{\sum_{i=1}^{n}
  \omega_i(\hat{\varepsilon}_{i}^{4}-6\hat{\sigma}_i^2\mu_2
  -3\hat{\sigma}_i^4)}{\mu_{2}^{2}\sum_{i=1}^n\omega_i}$. Using the above
truncation formula for $\hat\kappa^*_{\text{UC}}-1$ along with the fact that
$\hat\kappa^*_{\text{UC}}\geq \frac{\sum_{i=1}^{n}
  \omega_i(-6\hat{\sigma}_i^2\mu_2
  -3\hat{\sigma}_i^4)}{\mu_{2}^{2}\sum_{i=1}^n\omega_i}$ and the lower bound
$8\sum_{i}\omega_{i}^{2}(2\mu_{2}^{3}\sigma_{i}^{2} +
21\mu_{2}^{2}\sigma_{i}^{4} +
48\mu_{2}\sigma_{i}^{6}+12\sigma_{i}^{8})/\mu_{2}^{4}(\sum_{i}\omega_{i})^{2}$
on the variance yields
$V_{0}/m_{0}= -\frac{8\sum_{i}\omega_{i}^{2}(2\mu_{2}^{3}\sigma_{i}^{2} +
  21\mu_{2}^{2}\sigma_{i}^{4} + 48\mu_{2}\sigma_{i}^{6}+12\sigma_{i}^{8})}{
  \mu_{2}^{2}(\sum_{i}\omega_{i})\sum_{i=1}^{n}
  \omega_i(\mu_{2}^{2}+6\hat{\sigma}_i^2\mu_2 +3\hat{\sigma}_i^4)}$. To simplify
the trimming rule even further, we only use the leading term of $V_{0}/m_{0}$ as
$\mu_{2}\to 0$,
$V_{0}/m_{0}= -\frac{32\sum_{i}\omega_{i}^{2}\sigma_{i}^{8}}{
  \mu_{2}^{2}(\sum_{i}\omega_{i})\sum_{i=1}^{n}
  \omega_i\hat{\sigma}_i^4}+o(1/\mu_{2}^{2}) $. Plugging in
$\hat{\mu}_{2, \text{PMT}}$ in place of the unknown $\mu_{2}$ then gives the \ac{PMT}
estimator
\begin{equation*}
  \hat\kappa_{\text{PMT}} = \max\left\{\frac{\hat\mu_{4,\text{UC}}}{\hat\mu_{2,\text{PMT}}^2},
    1 + \frac{32 \sum_{i=1}^n \omega_i^2\hat\sigma_i^8}{\hat\mu_{2,\text{PMT}}^2\sum_{i=1}^n\omega_i \cdot \sum_{i=1}^n\omega_i\hat\sigma_i^4} \right\}.
\end{equation*}
The estimators in step~\ref{item:param-estimates} of our baseline implementation
in~\Cref{sec:baseline-implementation} correspond to $\hat{\mu}_{2, \text{PMT}}$
and $\hat{\kappa}_{\text{PMT}}$, due to their slightly simpler form relative to
the \ac{FPLIB} estimators. In unreported simulations based on the designs
described in \Cref{sec:sim} and \Cref*{sec:sim-heterosk}, we find that
\acp{EBCI} based on \ac{FPLIB} lead to even smaller finite-sample coverage
distortions than those based on the baseline implementation that uses \ac{PMT},
at the expense of slightly longer average length.

To implement the nonparametric estimates $\hat{\kappa}_{i}$ and $\hat{\mu}_{2i}$
in~\Cref{rem:nonparam}, we use the nearest-neighbor estimator that for each $i$
computes the \ac{PMT} estimates $\hat{\mu}_{2, \text{PMT}}$ and
$\hat{\kappa}_{\text{PMT}}$ described above, using only the $J$ observations
closest to $i$, rather than the full sample of $n$ observations. We define
distance as a Euclidean distance on $(X_{i}, \sigma_{i})$, after scaling elements
of this vector by their standard deviations. Under regularity conditions, the
resulting estimates will be consistent for $\mu_{2i}$ and $\kappa_{i}$, so long
as $J\to\infty$ and $J/n\to 0$. We select $J$ using leave-one-out
cross-validation, using the squared prediction error in predicting $\WW_{2i}$ as
the criterion. For simplicity, we use the same $J$ for estimating the kurtosis as
that used for estimating the second moment.

\subsection{Choice of weighting}\label{sec:weighting}

Under condition~\eqref{eq:moment_independence}, the weights $\omega_i$ used to
estimate $\mu_2$ and $\kappa$ can be any function of $X_i, \sigma_i$.
Furthermore, while $\hat\delta$ can be essentially arbitrary as long as it
converges in probability to some $\delta$ such that
\Cref{eq:moment_independence} holds, that equation will often be
motivated by the assumption that the conditional mean of $\theta_{i}$ is linear
in $X_{i}$,
\begin{equation}\label{eq:linear-mean}
  E[\theta_i-X_i'\delta\mid X_i, \sigma_i]=0.
\end{equation}
Under this condition, the weights $\omega_{i}$ used to estimate $\delta$ can
also be any function of $X_{i}, \sigma_{i}$.

Thus, under conditions~\eqref{eq:moment_independence}
and~\eqref{eq:linear-mean}, the choice of weighting can be guided by efficiency
concerns. In general, the optimal weights are different for each of the three
estimates of $\delta, \mu_{2}$, and $\kappa$, and implementing them requires
first stage estimates of the variances of $Y_{i}$, $\WW_{2i}$ and $\WW_{4i}$,
conditional on $(X_{i}, \sigma_{i})$ (with $\WW_{2i}$ and $\WW_{4i}$ defined in
\Cref{sec:finite_n}). To avoid estimation of these variances, consider the
limiting case where the signal-to-noise ratio goes to $0$,
$\mu_{2}/\min_{i}\sigma^{2}_{i}\to 0$. The resulting weights will be
near-optimal under a low signal-to-noise ratio, when precise estimation of these
parameters is relatively more important for accurate coverage (under a high
signal-to-noise ratio, shrinkage is limited, and estimation error in these
parameters has little effect on coverage). Let us also ignore estimation error
in $\delta$ for simplicity, and suppose that the $Y_{i}$'s are independent
conditional on $(\theta_{i}, X_{i}, \sigma_{i})$. Then, as
$\mu_{2}/\min_{i}\sigma_{i}^2\to 0$, the weights $\hat{\sigma}_{i}^{-2}$,
$\hat{\sigma}_{i}^{-4}$, and $\hat{\sigma}_{i}^{-8}$, for estimating $\delta$,
$\mu_{2}$, and $\mu_{4}$, respectively, become optimal. For simplicity, the
baseline implementation in \Cref{sec:baseline-implementation} uses the same
weights $\omega_i$ for each of the estimates; the choice
$\omega_{i}=\hat{\sigma}_{i}^{-2}$ targets optimal estimation of $\delta$.
However, one could relax this constraint, and use the weights
$\hat{\sigma}_{i}^{-4}$, and $\hat{\sigma}_{i}^{-8}$ for estimating $\mu_{2}$
and $\mu_{4}$ instead. The choice $\omega_{i}=1/n$ has the advantage of
simplicity; one may also motivate it by robustness concerns when
\Cref{eq:moment_independence} fails, though our preferred robustness check is to
use nonparametric moment estimates, as outlined in \Cref{rem:nonparam}.

\section{Computational details}\label{sec:comput}
To simplify the statement of the results below, let
$r_{0}(b, \chi)=r(\sqrt{b}, \chi)$, and put $m_{2}=\sigma^{2}/\mu_{2}$. The next
proposition shows that, if only a second moment constraint is imposed, the
maximal non-coverage probability $\rho(m_{2}, \chi)$, defined in
\Cref{eq:non-coverage-bound}, has a simple solution:
\begin{proposition}\label{theorem:bound-second-moment}
  Consider the problem in~\Cref{eq:non-coverage-bound}. The solution
  is given by
  \begin{equation*}
    \rho(m_{2}, \chi)=\begin{cases}
      r_{0}(0, \chi)+\frac{m_{2}}{t_{0}}(r_{0}(t_{0}, \chi)-r_{0}(0, \chi)) &
      \text{if $m_{2}<t_{0}$,} \\
      r_{0}(m_{2}, \chi) & \text{otherwise.}
    \end{cases}
  \end{equation*}
  Here $t_{0}=0$ if $\chi<\sqrt{3}$, otherwise $t_{0}$ is the unique solution to
  $r_{0}(t, \chi)+u\frac{\partial}{\partial u}r_{0}(u, \chi)=r_{0}(u, \chi)$.
\end{proposition}
The proof of \Cref{theorem:bound-second-moment} shows that $\rho(m_{2}, \chi)$
corresponds to the least concave majorant of the function $r_{0}$.

The next result shows that, if in addition to a second moment constraint, we
impose a constraint on the kurtosis, the maximal non-coverage probability can be
computed as a solution to two nested univariate optimizations:
\begin{proposition}\label{theorem:bound-fourth-moment}
  Suppose $\kappa>1$ and $m_{2}>0$. Then the solution to the problem
  \begin{equation*}
    \rho(m_{2}, \kappa, \chi)=\sup_{F}E_{F}[r(b, \chi)]\quad
    \text{s.t.}\quad E_{F}[b^{2}]=m_{2}, \, E_{F}[b^{4}]=\kappa m_{2}^{2},
  \end{equation*}
  is given by $\rho(m_{2}, \kappa, \chi)=r_{0}({m_{2}}, \chi)$ if
  $m_{2}\geq t_{0}$, with $t_{0}$ defined in~\Cref{theorem:bound-second-moment}.
  If $m_{2}< t_{0}$, then the solution is given by
  \begin{equation}\label{eq:dual-solution}
    \inf_{0<x_{0}\leq t_{0}}
        \left\{
      r_{0}(x_{0}, \chi)
      +(m_{2}-x_{0})r_{0}'(x_{0}, \chi)+
      ((x_{0}-m_{2})^{2}+(\kappa-1) m^{2}_{2})\sup_{0\leq x\leq t_{0}}\delta(x;x_{0})
    \right\},
  \end{equation}
  where $r_{0}'(x_{0}, \chi)=\partial r_{0}(x_{0}, \chi)/\partial x_{0}$,
  $\delta(x;x_{0})= \frac{r_{0}({x}, \chi)- r_{0}({x_{0}},
    \chi)-(x-x_{0})r_{0}'(x_{0}, \chi) }{(x-x_{0})^{2}}$ if $x\neq x_{0}$, and
  $\delta(x_{0};x_{0})=\lim_{x\to x_{0}}\delta(x;x_{0})=
  \frac{1}{2}\frac{\partial^{2}}{\partial x_{0}^{2}}r_{0}(x_{0}, \chi)$.
\end{proposition}

If $m_{2}\geq t_{0}$, then imposing a constraint on the kurtosis does not help to
reduce the maximal non-coverage probability, and
$\rho(m_{2}, \kappa, \chi)=\rho(m_{2}, \chi)$.

\begin{remark}[Least favorable distributions]\label{remark:lf-distro}
  It follows from the proof of these propositions that distributions maximizing
  \Cref{eq:non-coverage-bound}---the least favorable distributions for the
  normalized bias $b$---have two support points if $m_{2}\geq t_{0}$, namely
  $-\sqrt{m_{2}}$ and $\sqrt{m_{2}}$ (since the rejection probability
  $r(b, \chi)$ depends on $b$ only through its absolute value, any distribution
  with these two support points maximizes~\Cref{eq:non-coverage-bound}). If
  $m_{2}<t_{0}$, there are three support points, $b=0$, with probability
  $1-m_{2}/t_{0}$ and $b=\pm \sqrt{t_{0}}$ with total probability $m_{2}/t_{0}$
  (again, only the sum of the probabilities is uniquely determined). If the
  kurtosis constraint is also imposed, then there are four support points,
  $\pm \sqrt{x_{0}}$ and $\pm \sqrt{x}$, where $x$ and $x_{0}$
  optimize~\Cref{eq:dual-solution}.
\end{remark}

Finally, the characterization of the solution to the general program in
\Cref{eq:rho_general} depends on the form of the constraint function $g$. To
solve the program numerically, discretize the support of $F$ to turn the problem
into a finite-dimensional linear program, which can be solved using a standard
linear solver. In particular, we solve the problem
\begin{equation*}
  \rho_{g}(m, \chi)=\sup_{p_{1}, \dotsc, p_{K}}
  \sum_{k=1}^{K}p_{k}r(x_{k}, \chi)\quad \text{s.t.}\quad
  \sum_{k=1}^{K}p_{k}g(x_{k})=m, \quad \sum_{k=1}^{K}p_{k}=1,\quad p_{k}\geq 0.
\end{equation*}
Here $x_{1}, \dotsc, x_{K}$ denote the support points of $b$, with $p_{k}$
denoting the associated probabilities.

\section{Coverage results}\label{coverage_results_sec_append}

This \namecref{coverage_results_sec_append} provides coverage results that
generalize \Cref{thm:coverage_baseline}.
\Cref{setup_sec_append} introduces the general setup.
\Cref{general_coverage_sec_append} provides results for general shrinkage
estimators that satisfy an approximate normality assumption.
\Cref{indep_shrinkage_sec} considers a generalization of our baseline
specification in the \ac{EB} setting, and states a generalization of
\Cref{thm:coverage_baseline}.

\subsection{General setup and notation}\label{setup_sec_append}

Let $\hat\theta_{1}, \dotsc, \hat\theta_{n}$ be estimates of
$\theta_1, \dotsc, \theta_n$, with standard errors
$\text{se}_1, \dotsc, \text{se}_n$. The standard errors may be random variables
that depend on the data. We are interested in coverage properties of the
intervals $CI_i=\{\hat\theta_i\pm \text{se}_i\cdot \chi_i \}$ for some
$\chi_1, \dotsc, \chi_n$, which may be chosen based on the data. In some cases,
we will condition on some variable $\tilde{X}_{i}$ when defining \ac{EB}
coverage or average coverage. Let
$\tilde X^{(n)}=(\tilde{X}_1, \dotsc, \tilde{X}_n)'$ and let
$\chi^{(n)}=(\chi_{1}, \dotsc, \chi_{n})'$.

As discussed in \Cref{sec:cover-under-basel}, the average coverage criterion
does not require thinking of $\theta$ as random. To save on notation, we will
state most of our average coverage results and conditions in terms of a general
sequence of probability measures $\tilde P=\tilde P^{(n)}$ and triangular arrays
$\theta$ and $\tilde X^{(n)}$. We will use $E_{\tilde P}$ to denote expectation
under $\tilde{P}$. We can then obtain \ac{EB} coverage statements by considering
a distribution $P$ for the data and $\theta$, $\tilde X^{(n)}$ and an additional
variable $\nu$ such that these conditions hold for the measure
$\tilde P(\cdot)=P(\cdot\mid \theta, \nu, \tilde X^{(n)})$ for
$\theta, \nu, \tilde X^{(n)}$ in a probability one set. The variable $\nu$ is
allowed to depend on $n$, and can include nuisance parameters as well as
additional variables.

It will be useful to formulate a conditional version of the average coverage
criterion~\eqref{eq:aci_def}, to complement the conditional version of \ac{EB}
coverage discussed in the main text. Due to discreteness of the empirical
measure of the $\tilde X_i$'s, we consider coverage conditional on each set in
some family $\mathcal{A}$ of sets. To formalize this, let
$\mathcal{I}_{\mathcal{X}, n}=\{i\in \{1, \dotsc, n\}\colon
\tilde{X}_i\in\mathcal{X}\}$, and let
$N_{\mathcal{X}, n}=\#\mathcal{I}_{\mathcal{X}, n}$. The sample average
non-coverage on the set $\mathcal{X}$ is then given by
\begin{equation*}
  ANC_n(\chi^{(n)};\mathcal{X})
  = \frac{1}{N_{\mathcal{X}, n}}\sum_{i\in \mathcal{I}_{\mathcal{X}, n}}
  \1{\theta_i\notin \{\hat\theta\pm \text{se}_i\cdot \chi_i\}}
  = \frac{1}{N_{\mathcal{X}, n}}\sum_{i\in \mathcal{I}_{\mathcal{X}, n}} \1{\abs{Z_i}>\chi_{i}},
\end{equation*}
where $Z_i=(\hat\theta_i-\theta_i)/\text{se}_i$. We consider two notions of
average coverage control, conditional on the set $\mathcal{X}\in\mathcal{A}$:
\begin{equation}\label{eq:sample_ac_control}
  ANC_{n}(\chi;\mathcal{X})\le \alpha+o_{\tilde P}(1),
\end{equation}
and
\begin{equation}\label{eq:conditional_ac_control}
  \limsup_n E_{\tilde P}\left[ANC_n(\chi;\mathcal{X}) \right]
  = \limsup_n \frac{1}{N_{\mathcal{X}, n}}\sum_{i\in \mathcal{I}_{\mathcal{X}, n}}
  \tilde P(\abs{Z_i}> \chi_i)
  \le \alpha.
\end{equation}
Since $ANC_n(\chi;\mathcal{X})$ is uniformly
bounded,~\eqref{eq:sample_ac_control} implies~\eqref{eq:conditional_ac_control}.
Furthermore, if we integrate with respect to some distribution on
$\nu, \tilde X^{(n)}$ such that~\eqref{eq:conditional_ac_control} holds with
$\tilde P(\cdot)=P(\cdot\mid \theta, \nu, \tilde X^{(n)})$ almost surely, we get
(again by uniform boundedness)
$\limsup_n E\left[ANC_n(\chi;\mathcal{X}) \mid \theta \right]\le \alpha$, which,
if $\mathcal{X}$ contains all $\tilde X_i$'s with probability one, is
condition~\eqref{eq:aci_def} from the main text.

Now consider \ac{EB} coverage, as defined in \Cref{eq:ebci_def} in the main
text, but conditioning on $\tilde X_i$. We consider \ac{EB} coverage under a
distribution $P$ for the data, $\tilde X^{(n)}$, $\theta$ and $\nu$, where $\nu$
includes additional nuisance parameters and covariates, and where the average
coverage condition~\eqref{eq:conditional_ac_control} holds with
$P(\cdot\mid\theta, \nu, \tilde{X}^{(n)})$ playing the role of $\tilde P$ with
probability one. Suppose $\tilde X_i$ is discretely distributed under $P$, and
that the exchangeability condition
\begin{equation}\label{eq:conditional_exchangeability}
  P(\theta_i\in CI_{i}\mid \mathcal{I}_{\{\tilde x\}, n})
  = P(\theta_j\in CI_{j}\mid\mathcal{I}_{\{\tilde x\}, n})
  \;\text{for all}\;i, j\in\mathcal{I}_{\{\tilde x\}, n}
\end{equation}
holds with probability one. Then, for each $j$,
\begin{multline*}
  P(\theta_j\in CI_j\mid \tilde X_j=\tilde x)
    = P(\theta_j\in CI_j\mid j\in\mathcal{I}_{\{\tilde x\}, n})
    = E\left[P(\theta_j\in CI_j\mid
    \mathcal{I}_{\{\tilde x\}, n}) \mid j\in\mathcal{I}_{\{\tilde x\}, n} \right] \\
  = E\left[\textstyle\frac{1}{\mathcal{N}_{\{\tilde{x}\}, n}}
    \sum_{i\in\mathcal{I}_{\{\tilde x\}}}P(\theta_i\in CI_i\mid \mathcal{I}_{\{\tilde x\}})
    \, \Big|\, j\in\mathcal{I}_{\{\tilde x\}, n} \right].
\end{multline*}
Plugging in $P(\cdot\mid \theta, \nu, \tilde X^{(n)})$ for $\tilde P$ in the
coverage condition~\eqref{eq:conditional_ac_control}, taking the expectation
conditional on $\mathcal{I}_{\{\tilde x\}, n}$ and using uniform boundedness, it
follows that the $\liminf$ of the term in the conditional expectation is no less
than $1-\alpha$. Then, by uniform boundedness of this term,
\begin{equation}\label{eq:conditional_eb_coverage}
  \liminf_{n\to\infty} P(\theta_j\in CI_j\mid\tilde X_j=\tilde x) \ge 1-\alpha.
\end{equation}
This is a conditional version of the \ac{EB} coverage
condition~\eqref{eq:ebci_def} from the main text.

\subsection{Results for general shrinkage
  estimators}\label{general_coverage_sec_append}

We assume that $Z_i=(\hat\theta_i-\theta_i)/\text{se}_i$ is approximately normal
with variance one and mean $b_{i}$ under the sequence of probability measures
$\tilde P=\tilde P^{(n)}$. To formalize this, we consider a triangular array of
distributions satisfying the following conditions.

\begin{assumption}\label{Zi_assump}
  For some random variables $\tilde b_{i}$ and constants $b_{i, n}$,
  $Z_i-\tilde b_i$ satisfies
\begin{equation*}
 \lim_{n\to\infty} \max_{1\le i\le n} \abs*{\tilde P(Z_i-\tilde b_i\le t) - \Phi(t)} = 0
\end{equation*}
for all $t\in\mathbb{R}$ and, for all $\mathcal{X}\in \mathcal{A}$ and any
$\varepsilon>0$,
$ \frac{1}{N_{\mathcal{X}, n}} \sum_{i\in\mathcal{I}_{\mathcal{X}, n}}
\tilde{P}(\abs{\tilde{b}_i - b_{i, n}} \ge \varepsilon )\to 0$.
\end{assumption}

Note that, when applying the results with $\tilde P(\cdot)$ given by the
sequence of measures $P(\cdot\mid \theta, \nu, \tilde X^{(n)})$, the constants
$b_{i, n}$ will be allowed to depend on $\theta, \nu, \tilde X^{(n)}$.

Let $g\colon\mathbb{R}\to\mathbb{R}^p$ be a vector of moment functions. We consider
critical values $\hat\chi^{(n)}=(\hat\chi_1, \dotsc, \hat\chi_n)$ based on an
estimate of the conditional expectation of $g(b_{i, n})$ given $\tilde X_i$,
where the expectation is taken with respect to the empirical distribution of
$\tilde X_i, b_{i, n}$. Due to the discreteness of this measure, we consider the
behavior of this estimate on average over sets $\mathcal{X}\in\mathcal{A}$. We
assume that there exists a function $m: \mathcal{X}\to\mathbb{R}^p$ that plays
the role of the conditional expectation of $g(b_{i, n})$ given $\tilde X_i$,
along with estimates $\hat m_i$ of $m(\tilde X_i)$, which satisfy the following
assumptions.

\begin{assumption}\label{mu_consistency_assump}
  For all $\mathcal{X}\in\mathcal{A}$, $N_{\mathcal{X}, n}\to \infty$,
  $ \frac{1}{N_{\mathcal{X}, n}} \sum_{i\in\mathcal{I}_{\mathcal{X}, n}}
  (g(b_{n, i}) - m(\tilde X_i)) \to 0$, and, for all $\varepsilon>0$,
  $\frac{1}{N_{\mathcal{X}, n}} \sum_{i\in\mathcal{I}_{\mathcal{X}, n}}
  \tilde{P}(\norm{\hat m_i-m(\tilde X_i)}\ge \varepsilon) \to 0$.
\end{assumption}

\begin{assumption}\label{partition_assump}
  For every $\mathcal{X}\in\mathcal{A}$ and every $\varepsilon>0$, there is a
  partition $\mathcal{X}_1, \dotsc, \mathcal{X}_{J}\in\mathcal{A}$ of
  $\mathcal{X}$ and $m_1, \dotsc, m_J$ such that, for each $j$ and all
  $x\in \mathcal{X}_j$, $m(x)\in B_{\varepsilon}(m_j)$, where
  $B_{\varepsilon}(m)=\{\tilde m: \|\tilde m-m\|\le \varepsilon\}$.
\end{assumption}

\begin{assumption}\label{mu_set_assump}
For some compact set $M$
in the interior of the set of values of $\int g(b) dF(b)$ where $F$ ranges
over all probability measures on $\mathbb{R}$, we have $m(x)\in M$ for all $x$.
\end{assumption}

Let $\rho_g(m, \chi)$ and $\cva_{\alpha, g}(m)$
be defined as in \Cref{sec:extensions},
\begin{equation*}
  \cva_{\alpha, g}(m)=\inf \{\chi: \rho_g(m, \chi)\le \alpha\}
  \quad\text{where}\quad
  \rho_g(m, \chi)
  = \sup_{F} E_{F}[r(b, \chi)]\; \text{s.t.}\; E_{F}[g(b)] = m.
\end{equation*}
Let $\hat\chi_i=\cva_{\alpha, g}(\hat m_i)$. We will consider the average
non-coverage $ANC_n(\hat\chi^{(n)};\mathcal{X})$ of the collection of intervals
$\{\hat\theta_i\pm \text{se}_i\cdot \hat\chi_i\}$.

\begin{theorem}\label{high_level_conditional_coverage_thm}
  Suppose that
  \Cref{Zi_assump,mu_consistency_assump,,partition_assump,mu_set_assump} hold,
  and that, for some $j$,
  $\lim_{b\to\infty}g_j(b)=\lim_{b\to -\infty}g_j(b)=\infty$ and
  $\inf_b g_j(b)\ge 0$. Then, for all $\mathcal{X}\in \mathcal{A}$,
  \begin{equation*}
    E_{\tilde P}ANC_n(\hat\chi^{(n)};\mathcal{X})\le \alpha+o(1).
  \end{equation*}
  If, in addition, $Z_i-\tilde b_i$ is independent over $i$ under $\tilde P$,
  then $ANC_n(\hat\chi^{(n)};\mathcal{X})\le \alpha+o_{\tilde P}(1)$.
\end{theorem}

\subsection{Empirical Bayes shrinkage toward regression estimate}\label{indep_shrinkage_sec}

We now apply the general results in \Cref{general_coverage_sec_append} to the
\ac{EB} setting. As in \Cref{sec:pract-impl}, we consider unshrunk estimates
$Y_1, \dotsc, Y_n$ of parameters $\theta=(\theta_1, \dotsc, \theta_n)'$, along
with regressors $X^{(n)}=(X_1, \dotsc, X_n)$ and variables
$\tilde X^{(n)}=(\tilde X_1, \dotsc, \tilde{X}_{n})'$, which include $\sigma_i$,
and which play the role of the conditioning variables (the setting in
\Cref{sec:pract-impl} obtains as a special case with
$\tilde X_{i}=(X_i, \sigma_i)$). The initial estimate $Y_i$ has standard
deviation $\sigma_i$, and we observe an estimate $\hat\sigma_i$. We obtain
average coverage results by considering a triangular array of probability
distributions $\tilde P=\tilde P^{(n)}$, in which the $X_i$'s, $\sigma_i$'s and
$\theta_i$'s are fixed. \ac{EB} coverage can then be obtained for a distribution
$P$ of the data, $\theta$ and some nuisance parameter $\tilde\nu$ such that
these conditions hold almost surely with
$P(\cdot\mid \theta, \tilde{\nu}, \tilde X^{(n)}, X^{(n)})$ playing the role of
$\tilde{P}$.

We generalize the baseline specification in the main text, and consider
\begin{equation*}
  \hat\theta_i=\hat X_i'\hat\delta + w(\hat \gamma, \hat\sigma_i)(Y_i-\hat X_i'\hat\delta)
\end{equation*}
where $\hat X_i$ is an estimate of $X_i$ (this allows some elements of $X_i$ to
be estimated rather than observed directly, such as when $\sigma_i$ is included
in $X_i$), $\hat{\delta}$ is any random vector that depends on the data (such as
the \ac{OLS} estimator in a regression of $Y_i$ on $X_i$), and $\hat{\gamma}$ is
a tuning parameter that determines shrinkage and may depend on the data. This
leads to the standard error
$\text{se}_i=w(\hat \gamma, \hat\sigma_i)\hat\sigma_i$ so that the $t$-statistic
is
\begin{equation*}
  Z_i=\frac{\hat\theta_i-\theta_i}{\text{se}_i}
  =\frac{\hat X_i'\hat\delta + w(\hat \gamma, \hat\sigma_i)(Y_i-\hat X_i'\hat\delta)-\theta_i}{w(\hat \gamma, \hat\sigma_i)\hat\sigma_i}
  =\frac{Y_i-\theta_i}{\hat\sigma_i} +
  \frac{[w(\hat \gamma, \hat\sigma_i)-1](\theta_i-\hat X_i'\hat\delta)}{w(\hat \gamma, \hat\sigma_i)\hat\sigma_i}.
\end{equation*}
We use estimates of moments of the bias of positive integer order
$\ell_1<\dotsb<\ell_p$. Let $\hat\mu_\ell$ be an estimate of the $\ell$th moment
of $\theta_i-X_i'\delta$, and suppose that this moment is independent of
$\sigma_i$ in a sense formalized below. Then an estimate of the $\ell_{j}$th
moment of the bias is
$\hat{m}_{i, j}= \frac{[w(\hat\gamma,
  \hat\sigma_i)-1]^{\ell_j}\hat\mu_{\ell_j}}{w(\hat\gamma,
  \hat\sigma_i)^{\ell_j} \hat{\sigma}_i^{\ell_j}}$. Let
$\hat m_i=(\hat{m}_1, \dotsc, \hat{m}_p)'$. The \ac{EBCI} is then given by
$\hat\theta_i\pm w(\hat\gamma, \hat\sigma_i)\hat\sigma_i\cdot \cva_{\alpha,
  g}(\hat m_i)$ where $g_j(b)=b^{\ell_j}$. We obtain the baseline specification
in \Cref{sec:baseline-implementation} when $p=2$, $\ell_1=2$, $\ell_2=4$,
$\hat\gamma=\hat\mu_{2}$ and
$w(\hat \mu_{2}, \hat\sigma_i)=\hat\mu_{2}/(\hat\mu_{2}+\hat\sigma_i^2)$.

We make the following assumptions.

\begin{assumption}\label{indep_shrinkage_Zi_assump}
  $\lim_{n\to\infty} \max_{1\le i\le n}
  \abs{\tilde P\left((Y_i-\theta_i)/\hat \sigma_i \le t \right) - \Phi(t)} = 0$.
\end{assumption}

\Cref*{clt_primitive_conditions_sec} gives primitive conditions for
\Cref{indep_shrinkage_Zi_assump}, and verifies them in a linear fixed effects
panel data model. These conditions involve considering a triangular array of
parameter values such that sampling error and empirical moments of the parameter
value sequence are of the same order of magnitude, and defining $\theta_i$ to be
a scaled version of the corresponding parameter.

\begin{assumption}\label{indep_shrinkage_sigma_consistency_assump}
  The standard deviations $\sigma_i$ are bounded away from zero. In addition,
  for some $\delta$ and $\gamma$, $\hat\delta$ and $\hat\gamma$ converge to
  $\delta$ and $\gamma$ under $\tilde{P}$, and, for any $\varepsilon>0$,
\begin{equation*}
  \lim_{n\to\infty} \max_{1\le i\le n} \tilde P(\abs{\hat\sigma_i-\sigma_i}
  \ge \varepsilon)
  = 0
  \; \text{and}\;
  \lim_{n\to\infty} \max_{1\le i\le n} \tilde P(\abs{\hat X_i- X_i}
  \ge \varepsilon)
  = 0.
\end{equation*}
\end{assumption}

\begin{assumption}\label{indep_shrinkage_mu_consistency_assump}
  The variable $\tilde X_i$ takes values in
  $\mathcal{S}_1\times \dotsm \times \mathcal{S}_s$ where, for each $k$, either
  $\mathcal{S}_k=[\underline x_k, \overline x_k]$ (with
  $-\infty<\underline x_k<\overline x_k<\infty$) or $\mathcal{S}_k$ is a
  finitely discrete set with minimum element $\underline x_k$ and maximum
  element $\overline x_k$. In addition, $\tilde X_{i1}=\sigma_i$ (the first
  element of $\tilde X_i$ is given by $\sigma_i$). Furthermore, for some $\mu_0$
  such that $(\mu_{0,\ell_1}, \dotsc, \mu_{0,\ell_p})$ is in the interior of the
  set of values of $\int g(b)\, dF(b)$ where $F$ ranges over probability
  measures on $\mathbb{R}$ where $g_j(b)=b^{\ell_j}$ and some constant $K$, the
  following holds. Let $\mathcal{A}$ denote the collection of sets
  $\widetilde{\mathcal{S}}_1\times\dotsm \times \widetilde{\mathcal{S}}_s$ where
  $\widetilde{\mathcal{S}}_k$ is a positive Lebesgue measure interval contained
  in $[\underline x_k, \overline x_k]$ in the case where
  $\mathcal{S}_k=[\underline x_k, \overline x_k]$, and
  $\widetilde{\mathcal{S}}_k$ is a nonempty subset of $\mathcal{S}_k$ in the
  case where $\mathcal{S}_k$ is finitely discrete. For any
  $\mathcal{X}\in\mathcal{A}$, $N_{\mathcal{X}, n}\to\infty$ and
\begin{equation*}
  \frac{1}{N_{\mathcal{X}, n}}\sum_{i\in\mathcal{I}_{\mathcal{X}, n}} (\theta_i-X_i'\delta)^{\ell_j}\to \mu_{0,\ell_j}, \quad
      \frac{1}{N_{\mathcal{X}, n}}\sum_{i\in\mathcal{I}_{\mathcal{X}, n}} \abs{\theta_i}^{\ell_j}
      \le K, \quad \text{and}\quad
    \frac{1}{N_{\mathcal{X}, n}}\sum_{i\in\mathcal{I}_{\mathcal{X}, n}} \|X_i\|^{\ell_j}\le K.
\end{equation*}
In addition, the estimate $\hat \mu_{\ell_j}$ converges in probability to
$\mu_{0,\ell_j}$ under $\tilde P$ for each $j$.
\end{assumption}

\begin{theorem}\label{indep_shrinkage_coverage_thm}
  Let $\hat\theta_i$ and $\text{se}_i$ be given above and let
  $\hat\chi_i=\cva_{\alpha, g}(\hat m_i)$ where $\hat m_i$ is given above and
  $g(b)=(b^{\ell_1}, \dotsc, b^{\ell_p})$ for some positive integers
  $\ell_1, \dotsc, \ell_p$, at least one of which is even. Suppose that
  \Cref{indep_shrinkage_Zi_assump,indep_shrinkage_sigma_consistency_assump,,indep_shrinkage_mu_consistency_assump}
  hold, and that $w()$ is continuous in an open set containing
  $\{\gamma\}\times \mathcal{S}_1$ and is bounded away from zero on this set.
  Let $\mathcal{A}$ be as given in \Cref{indep_shrinkage_mu_consistency_assump}.
  Then, for all $\mathcal{X}\in \mathcal{A}$,
  $E_{\tilde P}ANC_n(\hat\chi^{(n)};\mathcal{X})\le \alpha+o(1)$. If, in
  addition, $(Y_i, \hat\sigma_i)$ is independent over $i$ under $\tilde P$, then
  $ANC_n(\hat\chi^{(n)};\mathcal{X})\le \alpha+o_{\tilde P}(1)$.
\end{theorem}

As a consequence of \Cref{indep_shrinkage_coverage_thm}, we obtain, under the
exchangeability condition~\eqref{eq:conditional_exchangeability}, conditional
\ac{EB} coverage, as defined in \Cref{eq:conditional_eb_coverage}, for any
distribution $P$ of the data and $\theta, \tilde \nu$ such that the conditions
of \Cref{indep_shrinkage_coverage_thm} hold with probability one with the
sequence of probability measures
$P(\cdot\mid \theta, \tilde \nu, X^{(n)}, \tilde X^{(n)})$ playing the role of
$\tilde P$. This follows from the arguments in \Cref{setup_sec_append}.

\begin{corollary}\label{general_eb_coverage_corollary}
  Let $\theta, \nu, X^{(n)}, \tilde X^{(n)}, Y_i$ follow a sequence of
  distributions $P$ such that the conditions of
  \Cref{indep_shrinkage_coverage_thm} hold with $\tilde X_i$ taking on finitely
  many values, and $P(\cdot\mid \theta, \nu, X^{(n)}, \tilde X^{(n)})$ playing
  the role of $\tilde P$ with probability one, and such that the exchangeability
  condition~\eqref{eq:conditional_exchangeability} holds. Then the intervals
  $CI_i=\{\hat\theta_i\pm w(\hat\gamma, \hat\sigma_i)\hat\sigma_i\cdot
  \cva_{\alpha, g}(\hat m_i)\}$ satisfy the conditional \ac{EB} coverage
  condition~\eqref{eq:conditional_eb_coverage}.
\end{corollary}

The first part of \Cref{thm:coverage_baseline} (average coverage) follows by
applying \Cref{indep_shrinkage_coverage_thm} with the conditional distribution
$P(\cdot\mid \theta)$ playing the role of $\tilde P$. The second part (\ac{EB}
coverage) follows immediately from \Cref{general_eb_coverage_corollary}.
\end{appendices}

\onehalfspacing
\phantomsection%
\addcontentsline{toc}{section}{References}
\bibliography{ebci_references}

\end{document}


\maketitle

\renewcommand{\theequation}{S\arabic{equation}}
\renewcommand{\thetable}{S\arabic{table}}
\renewcommand{\thefigure}{S\arabic{figure}}

\begin{appendices}
\crefalias{section}{sappsec}
\crefalias{subsection}{sappsubsec}
\crefalias{subsubsection}{sappsubsubsec}
\setcounter{section}{3}

This supplement is organized as follows. \Cref{sec:theor-deta-proofs} gives
proofs of the formal results in the main text and details on
\Cref*{indep_shrinkage_Zi_assump}. \Cref{sec:appendix_empirics} gives details
on the simulations. \Cref{sec:power-details} discusses the power of tests based
on our \acp{EBCI}, and \Cref{sec:appendix_general} works through examples of the
general shrinkage estimators in \Cref*{sec:general_shrinkage}.

\section{Theoretical details and proofs}\label{sec:theor-deta-proofs}
\Cref{clt_primitive_conditions_sec} gives technical details on
\Cref*{indep_shrinkage_Zi_assump}. The remainder of this
\namecref{sec:theor-deta-proofs} provides the proofs of all results in the main
paper and in this supplement.

\subsection{Primitive conditions for
  Assumption~\ref*{indep_shrinkage_Zi_assump}}\label{clt_primitive_conditions_sec}

To verify \Cref*{indep_shrinkage_Zi_assump}, we will typically have to
define $\theta_i$ to be scaled by a rate of convergence.  Let $\tilde Y_i$ be
an estimator of a parameter $\vartheta_{i, n}$ with rate of convergence $\kappa_{n}$
and asymptotic variance estimate $\hat \sigma_i^2$.
Suppose that
\begin{equation}\label{kappa_clt_eq}
  \lim_{n\to\infty} \max_{1\le i\le n}\sup_{t\in\mathbb{R}}
  \abs*{P\left(\frac{\kappa_n(\tilde Y_i-\vartheta_{i, n})}{\hat \sigma_i} \le t \right)
  - \Phi(t)} = 0.
\end{equation}
Then \Cref*{indep_shrinkage_Zi_assump} holds with $\theta_i=\kappa_n
\vartheta_{i, n}$ and $Y_i=\kappa_n\tilde Y_i$.  Consider an affine estimator
$\hat\vartheta_i=a_i/\kappa_n+w_i \tilde Y_i = (a_i + w_i Y_i)/\kappa_n$ with standard error
$\widetilde{\text{se}}_i=w_i \hat\sigma_i/\kappa_n$.
The corresponding affine estimator of $\theta_i$ is $\hat\theta_i=\kappa_n
\hat\vartheta_i=a_i + w_i Y_i$ with standard error
$\text{se}_i=\kappa_n\cdot \widetilde{\text{se}}_i=w_i \hat\sigma_i$.
Then
$\vartheta_{i, n}\in \{\hat\vartheta_i\pm \widetilde{\text{se}}_i\cdot \hat\chi_i\}$
iff.
$\theta_{i}\in \{\hat\theta_i\pm \text{se}_i\cdot \hat\chi_i\}$.
Thus, \Cref*{indep_shrinkage_coverage_thm} guarantees average coverage of the intervals
$\{\hat\vartheta_i\pm \widetilde{\text{se}}_i\cdot \hat\chi_{i}\}$ for $\vartheta_{i, n}$.
Note that, in order for the moments of $\theta_i$ to converge to a non-degenerate
constant, we will need to consider triangular arrays $\vartheta_{i, n}$ that converge
to zero at a $\kappa_n$ rate.

As an example, we now verify \Cref*{indep_shrinkage_Zi_assump} for the linear
fixed effects panel data model
\begin{equation*}
  W_{it} = \vartheta_{i, n} + X_{it}'\beta + u_{it},
  \quad i=1,\ldots, n, t=1,\ldots, T_i,
\end{equation*}
where $X_{it}$ are covariates in the fixed effects regression.\footnote{We note
  that, despite the similarity in notation, we do not make any assumption about
  the relation between the individual level prediction variables $X_i$ used in
  the individual level predictive regression and the covariates $X_{it}$ used in
  the fixed effects regression.} We assume that the $T_i$s increase at the same
rate so that, letting $\bar{T}=\frac{1}{n}\sum_{i=1}^n T_i$, we can apply the
approach described above with $\kappa_n=\sqrt{\bar{T}}$ to verify
\Cref*{indep_shrinkage_Zi_assump} with
$\theta_i=\sqrt{\bar{T}}\vartheta_{i, n}$. We consider the fixed effects
estimate of $\vartheta_{i, n}$ formed by regressing $W_{it}$ on $X_{it}$ and
indicator variables for each individual $i$, along with the heteroskedasticity
robust variance estimate from this regression. To give the formulas for these
estimates, we first define some notation.
Let $\bar W_i=\frac{1}{T_i}\sum_{t=1}^{T_i} W_{it}$,
$\bar X_i=\frac{1}{T_i}\sum_{t=1}^{T_i} X_{it}$, $\ddot X_{it}=X_{it}-\bar X_i$,
$\ddot W_{it}=W_{it}-\bar W_i$, $\bar u_i=\frac{1}{T_i}\sum_{t=1}^{T_i} u_{it}$
and $\bar T=\frac{1}{n}\sum_{i=1}^n T_i$. Letting
$\hat Q_{XX}=\frac{1}{n\bar T} \sum_{i=1}^n\sum_{t=1}^{T_i} \ddot{X}_{it}
\ddot{X}_{it}'$, the fixed effect estimate of $\beta$ is given by
$\hat\beta=\hat Q_{XX}^{-1}\sum_{i=1}^n \sum_{t=1}^{T_i}
\ddot{X}_{it}W_{it}/(n\bar T)$, and the fixed effect estimate of
$\vartheta_{i, n}$ is given by
\begin{equation} \label{eq:fe_y}
  \tilde Y_i = \bar W_i-\bar X_i'\hat\beta
  = \sum_{j=1}^n\sum_{t=1}^{T_j}\left(\1{i=j}\frac{1}{T_i} - \frac{1}{n\bar{T}}\bar X_i'\hat Q_{XX}^{-1}\ddot X_{it} \right)W_{it}.
\end{equation}
We assume that the $T_i$s grow at the same rate, so that all $\tilde{Y}_{i}$'s
converge at the same rate $1/\sqrt{\bar T}$. An estimate of the variance of
$\sqrt{\bar T}(\tilde Y_i-\vartheta_{i, n})$ that is robust to heteroskedasticity
in $u_{it}$ is given by
\begin{equation} \label{eq:fe_sigma}
  \hat\sigma_i^2=\bar T\sum_{j=1}^n\sum_{t=1}^{T_i}\left(\1{i=j}\frac{1}{T_i} - \frac{1}{n\bar T}\bar X_i'\hat Q_{XX}^{-1}\ddot X_{jt} \right)^2\hat u_{jt}^2,
\end{equation}
where $\hat u_{it}=W_{it}-X_{it}'\hat\beta-\tilde Y_i$.

We consider ``large $n$ large $T$'' asymptotics in which the $T_i$'s are
implicitly indexed by $n$. We make the following assumptions about the $T_i$'s
and the distribution $\tilde{P}=\tilde P^{(n)}$ of
$\{X_{it}, u_{it}\}_{i=1,\ldots, n,\, t=1,\ldots, T_i}$.

\begin{assumption}\label{fe_asymptotics_assump}
  For some constants $\gamma>0$ and $K>0$,
  \begin{enumerate}
  \item $u_{it}$ is mean zero and independent across $i$ and $t$ with
    $1/K\le E_{\tilde P} u_{it}^2$ and $E_{\tilde P} \abs{u_{it}}^{2+\gamma}\le K$.

  \item $\abs{X_{it}}\le K$ for all $i, t$.

  \item $n\to\infty$ and $\min_{1\le i\le n} T_i\to \infty$ and $T_i/T_j\le K$ for all $i, j\le
    n$.

  \item Under $\tilde P$, $\sqrt{n\bar T}(\hat\beta-\beta)=\mathcal{O}(1)$ and
    the minimum eigenvalue of
    $\hat Q_{XX}$ is
    greater than $1/K$ with probability approaching one as $n\to \infty$.
  \end{enumerate}

\end{assumption}

\Cref{fe_asymptotics_assump} is meant to give a simple set of sufficient
conditions, and it could be modified for other settings, so long as large $n$
and $T$ asymptotics allow for valid inference on the individual fixed effects.
For example, one could relax the independence assumption on the $u_{it}$'s and
modify the standard errors to take into account dependence, so long as one puts
enough structure on the dependence that consistent variance estimation is
possible as $n$ and $T$ increase. The assumption of bounded covariates is made
for simplicity, and could be relaxed, at the possible expense of strengthening
the moment condition on $u_{it}$.
The convergence rate assumption on $\hat\beta$ follows from standard arguments under
appropriate conditions on $u_{it}$ and $X_{it}$
\citep[see, e.g.,][]{stock_heteroskedasticity-robust_2008}.

\begin{theorem}\label{fe_primitive_conditions_thm}
  Consider the fixed effects setting given above, and suppose
  \Cref{fe_asymptotics_assump} holds. Then \Cref*{indep_shrinkage_Zi_assump}
  holds with $\theta_i=\sqrt{\bar T}\vartheta_{i, n}$,
  $Y_i=\sqrt{\bar T}\tilde Y_i$ where $\tilde Y_i$ is the fixed effects
  estimator defined in \Cref{eq:fe_y}, and $\hat{\sigma}_i^2$ is the variance
  estimate defined in \Cref{eq:fe_sigma}.
\end{theorem}

To prove \Cref{fe_primitive_conditions_thm}, we first prove a series of lemmas.

\begin{lemma}\label{tildeY_baru_lemma_fe}
  For any $\eta>0$,
  $\max_{1\le i\le n}\tilde P\left(\sqrt{\bar{T}}
    \abs{\tilde{Y}_i-\vartheta_{i, n}-\bar u_i} > \eta \right)\to 0$.
\end{lemma}
\begin{proof}
  The result is immediate from \Cref{fe_asymptotics_assump}
  since
  $\tilde Y_i-\vartheta_{i, n}-\bar u_i=\bar X_{it}'(\beta-\hat\beta)$.
\end{proof}

\begin{lemma}\label{uhat_lemma_fe}
  For any $\eta>0$,
  $\max_{1\le i\le n}\tilde P\left(\left| \frac{1}{T_i}\sum_{t=1}^{T_i}
      (\hat{u}_{it}^2-u_{it}^2) \right| > \eta \right)\to 0$. Furthermore, if
  $A_{it, n}$ is a triangular array of random variables that are bounded almost
  surely uniformly in $n$ and $i, t$, then, for any $\eta>0$, there exists $C$
  such that
  $\max_{1\le i\le n}\tilde P\left(\left| \frac{1}{T_i}\sum_{t=1}^{T_i}
      A_{it,n}\hat u_{it}^2 \right| > C \right)< \eta$ and
  $\tilde P\left( \left| \frac{1}{n\bar T}\sum_{i=1}^n\sum_{t=1}^{T_i}
      A_{it,n}\hat u_{it}^2 \right| > C \right)<\eta$ for large enough $n$.
\end{lemma}
\begin{proof}
Some algebra shows that $\hat u_{it}=\ddot X_{it}'(\beta-\hat\beta)+u_{it}-\bar u_i$.
Thus,
\begin{align}\label{hat_u_eq}
  \hat u_{it}^2
  =u_{it}^2
  +(\beta-\hat\beta)'\ddot X_{it}\ddot X_{it}'(\beta-\hat\beta)+\bar u_i^2
  +2u_{it}\ddot X_{it}'(\beta-\hat\beta)
  -2\bar u_i \ddot X_{it}'(\beta-\hat\beta)
  -2\bar u_{i}u_{it}.
\end{align}
It follows that
$\left| \frac{1}{n\bar{T}} \sum_{i=1}^n\sum_{t=1}^{T_i}A_{it, n}\hat u_{it}^2
\right|$ is bounded by $\max_{i, t,n}\abs{A_{it, n}}$ times
\begin{align*}
  &\frac{1}{n\bar T} \sum_{i=1}^n \sum_{t=1}^{T_i}u_{it}^2
    +(\beta-\hat\beta)'  \hat Q_{XX}(\beta-\hat\beta)
    +\frac{1}{n\bar T}\sum_{i=1}^n\sum_{t=1}^{T_i}\bar u_i^2  \\
  &+\frac{1}{n\bar T} \sum_{i=1}^n \sum_{t=1}^{T_i}2 \abs{u_{it}}\cdot
    \abs{\ddot{X}_{it}'(\beta-\hat\beta)}
    -\frac{1}{n\bar T} \sum_{i=1}^n \sum_{t=1}^{T_i}2|\bar u_i| |\ddot X_{it}'(\beta-\hat\beta)|
    -\frac{1}{n\bar T} \sum_{i=1}^n \sum_{t=1}^{T_i}2|\bar u_{i}u_{it}|.
\end{align*}
The second term converges in probability to zero by the assumptions on $X_{it}$
and $\hat \beta$. The remaining terms are bounded by a constant times
$\frac{1}{n\bar T}\sum_{i=1}^n\sum_{t=1}^{T_i} (
u_{it}^2+\bar{u}_i^2+|u_{it}|+|\bar u_i|+|\bar u_i u_{it}|)$. By Jensen's
inequality, we have $\bar u_i^2\le \frac{1}{T_i}\sum_{i=1}^{T_i}u_{it}^2$,
$\abs{\bar{u}_i}\le \frac{1}{T_i}\sum_{i=1}^{T_i}|u_{it}|$ and
\begin{align*}
  \sum_{t=1}^{T_i}|\bar u_i| |u_{it}|
  = |\bar u_i| \sum_{t=1}^{T_i} |u_{it}|
  \le \frac{1}{T_i}\left[ \sum_{t=1}^{T_i} |u_{it}| \right]^2
  \le T_i \frac{1}{T_i}\sum_{t=1}^{T_i} u_{it}^2
  = \sum_{t=1}^{T_i} u_{it}^2.
\end{align*}
This gives a bound of a constant times $\frac{1}{n\bar T} \sum_{i=1}^n
\sum_{t=1}^{T_i}(u_{it}^2+|u_{it}|)$.  The last statement in the lemma then follows by
Markov's inequality.  The second statement in the lemma follows from similar
arguments.

For the first statement in the lemma, it follows from~\eqref{hat_u_eq} that
$\frac{1}{T_i}\sum_{t=1}^{T_i} (u_{it}^2-\hat u_{it}^2)$ is equal to
\begin{align*}
  (\hat\beta-\beta)'\left( \frac{1}{T_i}\sum_{t=1}^{T_i}\ddot X_{it}\ddot X_{it}'
\right)(\hat\beta-\beta)-\bar u_i^2
  +2\frac{1}{T_i}\sum_{t=1}^{T_i} u_{it}\ddot X_{it}'(\beta-\hat\beta)
  -2\frac{\bar u_i }{T_i}\sum_{t=1}^{T_i} \ddot X_{it}'(\beta-\hat\beta).
\end{align*}
The first term is bounded by a constant that does not depend on $i$ times $|\hat\beta-\beta|^2$ (the squared Euclidean norm), which
converges in probability to $0$ by assumption.  The second term has expectation
bounded by $\bar T^{-1}$ times a constant that does not depend on $i$.  From the
bounds on the support of $X_{it}$ and the first moment of $u_{it}$ it follows
that the last two terms are bounded by $|\hat\beta-\beta|$ times a constant that
does not depend on $i$.  This gives the first statement of the lemma.
\end{proof}

\begin{lemma}\label{sigma_consistency_lemma_fe}
  Let $\sigma_i^2=\frac{\bar T}{T_i^2}\sum_{t=1}^{T_i}E_{\tilde P} u_{it}^2$.
  For any $\eta>0$,
   $\max_{1\le i\le n}\tilde P\left( \left| \hat\sigma_i^2 -
       \sigma_i^2  \right| > \eta \right)\to 0$.
\end{lemma}
\begin{proof}
We have $\hat\sigma_i^2=I+II+III$ where $I=\frac{\bar{T}}{T_i^2}\sum_{t=1}^{T_i}\hat u_{it}^2$,
\begin{align*}
  II=\frac{1}{n^2\bar T}\sum_{j=1}^n\sum_{t=1}^{T_i}\bar X_i'\hat Q_{XX}^{-1}\ddot X_{jt}\ddot X_{jt}'\hat Q_{XX}^{-1}\bar X_i \hat u_{jt}^2
  =\frac{1}{n}\bar X_i'\hat Q_{XX}^{-1}\hat Q_{XXu}\hat Q_{XX}^{-1}\bar X_i,
\end{align*}
where $\hat Q_{XXu}=\frac{1}{n\bar T}\sum_{j=1}^n\sum_{t=1}^{T_i}
\ddot X_{jt}\ddot X_{jt}'\hat u_{it}^2$,
and
\begin{align*}
  III=-2\frac{1}{nT_i}\sum_{t=1}^{T_i}\bar X_i'\hat Q_{XX}^{-1}\ddot X_{it}\hat u_{it}^2
  =-2\frac{1}{n}\bar X_i'\hat Q_{XX}^{-1}\hat Q_{Xu, i},
\end{align*}
where $\hat Q_{Xu,i}=\frac{1}{T_i}\sum_{i=1}^n \ddot X_{it}\hat u_{it}^2$. By
\Cref{uhat_lemma_fe} and the condition on the minimum eigenvalue of
$\hat Q_{XX}$, it follows that
$\max_{1\le i\le n}\tilde P(|II+III|>\eta/3)\to 0$. It also follows from
\Cref{uhat_lemma_fe} that
$\max_{1\le i\le n}\tilde P\left(\left| I - \frac{\bar{T}}{T_i^2}
    \sum_{t=1}^{T_i}u_{it}^2 \right|>\eta/3\right)\to 0$. It now suffices to
show that
$\max_{1\le i\le n} \tilde P\left(\left| \frac{\bar T}{T_i^2}\sum_{t=1}^{T_i}
    \left( u_{it}^2-E_{\tilde P}u_{it}^2 \right) \right|>\eta/3 \right)\to 0$.
By \citet[][Theorem 3]{von_bahr_inequalities_1965},
\begin{align*}
E_{\tilde P}\left| \frac{\bar T}{T_i^2}\sum_{t=1}^{T_i} \left( u_{it}^2-E_{\tilde P}u_{it}^2 \right) \right|^{1+\gamma/2}
  \le 2(\bar T/T_i^2)^{1+\gamma/2} \sum_{t=1}^{T_i}E_{\tilde P}\left| u_{it}^2-E_{\tilde P}u_{it}^2 \right|^{1+\gamma/2},
\end{align*}
which is bounded by a constant times $\bar T^{-\gamma/2}$ by the moment bound on
$u_{it}$ and the bound on $T_i/T_j$.  The result now follows from Markov's inequality.
\end{proof}

Let $\tilde Z_i=\sqrt{\bar{T}}\bar u_i/\sigma_i$,
$R_{1,i}=\sqrt{\bar{T}_i}( \tilde Y_i - \vartheta_{i, n} - \bar u_i)/\sigma_i$
and $R_{2,i}=\hat\sigma_i-\sigma_i$. We have
\begin{align*}
  &\frac{\sqrt{\bar T}(\tilde Y_i-\vartheta_{i, n})}{\hat\sigma_i}
    = \left( \tilde Z_i+ R_{1,i}  \right)\frac{\sigma_i}{\sigma_i+R_{2,i}}
    = \tilde Z_i - \tilde Z_i \frac{R_{2,i}}{\sigma_i+R_{2,i}}
      + R_{1,i}\frac{\sigma_i}{\sigma_i+R_{2,i}}.
\end{align*}
It follows from the Lyapounov Central Limit Theorem (applied to $Z_{i_n}$ for
arbitrary sequences $i_n\le n$) that
$\lim_{n\to\infty} \max_{1\le i\le n}\sup_{t\in\mathbb{R}}
\abs*{P\left(\tilde{Z}_i \le t \right) - \Phi(t)} = 0$. The conclusion of
\Cref{fe_primitive_conditions_thm} then follows so long as
$\max_{1\le i\le n} P\left(\left| \tilde Z_i
    \frac{R_{2,i}}{\sigma_i+R_{2,i}}\right| +
  \left|R_{1,i}\frac{\sigma_i}{\sigma_i+R_{2,i}} \right|>\eta \right) \to 0$ for
any $\eta>0$. But this follows by
\Cref{tildeY_baru_lemma_fe,sigma_consistency_lemma_fe} and the fact that
$\sigma_i$ is bounded from above and from below away from zero by the moment
assumptions on $u_{it}$.

\subsection{Proof of Lemma~\ref*{thm:eb_param}}\label{sec:proof-lemma-refthm}
We first show that the non-coverage probability is weakly decreasing in $w_{EB,i}$. Let $\Gamma(m)$ denote the space of probability measures on $\mathbb{R}$ with
second moment bounded above by $m>0$. Abbreviating $z_{1-\alpha/2}$ by $z$, let
$\tilde{\rho}(w) = \rho(1/w-1,z/\sqrt{w})$ denote the maximal undercoverage when $w_{EB,i}=w$. By
definition of $\rho$,
\begin{equation}\label{eq:sup}
  \tilde{\rho}(w) = \sup_{F \in \Gamma(1/w-1)} E_{b \sim F}\left[P(\abs{b-Z}
    > z/\sqrt{w} \mid b) \right]
  = \sup_{F \in \Gamma(1/w-1)} P_{b \sim F}\big(\sqrt{w}\abs{b-Z}
  > z\big),
\end{equation}
where $Z$ denotes a $N(0,1)$ variable that is independent of $b$.

Consider any $w_{0}, w_{1}$ such that $0<w_{0}\leq w_{1}<1$. Let
$F_1^* \in \Gamma(1/w_1-1)$ denote the least-favorable distribution---i.e., the
distribution that achieves the supremum~\eqref{eq:sup}---when $w=w_1$. (\Cref*{theorem:bound-second-moment} implies that the supremum is in fact attained at a particular discrete
distribution.) Let $\tilde{F}_0$ denote the distribution of the linear
combination
\[\sqrt{\frac{w_1}{w_0}}b - \sqrt{\frac{w_1-w_0}{w_0}}Z\]
when $b \sim F_1^*$ and $Z \sim N(0,1)$ are independent. Note that the second
moment of this distribution is
$\frac{w_1}{w_0} \cdot \frac{1-w_1}{w_1} +
\frac{w_1-w_0}{w_0}=\frac{1-w_0}{w_0}$, so $\tilde{F}_0 \in \Gamma(1/w_0-1)$.
Thus, if we let $\tilde{Z}$ denote another $N(0,1)$ variable that is independent
of $(b, Z)$, then
\begin{multline*}
    \tilde{\rho}(w_0) \geq P_{b \sim \tilde{F}_0}\big(\sqrt{w_0}\abs{b - Z} > z\big)
    = P_{b \sim F_1^*}\left(\sqrt{w_0}\left|\sqrt{\frac{w_1}{w_0}}b - \sqrt{\frac{w_1-w_0}{w_0}}\tilde{Z} - Z\right| > z\right) \\
    = P_{b \sim F_1^*}\bigg(\big|\sqrt{w_1}b - \underbrace{(\sqrt{w_1-w_0}\tilde{Z} + \sqrt{w_0}Z)}_{\sim N(0,w_1)}\big| > z\bigg) = P_{b \sim F_1^*}\left(\sqrt{w_1}|b - Z| > z\right) = \tilde{\rho}(w_1).
  \end{multline*}

Next, we derive the limit of the non-coverage probability as $w_{EB,i} \to 0$. It follows from \Cref*{theorem:bound-second-moment} that
\begin{equation*}
\rho(t, \chi) = \sup_{0 \leq \lambda \leq 1} (1-\lambda)r(0,\chi) + \lambda r((t/\lambda)^{1/2}, \chi).
\end{equation*}
Note that $r(0,z/\sqrt{w})\to 0$ as $w \to 0$. Thus,
\begin{equation*}
  \lim_{w\to 0}\; \tilde{\rho}(w) =
  \lim_{w\to 0}\; \rho\left(1/w-1,z/\sqrt{w}\right) =
  \lim_{w\to 0} \sup_{0 \leq \lambda \leq 1} \lambda r\left(\lambda^{-1/2}(1/w-1)^{1/2}, zw^{-1/2}\right),
\end{equation*}
provided the latter limit exists. We will first show that the supremum above is bounded below by an expression that tends to $1/\max\lbrace z^2, 1\rbrace$. Then we will show that the supremum is bounded above by an expression that tends to $1/z^2$ (and the supremum is obviously also bounded above by 1).

Let $\varepsilon(w) \geq 0$ be any function of $w$ such that $\varepsilon(w) \to 0$ and $\varepsilon(w)(1/w-1)^{1/2} \to \infty$ as $w \to 0$. Let $\tilde{z} = \max\lbrace z,1\rbrace$. Note first that, by setting $\lambda=(\tilde{z}(1-w)^{-1/2}+\varepsilon(w))^{-2} \in [0,1]$,
\[\sup_{0 \leq \lambda \leq 1} \lambda r\left(\lambda^{-1/2}(1/w-1)^{1/2}, zw^{-1/2}\right) \geq \frac{r\left((\tilde{z}(1-w)^{-1/2}+\varepsilon(w))(1/w-1)^{1/2}, zw^{-1/2}\right)}{(\tilde{z}(1-w)^{-1/2}+\varepsilon(w))^{2}} \to \frac{1}{\tilde{z}^2}\]
as $w \to 0$, since $r(b, \chi)\to 1$ when $(b-\chi)\to\infty$, and
\begin{equation*}
  \begin{split}
    (\tilde{z}(1-w)^{-1/2}+\varepsilon(w))(1/w-1)^{1/2} - zw^{-1/2}
    &\geq (z(1-w)^{-1/2}+\varepsilon(w))(1/w-1)^{1/2} - zw^{-1/2} \\
    &= \varepsilon(w)(1/w-1)^{1/2} \to \infty.
  \end{split}
\end{equation*}
Second,
\begin{multline*}
\sup_{0 \leq \lambda \leq 1} \lambda r\left(\lambda^{-1/2}(1/w-1)^{1/2}, zw^{-1/2}\right) \\
\leq \Phi\left(-zw^{-1/2}\right) + \sup_{0 \leq \lambda \leq 1} \lambda \Phi\left(\lambda^{-1/2}(1/w-1)^{1/2} - zw^{-1/2}\right).
\end{multline*}
The first term above tends to 0 as $w\to 0$. The second term equals
\begin{multline*}
  \max\bigg\lbrace \sup_{0 \leq \lambda \leq (z-\varepsilon(w))^{-2}} \lambda \Phi\left(\lambda^{-1/2}(1/w-1)^{1/2} - zw^{-1/2}\right), \\
  \sup_{(z-\varepsilon(w))^{-2} < \lambda \leq 1}\lambda
  \Phi\left(\lambda^{-1/2}(1/w-1)^{1/2} - zw^{-1/2}\right)
  \bigg\rbrace,
\end{multline*}
where the first argument is bounded above by
$\sup_{0 \leq \lambda \leq (z-\varepsilon(w))^{-2}}\lambda =
(z-\varepsilon(w))^{-2} \to \frac{1}{z^2}$. The second argument tends to 0 as
$w \to 0$, since
\begin{equation*}
  \lambda^{-1/2}(1/w-1)^{1/2} - zw^{-1/2} \leq (\lambda^{-1/2}-z)(1/w-1)^{1/2}
  \leq -\varepsilon(w)(1/w-1)^{1/2}
\end{equation*}
for all $\lambda > (z-\varepsilon(w))^{-2}$, and the far right-hand side above
tends to $-\infty$ as $w \to 0$.

\subsection{Proof of Proposition~\ref*{theorem:bound-second-moment}}
Since $r(b, \chi)$ is symmetric in $b$, \Cref*{eq:non-coverage-bound} is
equivalent to maximizing $E_{F}[r_{0}(t, \chi)]$ over distributions $F$ of $t$
with $E_{F}[t]=m_{2}$. Let $\overbar{r}(t, \chi)$ denote the least concave
majorant of $r_{0}(t, \chi)$. We first show that
$\rho(m_{2}, \chi)=\overbar{r}(m_{2}, \chi)$.

Observe that $\rho(m_{2}, \chi)\leq \overbar{\rho}(m_{2}, \chi)$, where
$\overbar{\rho}(m_{2}, \chi)$ denotes the value of the
problem
\begin{equation*}
  \overbar{\rho}(m_{2}, \chi)=
  \sup_{F}E_{F}[\overbar{r}(t, \chi)]\quad
  \text{s.t.}\quad E_{F}[t]=m_{2}.
\end{equation*}
Furthermore, since $\overbar{r}$ is concave, by Jensen's inequality, the optimal
solution $F^{*}$ to this problem puts point mass on $m_{2}$, so that
$\overbar{\rho}(m_{2}, \chi)=\overbar{r}(m_{2}, \chi)$, and hence
$\rho(m_{2}, \chi)\leq \overbar{r}(m_{2}, \chi)$.

Next, we show that the reverse inequality holds,
$\rho(m_{2}, \chi)\geq \overbar{r}(m_{2}, \chi)$. By Corollary 17.1.4 on page
157 in \citet{rockafellar70}, the majorant can be written as
\begin{equation}\label{eq:majorant}
  \overbar{r}(t, \chi)=\sup\{\lambda r_{0}(x_{1}, \chi)+
  (1-\lambda)r_{0}(x_{2}, \chi)\colon \lambda
  x_{1}+(1-\lambda)x_{2}=t, \;0\leq x_{1}\leq x_{2}, \lambda\in[0,1]\},
\end{equation}
which corresponds to the problem in~\Cref*{eq:non-coverage-bound}, with the
distribution $F$ constrained to be a discrete distribution with two support
points. Since imposing this additional constraint on $F$ must weakly decrease
the value of the solution, it follows that
$\rho(m_{2}, \chi)\geq \overbar{r}(m_{2}, \chi)$. Thus,
$\rho(m_{2}, \chi)= \overbar{r}(m_{2}, \chi)$. The
\namecref{theorem:bound-second-moment} then follows by~\Cref{theorem:majorant}
below.
\begin{lemma}\label{theorem:rczero}
  Let $r_{0}(t, \chi)=r(\sqrt{t}, \chi)$. If $\chi\leq \sqrt{3}$, then $r_{0}$
  is concave in $t$. If $\chi>\sqrt{3}$, then its second derivative is positive
  for $t$ small enough, negative for $t$ large enough, and crosses zero exactly
  once, at some $t_{1}\in[{\chi^{2}-3}, (\chi-1/\chi)^{2}]$.
\end{lemma}
\begin{proof}
  Letting $\phi$ denote the standard normal density, the first and second
  derivative of $r_{0}(t)=r_{0}(t, \chi)$ are given by
  \begin{align*}
    r_{0}'(t)&=\frac{1}{2 \sqrt{t}}\left[\phi(\sqrt{t}-\chi)
      -\phi(\sqrt{t}+\chi)\right]\geq 0,\\
    r''_{0}(t)&= \frac{\phi(\chi-\sqrt{t})(\chi\sqrt{t}-{t}-1)
      +\phi(\chi+\sqrt{t})(\chi\sqrt{t}+{t}+1)}{4t^{3/2}}\\
             &
               =\frac{\phi(\chi+\sqrt{t})}{4t^{3/2}}
               \left[e^{2\chi \sqrt{t}} (\chi\sqrt{t}-{t}-1)
               +(\chi\sqrt{t}+{t}+1)\right]=
               \frac{\phi(\chi+\sqrt{t})}{4t^{3/2}}f(\sqrt{t}),
  \end{align*}
  where the last line uses $\phi(a+b)e^{-2ab}=\phi(a-b)$, and
  \begin{equation*}
    f(u)=(\chi u+u^{2}+1)-e^{2\chi u} (u^{2}-\chi u+1).
  \end{equation*}
  Thus, the sign of $r''_{0}(t)$ corresponds to that of $f(\sqrt{t})$, with
  $r''_{0}(t)=0$ if and only if $f(\sqrt{t})=0$. Observe $f(0)=0$, and $f(u)<0$
  is negative for $u$ large enough, since the term $-u^{2}e^{2\chi u}$
  dominates. Furthermore,
  \begin{align*}
    f'(u)&=2u+\chi-e^{2\chi u}(2\chi(u^{2}-\chi u+1) +2u-\chi)
    & f'(0)&=0\\
    f''(u)&=e^{2 \chi u} (4 \chi^3 u - 4 \chi^2 u^2 - 8 \chi u - 2) + 2
    & f''(0)&=0
    \\
    f^{(3)}(u)&= 4 \chi e^{2 \chi u} (2 \chi^3 u + \chi^2 (1 - 2 u^2) - 6 \chi u - 3)
    & f^{(3)}(0) &=4 \chi(\chi^{2}-3).
  \end{align*}
  Therefore for $u>0$ small enough, $f(u)$, and hence $r''_{0}(u^{2})$ is
  positive if $\chi^{2}\geq 3$, and negative otherwise.

  Now suppose that $f(u_{0})=0$ for some $u_{0}>0$, so that
  \begin{equation}\label{eq:f0}
    \chi u_{0}+u_{0}^{2}+1=e^{2\chi u_{0}} (u_{0}^{2}-\chi u_{0}+1)
  \end{equation}
  Since $\chi u+u^{2}+1$ is strictly positive, it must be the case that
  $u_{0}^{2}-\chi u_{0}+1>0$. Multiplying and dividing the expression for
  $f'(u)$ above by $u_{0}^{2}-\chi u_{0}+1$ and plugging in the identity in
  \Cref{eq:f0} and simplifying the expression yields
  \begin{equation}\label{eq:f0prime}
    \begin{split}
      f'(u_{0}) &= \frac{(u_{0}^{2}-\chi u_{0}+1)(2u_{0} +\chi)-(\chi
        u_{0}+u_{0}^{2}+1)(2\chi(u_{0}^{2}-\chi u_{0}+1) +2u_{0}-\chi)}
      {u_{0}^{2}-\chi u_{0}+1}\\
      &=\frac{2 u^{2}_{0} \chi (\chi^2 - 3-u_{0}^{2})}{u_{0}^{2}-\chi u_{0}+1}.
    \end{split}
  \end{equation}
  Suppose $\chi^{2}<3$. Then $f'(u_{0})<0$ at all positive roots $u_{0}$ by
  \Cref{eq:f0prime}. But if $\chi^{2}<3$, then $f(u)$ is initially negative, so
  by continuity it must be that $f'(u_{1})\geq 0$ at the first positive root
  $u_{1}$. Therefore, if $\chi^{2}\leq 3$, $f$, and hence $r''_{0}$, cannot have
  any positive roots. Thus, if $\chi^{2}\leq 3$, $r_{0}$ is concave as claimed.

  Now suppose that $\chi^{2}\geq 3$, so that $f(u)$ is initially positive. By
  continuity, this implies that $f'(u_{1})\leq 0$ at its first positive root
  $u_{1}$. By \Cref{eq:f0prime}, this implies $u_{1}\geq \sqrt{\chi^{2}-3}$. As
  a result, again by \Cref{eq:f0prime}, $f(u_{i})\leq 0$ for all remaining
  positive roots. But since by continuity, the signs of $f'$ must alternate at
  the roots of $f$, this implies that $f$ has at most a single positive root.
  Since $f$ is initially positive, and negative for large enough $u$, it follows
  that it has a single positive root $u_{1}\geq \sqrt{\chi^{2}-3}$. Finally, to
  obtain an upper bound for $t_{1}=u^{2}_{1}$, observe that if $f(u_{1})=0$,
  then, by Taylor expansion of the exponential function,
    \begin{equation*}
      1+\frac{2\chi u_{1}}{\chi u_{1}+u_{1}^{2}+1}
      =e^{2\chi u_{1}}\geq 1+2\chi u_{1}+2(\chi u_{1})^{2},
    \end{equation*}
    which implies that $1\geq (1+\chi u_{1})(\chi u_{1}+u_{1}^{2}+1)$, so that
    $u_{1}\leq \chi-1/\chi$.
\end{proof}

\begin{lemma}\label{theorem:majorant}
  The problem in \Cref{eq:majorant} can be written as
  \begin{equation}
    \overbar{r}(t, \chi)=\sup_{u\geq t}\{(1-t/u) r_{0}(0,\chi)+
    \frac{t}{u}r_{0}(u, \chi)\}.\label{eq:majorant_simple}
  \end{equation}
  Let $t_{0}=0$ if $\chi\leq \sqrt{3}$, and otherwise let $t_{0}>0$ denote the
  solution to
  $r_{0}(0, \chi)-r_{0}(u, \chi)+u\frac{\partial}{\partial u}r_{0}(u, \chi)=0$.
  This solution is unique, and the optimal $u$ solving \Cref{eq:majorant_simple}
  satisfies $u=t$ for $t> t_{0}$ and $u=t_{0}$ otherwise.
\end{lemma}
\begin{proof}
  If in the optimization problem in~\Cref{eq:majorant}, the constraint on $x_{2}$
  binds, or either constraint on $\lambda$ binds, then the optimum is achieved
  at $r_{0}(t)=r_{0}(t, \chi)$, with $x_{1}=t$ and $\lambda=1$ and $x_{2}$
  arbitrary; $x_{2}=t$ and $\lambda=0$ and $x_{1}$ arbitrary; or else
  $x_{1}=x_{2}$ and $\lambda$ arbitrary. In any of these cases $\overbar{r}$
  takes the form in \Cref{eq:majorant_simple} as claimed. If, on the other hand,
  these constraints do not bind, then $x_{2}>t>x_{1}$, and substituting
  $\lambda=(x_{2}-t)/(x_{2}-x_{1})$ into the objective function yields the
  first-order conditions
  \begin{align}
    r_{0}(x_{2}) - (x_{2}-x_{1})r_{0}'(x_{1})
    -r_{0}(x_{1})
    & =\mu\frac{(x_{2}-x_{1})^{2}}{(x_{2}-t)},\label{eq:foc1}\\
    r_{0}(x_{2})
    +(x_{1}-x_{2})r_{0}'(x_{2})-r_{0}(x_{1})
    & =0,\label{eq:foc2}
  \end{align}
  where $\mu\geq 0$ is the Lagrange multiplier on the constraint that
  $x_{1}\geq 0$. Subtracting \Cref{eq:foc2} from \Cref{eq:foc1} and applying the
  fundamental theorem of calculus then yields
\begin{equation}\label{eq:mugt0}
  \mu\frac{x_{2}-x_{1}}{(x_{2}-t)}
  =r_{0}'(x_{2})-r_{0}'(x_{1})=\int_{x_{1}}^{x_{2}}r_{0}''(t)\, d t
  >0,
\end{equation}
which implies that $\mu>0$. Here the last inequality follows because by Taylor's
theorem, \Cref{eq:foc2} implies that
$\int_{x_{1}}^{x_{2}}r_{0}''(t)(t-x_{1})\, d t=0$. Since $r''_{0}$ is positive
for $t\leq t_{1}$ and negative for $t\geq t_{1}$ by \Cref{theorem:rczero}, it
follows that $x_{1}\leq t_{1}\leq x_{2}$, and hence that
\begin{multline*}
  0= \int_{x_{1}}^{t_{1}} r''_{0}(t)(t-x_{1})\, d t
  +\int_{t_{1}}^{x_{2}} r''_{0}(t)(t-x_{1})\, d t\\
  < (t_{1}-x_{1})\int_{x_{1}}^{t_{1}} r''_{0}(t)\, d t
  +(t_{1}-x_{1})\int_{t_{1}}^{x_{2}} r''_{0}(t)\, d t
  =(t_{1}-x_{1})\int_{x_{1}}^{x_{2}}r''_{0}(t)\, d t.
\end{multline*}
Finally \Cref{eq:mugt0} implies that $\mu>0$, so that $x_{1}=0$ at the optimum.
Consequently, the problem in \Cref{eq:majorant} takes the form
in~\Cref{eq:majorant_simple} as claimed.

To show the second part of \Cref{theorem:majorant}, note that
by~\Cref{theorem:rczero}, if $\chi\leq \sqrt{3}$, $r_{0}$ is concave, so that we
can put $u=t$ in \Cref{eq:majorant_simple}. Otherwise, let $\mu\geq 0$ denote the
Lagrange multiplier associated with the constraint $u\geq t$ in the optimization
problem in~\Cref{eq:majorant_simple}. The first-order condition is then given by
\begin{equation*}
  r_{0}(0)-r_{0}(u)+ur_{0}'(u)=\frac{-\mu u^{2}}{t}.
\end{equation*}
Let $f(u)=r_{0}(0)-r_{0}(u)+ur_{0}'(u)$. Since $f'(u)=ur_{0}''(u)$, it follows
from~\Cref{theorem:rczero} that $f(u)$ is increasing for $u\leq t_{1}$ and
decreasing for $u\geq t_{1}$. Since $f(0)=0$ and
$\lim_{u\to \infty}f(u)<r_{0}(0)-1<0$, it follows that $f(u)$ has exactly one
positive zero, at some $t_{0}>t_{1}$. Thus, if $t<t_{0}$, $u=t_{0}$ is the
unique solution to the first-order condition. If $t>t_{0}$, $u=t$ is the unique
solution.
\end{proof}

\subsection{Proof of Proposition~\ref*{theorem:bound-fourth-moment}}

Since $r(b, \chi)$ is symmetric in $b$, letting $t=b^{2}$, we can equivalently
write the optimization problem as
\begin{equation}\label{eq:primal}
  \rho(m_{2}, \kappa, \chi)=\sup_{F}E_{F}[r_{0}(t, \chi)]\quad
  \text{s.t.}\quad E_{F}[t]=m_{2}, \, E_{F}[t^{2}]=\kappa m_{2}^{2},
\end{equation}
where $r_{0}(t, \chi)=r(\sqrt{t}, \chi)$, and the supremum is over all
distributions supported on the positive part of the real line. The dual of this
problem is
\begin{equation*}
  \min_{\lambda_{0}, \lambda_{1}, \lambda_{2}}
  \lambda_{0}+\lambda_{1}m_{2}+\lambda_{2}\kappa m^{2}_{2}\qquad
  \text{s.t.}\quad \lambda_{0}+\lambda_{1}t+\lambda_{2}t^{2}\geq r_{0}(t), \quad 0\leq t< \infty,
\end{equation*}
where $\lambda_{0}$ the Lagrange multiplier associated with the implicit
constraint that $E_{F}[1]=1$, and $r_{0}(t)=r_{0}(t, \chi)$. So long as
$\kappa>1$ and $m_{2}>0$, so that the moments $(m_{2}, \kappa m^{2}_{2})$ lie in
the interior of the space of possible moments of $F$, by the duality theorem in
\citet{smith95}, the duality gap is zero, and if $F^{*}$ and
$\lambda^{*}=(\lambda_{0}^{*}, \lambda_{1}^{*}, \lambda_{2}^{*})$ are optimal
solutions to the primal and dual problems, then $F^{*}$ has mass points only at
those $t$ with
$\lambda^{*}_{0}+\lambda^{*}_{1}t+\lambda^{*}_{2}t^{2}= r(\sqrt{t}, \chi)$.

Define $t_{0}$ as in \Cref{theorem:majorant}. First, we claim that if
$m_{2}\geq t_{0}$, then $\rho(m_{2}, \kappa, \chi)=\rho(m_{2}, \chi)$, the value
of the objective function in \Cref*{theorem:bound-second-moment}. The reason that
adding the constraint $E_{F}[t^{2}]=\kappa m_{2}^{2}$ does not change the optimum
is that it follows from the proof of \Cref*{theorem:bound-second-moment} that the
distribution achieving the rejection probability $\rho(m_{2}, \chi)$ is a point
mass on $m_{2}$. Consider adding another support point $x_{2}=\sqrt{n}$ with
probability $\kappa m_{2}^{2}/n$, with the remaining probability on the support
point $m_{2}$. Then, as $n\to\infty$, the mean of this distribution converges to
$m_{2}$, and its second moment converges to $\kappa m^{2}_{2}$, so that the
constraints in~\Cref{eq:primal} are satisfied, while the rejection probability
converges to $\rho(m_{2}, \chi)$. Since imposing the additional constraint
$E_{F}[t^{2}]=\kappa m_{2}^{2}$ cannot increase optimum, the claim follows.

Suppose that $m_{2}< t_{0}$. At optimum, the majorant
$g(x)=\lambda_{0}+\lambda_{1}t+\lambda_{2}t^{2}$ in the dual constraint must
satisfy $g(x_{0})=r_{0}(x_{0})$ for at least one $x_{0}>0$. Otherwise, if the
constraint never binds, we could lower the value of the objective function by
decreasing $\lambda_{0}$; furthermore, $x_{0}=0$ cannot be the unique point at
which the constraint binds, since by the duality theorem, this would imply that
the distribution that puts point mass on $0$ maximizes the primal, which cannot
be the case.

At such $x_{0}$, we must also have $g'(x_{0})=r_{0}'(x_{0})$, otherwise the
constraint would be locally violated. Using this fact together with the equality
$g(x_{0})=r_{0}(x_{0})$, we therefore have that
$\lambda_{0}=r_{0}(x_{0})-\lambda_{1}x_{0}-\lambda_{2}x_{0}^{2}$ and
$\lambda_{1}=r_{0}'(x_{0})-2\lambda_{2}x_{0}$, so that the dual problem may be
written as
\begin{multline}\label{eq:dual-rewritten}
  \min_{x_{0}>0, \lambda_{2}}
  r_{0}(x_{0})
  +r_{0}'(x_{0})(m_{2}-x_{0})
  +\lambda_{2}((x_{0}-m_{2})^{2}+(\kappa-1) m^{2}_{2})\\\text{s.t.}\quad
  r_{0}(x_{0})
  +r_{0}'(x_{0})(x-x_{0})+\lambda_{2}(x-x_{0})^{2}\geq r_{0}(x).
\end{multline}
Since $\kappa>1$, the objective is increasing in $\lambda_{2}$. Therefore, given
$x_{0}$, the optimal value of $\lambda_{2}$ is as small as possible while still
satisfying the constraint,
\begin{equation*}
  \lambda_{2}=\sup_{x>0}\delta(x;x_{0}), \qquad
  \delta(x;x_{0})=
  \frac{r_{0}(x)- r_{0}(x_{0})-r_{0}'(x_{0})(x-x_{0})}{(x-x_{0})^{2}}.
\end{equation*}
Next, we claim that the dual constraint cannot bind for $x_{0}>t_{0}$. Observe
that $\lambda_{2}\geq 0$, otherwise the constraint would be violated for $t$
large enough. However, setting $\lambda_{2}=0$ still satisfies the constraint.
This is because the function
$h(x)=r_{0}(x_{0}) +r_{0}'(x_{0})(x-x_{0})-r_{0}(x)$ is minimized at $x=x_{0}$,
with its value equal to $0$. To see this, note that its derivative equals
zero if $r_{0}'(x_{0})=r'(x)$. By \Cref{theorem:rczero}, $r_{0}'(t)$ is
increasing for $t\leq t_{0}$ and decreasing for $t>t_{0}$. Therefore, if
$r_{0}'(x_{0})<r_{0}'(0)$, $h'(x)=0$ has a unique solution, $x=x_{0}$. If
$r_{0}'(x_{0})>r_{0}'(0)$, there is another solution at some
$x_{1}\in[0,t_{0}]$. However, $h''(x_{1})=-r_{0}''(x_{1})<0$, so $h(x)$ achieves
a local maximum here. Since $h(0)>0$ by arguments in the proof of
\Cref{theorem:rczero}, it follows that the maximum of $h(x)$ occurs at
$x=x_{0}$, and equals $0$. However, \Cref{eq:dual-rewritten} cannot be maximized
at $(x_{0},0)$, since by \Cref*{theorem:bound-second-moment}, setting
$(x_{2}, \lambda_{2})=(t_{0},0)$ achieves a lower value of the objective
function, which proves the claim.

Therefore, \Cref{eq:dual-rewritten} can be written as
\begin{equation*}
  \min_{0<x_{0}\leq t_{0}}
  r_{0}(x_{0})
  +r_{0}'(x_{0})(m_{2}-x_{0})+
((x_{0}-m_{2})^{2}+(\kappa-1) m^{2}_{2})\sup_{x\geq 0}\delta(x;x_{0}),
\end{equation*}
To finish the proof of the \namecref{theorem:bound-fourth-moment}, it remains to
show that $\delta$ cannot be maximized at $x>t_{0}$. This follows from observing
that the dual constraint in \Cref{eq:dual-rewritten} binds at any $x$ that
maximizes $\delta$. However, by the claim above, the constraint cannot bind for
$x>t_{0}$.

\subsection{Proof of Theorem~\ref*{high_level_conditional_coverage_thm}}
To prove this theorem, we begin with some lemmas.

\begin{lemma}\label{ANC_approx_lemma}
Under \Cref*{Zi_assump}, we have, for any deterministic $\chi_1, \dotsc, \chi_n$,
and any $\mathcal{X}\in \mathcal{A}$ with $N_{\mathcal{X}, n}\to \infty$,
\begin{equation*}
  \lim_{n\to\infty} \frac{1}{N_{\mathcal{X}, n}} \sum_{i\in\mathcal{I}_{\mathcal{X}, n}} \tilde P\left(\abs{Z_i}>\chi_i \right)
  - \frac{1}{N_{\mathcal{X}, n}} \sum_{i\in\mathcal{I}_{\mathcal{X}, n}} r(b_{i, n}, \chi_i)
  = 0.
\end{equation*}
Furthermore, if $Z_i-\tilde b_i$ is independent over $i$ under $\tilde P$, then
\begin{equation*}
  \frac{1}{N_{\mathcal{X}, n}} \sum_{i\in\mathcal{I}_{\mathcal{X}, n}} \1{\abs{Z_i}>\chi_i}
  - \frac{1}{N_{\mathcal{X}, n}} \sum_{i\in\mathcal{I}_{\mathcal{X}, n}} r(b_{i, n}, \chi_i)
  =o_{\tilde P}(1).
\end{equation*}
\end{lemma}
\begin{proof}
For any $\varepsilon>0$, $\frac{1}{N_{\mathcal{X}, n}}
\sum_{i\in\mathcal{I}_{\mathcal{X}, n}} \1{\abs{Z_i}>\chi_i}$ is bounded
from above by
\begin{equation*}
  \frac{1}{N_{\mathcal{X}, n}} \sum_{i\in\mathcal{I}_{\mathcal{X}, n}}
  \1{\abs{Z_i-\tilde b_{i} + b_{i, n}}
  >\chi_i - \varepsilon}
  + \frac{1}{N_{\mathcal{X}, n}} \sum_{i\in\mathcal{I}_{\mathcal{X}, n}} \1{\abs{\tilde b_i - b_{i, n}}
  \ge \varepsilon}.
\end{equation*}
The expectation under $\tilde P$ of the second term converges to zero by
\Cref*{Zi_assump}. The expectation under $\tilde P$ of the first term is
$\frac{1}{N_{\mathcal{X}, n}} \sum_{i\in\mathcal{I}_{\mathcal{X}, n}}
\tilde{r}_{i, n}(b_{i, n}, \chi_i-\varepsilon)$ where
$\tilde r_{i, n}(b, \chi)=\tilde P(Z_i - \tilde b_{i} < -\chi - b) + 1 -
\tilde{P}(Z_i - \tilde b_{i} \le \chi - b)$. Note that $r_{i, n}(b, \chi)$
converges to $r(b, \chi)$ uniformly over $b, \chi$ under \Cref*{Zi_assump}, using
the fact that the convergence in \Cref*{Zi_assump} is uniform in $t$ by Lemma
2.11 in \citet{van_der_vaart_asymptotic_1998}, and the fact that
$\tilde{P}(Z_i - \tilde b_{i} < -\chi - b)=\lim_{t\uparrow -\chi - b} P(Z_i -
\tilde b_{i} \le t)$. It follows that the expectation of the above display
under $\tilde P$ is bounded by
$\frac{1}{N_{\mathcal{X}, n}} \sum_{i\in\mathcal{I}_{\mathcal{X}, n}}
\tilde{r}(b_{i, n}, \chi_i-\varepsilon)+o(1)$. If $Z_i-\tilde b_i$ is independent
over $i$, the variance of each term in the above display converges to zero, so
that the above display equals
$\frac{1}{N_{\mathcal{X}, n}} \sum_{i\in\mathcal{I}_{\mathcal{X}, n}}
\tilde{r}(b_{i, n}, \chi_i-\varepsilon)+o_{\tilde P}(1)$. Taking
$\varepsilon\to 0$ and noting that $r(b, \chi)$ is uniformly continuous in both
arguments, and using an analogous argument with a lower bound, gives the result.
\end{proof}

\begin{lemma}\label{continuity_lemma}
$\rho_g(\chi;m)$ is continuous in $\chi$.
Furthermore, for any $m^*$ in the interior of the set of values of $\int g(b)\, dF(b)$,
where $F$ ranges over all probability measures on $\mathbb{R}$,
$\rho_g(\chi;m)$ is continuous with respect to $m$ at $m^*$.
\end{lemma}
\begin{proof}
To show continuity with respect to $\chi$, note that
\begin{equation*}
  \abs{\rho_g(\chi;m)-\rho_g(\tilde\chi;m)}
  \le \sup_F \abs*{\int [r(b, \chi)-r(b, \tilde\chi)]\, dF(b)}
  \quad \text{s.t.}\; \int g(b)\, dF(b) = m,
\end{equation*}
where we use the fact that the difference between suprema of two functions over
the same constraint set is bounded by the supremum of the absolute difference of
the two functions. The above display is bounded by
$\sup_b\abs{r(b, \chi)-r(b, \tilde\chi)}$, which is bounded by a constant times
$\abs{\tilde\chi - \chi}$ by uniform continuity of the standard normal CDF\@.

To show continuity with respect to $m$, note that, by \Cref{interior_set_lemma}
below, the conditions for the Duality Theorem in \citet[][p. 812]{smith95} hold
for $m$ in a small enough neighborhood of $m^*$, so that
\begin{equation*}
  \rho_g(\chi; m)
  = \inf_{\lambda_0,\lambda} \lambda_0 + \lambda'm
  \quad \text{s.t.}\quad \lambda_0 + \lambda'g(b)\ge r(b, \chi)\;\text{for all}\;b\in\mathbb{R}
\end{equation*}
and the above optimization problem has a finite solution.  Thus, for $m$ in
this neighborhood of $m^*$,
$\rho_g(\chi; m)$ is the infimum of a collection of affine functions of $m$,
which implies that it is concave function of $m$ \citep[][p. 81]{boyd_convex_2004}.  By concavity,
$\rho_g(\chi;m)$ is also continuous as a function of $m$ in this
neighborhood of $m^*$.
\end{proof}

\begin{lemma}\label{interior_set_lemma}
Suppose that $\mu$ is in the interior of the set of values of $\int g(b)\,
dF(b)$ as $F$ ranges over all probability measures with respect to the Borel
sigma algebra, where $g:\mathbb{R}\to\mathbb{R}^p$.  Then $(1,\mu')'$ is in the
interior of the set of values of $\int (1,g(b)')'\, dF(b)$ as $F$ ranges over
all measures with respect to the Borel sigma algebra.
\end{lemma}
\begin{proof}
Let $\mu$ be in the interior of the set of values of $\int g(b)\,
dF(b)$ as $F$ ranges over all probability measures with respect to the Borel
sigma algebra.  We need to show that, for any $a, \tilde\mu$ with
$(a, \tilde\mu')'$ close enough to $(1,\mu')$, there exists a measure $F$ such
that $\int (1,g(b)')d F(b)=(a, \tilde\mu')'$.
To this end, note that, $\tilde\mu/a$ can be made arbitrarily close to $\mu$ by
making $(a, \tilde\mu')'$ close to $(1,\mu')$.  Thus, for $(a, \tilde\mu')'$ close
enough to $(1,\mu')$, there exists a probability measure $\tilde F$ with $\int
g(b)\, d \tilde F(b)=\tilde\mu/a$.  Let $F$ be the measure defined by $F(A)=a
\tilde F(A)$ for any measurable set $A$.  Then $\int (1,g(b)')' dF(b) = a \int
(1,g(b)')' d \tilde F(b) = (a, \tilde\mu)$.  This completes the proof.
\end{proof}

\begin{lemma}\label{rho_uniform_continuity_lemma}
  Let $M$ be a compact subset of the interior of the set of values of
  $\int g(b)\, dF(b)$, where $F$ ranges over all measures on $\mathbb{R}$ with
  the Borel $\sigma$-algebra. Suppose
  $\lim_{b\to\infty}g_j(b)=\lim_{b\to -\infty}g_j(b)=\infty$ and that
  $\inf_b g_j(b)\ge 0$ for some $j$. Then
  $\lim_{\chi\to\infty} \sup_{m\in M} \rho_g(\chi;m) = 0$ and $\rho_g(\chi;m)$
  is uniformly continuous with respect to $(\chi, m')'$ on the set
  $\hor{0,\infty}\times M$.
\end{lemma}
\begin{proof}
  The first claim (that $\lim_{\chi\to\infty} \sup_{m\in M} \rho_g(\chi;m) = 0$)
  follows by Markov's inequality and compactness of $M$. Given $\varepsilon>0$,
  let $\overline\chi$ be large enough so that $\rho_g(\chi;m)<\varepsilon$ for
  all $\chi\in \hor{\overline\chi, \infty}$ and all $m\in M$. By
  \Cref{continuity_lemma}, $\rho_g(\chi;m)$ is continuous on
  $[0,\overline\chi+1]\times M$, so, since $[0,\overline\chi+1]\times M$ is
  compact, it is uniformly continuous on this set. Thus, there exists $\delta$
  such that, for any $\chi, m$ and $\tilde\chi, \tilde m$ with
  $\chi, \tilde\chi\le \overline\chi+1$ and
  $\|(\tilde\chi, \tilde m')'-(\chi, m')'\|\le \delta$, we have
  $\abs{\rho_g(\chi;m)-\rho_g(\tilde\chi;\tilde m)}<\varepsilon$. If we also set
  $\delta<1$, then, if either $\chi\ge \overline\chi+1$ or
  $\tilde\chi\ge \overline\chi+1$ we must have both $\chi\ge\overline\chi$ and
  $\tilde\chi\ge\overline\chi$, so that
  $\rho_g(\tilde\chi;\tilde m)<\varepsilon$ and $\rho_g(\chi;m)<\varepsilon$,
  which also implies
  $\abs{\rho_g(\chi;m)-\rho_g(\tilde\chi;\tilde m)}<\varepsilon$. This completes
  the proof.
\end{proof}

For any $\varepsilon>0$, let
\begin{equation*}
\overline\rho_g(\chi; m, \varepsilon)
  = \sup_{\tilde m\in B_{\varepsilon}(m)} \rho_g(\chi; \tilde m)\quad
\text{and}\quad
\underline\rho_g(\chi; m, \varepsilon)
  = \inf_{\tilde m\in B_{\varepsilon}(m)} \rho_g(\chi; \tilde m).
\end{equation*}

\begin{lemma}\label{overline_underline_rho_diff_lemma}
  Let $M$ be a compact subset of the interior of the set of values of
  $\int g(b)\, dF(b)$, where $F$ ranges over all measures on $\mathbb{R}$ with
  the Borel $\sigma$-algebra. Suppose
  $\lim_{b\to\infty}g_j(b)=\lim_{b\to -\infty}g_j(b)=\infty$ and
  $\inf_b g_j(b)\ge 0$ for some $j$. Then, for $\varepsilon$ smaller than a
  constant that depends only on $M$, the functions
  $\overline\rho_g(\chi;m, \varepsilon)$ and
  $\underline\rho_g(\chi;m, \varepsilon)$ are continuous in $\chi$. Furthermore,
  we have
  $\lim_{\varepsilon\to 0} \sup_{\chi\in\hor{0, \infty}, m\in M}
  [\overline\rho_g(\chi; m, \varepsilon) - \underline\rho_g(\chi; m,
  \varepsilon)]=0$.
\end{lemma}
\begin{proof}
  For $\varepsilon$ smaller than a constant that depends only on $M$, the set
  $\cup_{m\in M} B_{\varepsilon}(m)$ is contained in another compact subset of
  the interior of the set of values of $\int g(b)\, dF(b)$, where $F$ ranges
  over all measures on $\mathbb{R}$ with the Borel $\sigma$-algebra. The result
  then follows from \Cref{rho_uniform_continuity_lemma}, where, for the first
  claim, we use the fact that
  $\abs{\overline\rho_g(\chi;m, \varepsilon)-\overline\rho_g(\tilde\chi;m, \varepsilon)}
  \le \sup_{\tilde m\in B_\varepsilon(m)}\abs{\rho_g(\chi;\tilde{m})
    -\rho_{g}(\tilde\chi;\tilde m)}$ and similarly for $\underline\rho_{g}$.
\end{proof}

We now prove \Cref*{high_level_conditional_coverage_thm}. Given $\mathcal{X}\in\mathcal{A}$ and $\varepsilon>0$, let $m_1, \dotsc, m_J$
and $\mathcal{X}_1, \dotsc, \mathcal{X}_J$ be as in \Cref*{partition_assump}. Let
$\underline \chi_j=\min \{\chi\colon \underline{\rho}_g(\chi;m_j,2\varepsilon)\le
\alpha\}$. For $\hat m_i\in B_{2\varepsilon}(m_j)$, we have
$\underline \rho_g(\chi;m_j,2\varepsilon)\le \rho_g(\chi;\hat m_i)$ for all
$\chi$, so that, using the fact that $\underline \rho_g(\chi;m_j,2\varepsilon)$
and $\rho_g(\chi;\hat m_i)$ are weakly decreasing in $\chi$, we have
$\underline \chi_j\le \hat \chi_i$. Thus, letting $\tilde{\underline\chi}^{(n)}$
denote the sequence with $i$th element equal to $\underline\chi_j$ when
$\tilde X_i\in \mathcal{X}_j$, we have
\begin{multline*}
  ANC_n(\hat\chi^{(n)};\mathcal{X})\le
  \max_{1\le j\le J} ANC_n(\underline{\tilde\chi}^{(n)};\mathcal{X}_j) \\
  \le \max_{1\le j\le J} \left[\frac{1}{N_{\mathcal{X}_j, n}}
    \sum_{i\in\mathcal{I}_{\mathcal{X}_j, n}} \1{\hat m_i \notin
    B_{2\varepsilon}(m_j)} +\frac{1}{N_{\mathcal{X}_j, n}}
    \sum_{i\in\mathcal{I}_{\mathcal{X}_j, n}} \1{\abs{Z_i}>\underline\chi_{j}}
  \right].
\end{multline*}
The first term is bounded by
$\frac{1}{N_{\mathcal{X}_j, n}} \sum_{i\in\mathcal{I}_{\mathcal{X}_j, n}} \1{\|\hat m_i-m(\tilde X_i)\| > \varepsilon}$ since, for
$i\in \mathcal{I}_{\mathcal{X}_j, n}$, we have
$\|\hat m_i-m_j\|\le \varepsilon+\|\hat m_i-m(\tilde X_i)\|$. This converges in
probability (and expectation) to zero under $\tilde P$ by
\Cref*{mu_consistency_assump}. By \Cref{ANC_approx_lemma}, the second term is
equal to, letting $F_{j, n}$ denote the empirical distribution of the $b_{i, n}$'s
for $i$ with $x_i\in\mathcal{X}_j$,
\begin{equation*}
  \int r(b, \underline\chi_j)\, dF_{j, n}(b) + R_n
  \le \overline\rho_g(\underline\chi_j;\mu_j,2\varepsilon) + R_n
\end{equation*}
where $R_n$ is a term such that $E_{\tilde P}R_n\to 0$ and such that, if
$Z_i-\tilde{b}_i$ is independent over $i$ under $\tilde P$, then $R_n$ converges
in probability to zero under $\tilde P$. The result will now follow if we can
show that
$\max_{1\le j\le
  J}[\overline\rho_g(\underline\chi_j;\mu_j,2\varepsilon)-\alpha]$ can be made
arbitrarily small by making $\varepsilon$ small. This holds by
\Cref{overline_underline_rho_diff_lemma} and the fact that
$\underline\rho_g(\underline\chi_j;\mu_j,2\varepsilon)\le \alpha$ by
construction.

\subsection{Proof of Theorem~\ref*{indep_shrinkage_coverage_thm}}\label{sec:proof_thm_C2}

To prove \Cref*{indep_shrinkage_coverage_thm}, we will verify the conditions of
\Cref*{high_level_conditional_coverage_thm} with $\mathcal{A}$ given
in~\Cref*{indep_shrinkage_mu_consistency_assump},
$m_j(\tilde X_i)=c(\gamma, \sigma_i)^{\ell_j}\mu_{0,\ell_j}$, $\tilde b_i=c(\hat \gamma, \hat\sigma_i)(\theta_i-\hat X_i'\hat\delta)$ and
$b_{i, n}=c(\gamma, \sigma_i)(\theta_i-\hat X_i'\delta)$
where $c(\gamma, \sigma)=\frac{w(\gamma, \sigma)-1}{w(\gamma, \sigma)\sigma}$. The
first part of \Cref*{Zi_assump} is immediate from
\Cref*{indep_shrinkage_Zi_assump} since
$Z_i-\tilde b_i=(Y_i-\theta_i)/\hat\sigma_i$. For the second part, we have
\begin{multline*}
\tilde b_i - b_{i, n}
=c(\hat\gamma, \hat\sigma_i)(\theta_i-\hat X_i'\hat\delta) - c(\gamma, \sigma_i)(\theta_i-X_i'\delta) \\
=[c(\hat\gamma, \hat\sigma_i) - c(\gamma, \sigma_i)](\theta_i-X_i'\delta)
  + c(\hat\gamma, \hat\sigma_i)\cdot [(\hat X_i-X_i)'\hat\delta-X_i'(\delta-\hat\delta)].
\end{multline*}
For $\|\theta_i\|+\|X_i\|\le C$, the above expression is bounded by
\begin{equation*}
  [c(\hat\gamma, \hat\sigma_i) - c(\gamma, \sigma_i)]\cdot (\|\delta\|+ 1)\cdot C
  + c(\hat\gamma, \hat\sigma_i) \left[\|\hat\delta-\delta\|\cdot C +
    \|\hat X_i-X_i\| \cdot (C+\|\hat\delta-\delta\|) \right].
\end{equation*}
By uniform continuity of $c()$ on an open set containing
$\{\gamma\}\times \mathcal{S}_1$, for every $\varepsilon>0$ there exists
$\eta>0$ such that
$\|(\hat\sigma_i-\sigma_i, \hat{\gamma}-\gamma, \hat{\delta}'-\delta',
\hat{X}_i'-X_i')'\|\le \eta$ implies that the absolute value of the above
display is less than $\varepsilon$. Thus, for any $\mathcal{X}\in\mathcal{A}$,
\begin{multline*}
   \lim_{n\to\infty} \frac{1}{N_{\mathcal{X}, n}} \sum_{i\in\mathcal{I}_{\mathcal{X}, n}} \tilde P(|\tilde b_i - b_{i, n}|\ge \varepsilon) \\
  \le \lim_{n\to\infty} \frac{1}{N_{\mathcal{X}, n}}
    \sum_{i\in\mathcal{I}_{\mathcal{X}, n}} \tilde{P}
    (\|(\hat\sigma_i-\sigma_i, \hat{\gamma}-\gamma,
    \hat{\delta}'-\delta', \hat X_i'-X_i')'\|> \eta)\1{\|\theta_i\|+\|X_i\|\le C} \\
  + \limsup_{n\to\infty}\frac{1}{N_{\mathcal{X}, n}} \sum_{i\in\mathcal{I}_{\mathcal{X}, n}} \1{\|\theta_i\|+\|X_i\|> C}.
\end{multline*}
The first limit is zero by \Cref*{indep_shrinkage_sigma_consistency_assump}.
The last limit converges to zero as $C\to \infty$ by the second part of \Cref*{indep_shrinkage_mu_consistency_assump} and
Markov's inequality.
This completes the verification of \Cref*{indep_shrinkage_Zi_assump}.

We now verify \Cref*{mu_consistency_assump}. Given $\mathcal{X}\in \mathcal{A}$
and given $\varepsilon>0$, we can partition $\mathcal{X}$ into sets
$\mathcal{X}_1, \dotsc, \mathcal{X}_J$ such that, for some $c_1, \dotsc, c_J$,
we have $\abs{c(\gamma, \sigma_i)^{\ell_k}-c_j^{\ell_k}} <\varepsilon$ for all
$k=1, \dotsc, p$ whenever $i\in \mathcal{I}_{\mathcal{X}_j, n}$ for some $j$.
Thus, for each $j$ and $k$,
\begin{multline*}
  \frac{1}{N_{\mathcal{X}_j, n}} \sum_{i\in \mathcal{I}_{\mathcal{X}_j}, n}
  b_{i, n}^{\ell_k}-m_k(\tilde X_i) =\frac{1}{N_{\mathcal{X}_j, n}} \sum_{i\in
    \mathcal{I}_{\mathcal{X}_j}, n}
  c(\gamma, \sigma_i)^{\ell_k}\left[(\theta_i-X_i'\delta)^{\ell_k}-\mu_{0,\ell_k}\right] \\
  =c_j^{\ell_k}\cdot \frac{1}{N_{\mathcal{X}_j, n}} \sum_{i\in
    \mathcal{I}_{\mathcal{X}_j}, n}
  \left[(\theta_i-X_i'\delta)^{\ell_k}-\mu_{0,\ell_k}\right]\\
  +\frac{1}{N_{\mathcal{X}_j, n}} \sum_{i\in \mathcal{I}_{\mathcal{X}_j}, n}
  [c(\gamma, \sigma_i)^{\ell_k}-c_j^{\ell_k}]\left[(\theta_i-X_i'\delta)^{\ell_k}-\mu_{0,\ell_k}\right].
\end{multline*}
Under \Cref*{indep_shrinkage_mu_consistency_assump},
the first term converges to $0$ and the second term is bounded up to an $o(1)$
term by $\varepsilon$ times a constant that depends only on $K$.  Since the
absolute value of
$\frac{1}{N_{\mathcal{X}, n}} \sum_{i\in \mathcal{I}_{\mathcal{X}}, n}
b_{i, n}^{\ell_k}-m_k(\tilde X_i)$ is bounded by the maximum over $j$ of the
absolute value of the above display, and since $\varepsilon$ can be chosen
arbitrarily small, the first part of \Cref*{mu_consistency_assump} follows.

For the second part of \Cref*{mu_consistency_assump}, we have
$\hat m_{i, k}-m_k(\tilde{X}_i)= c(\gamma, \sigma_i)\hat\mu_{\ell_j}-c(\gamma,
\sigma_i)^{\ell_j}\mu_{0, \ell_j}$. By uniform continuity of
$(\tilde\gamma', \sigma, \mu_{\ell_1}, \dotsc, \mu_{\ell_p})'\mapsto
(c(\gamma, \sigma_i)^{\ell_1}\mu_{\ell_1}, \dotsc,
c(\gamma, \sigma_i)^{\ell_p}\mu_{\ell_p})'$ in an open set containing
$\{\gamma\}\times \mathcal{S}_1\times \{(\mu_{0,\ell_1}, \dotsc,
\mu_{0,\ell_p})'\}$, for any $\varepsilon>0$, there exists $\eta>0$ such that
$\|(\hat\gamma'-\gamma', \hat\sigma_i-\sigma, \hat\mu_{\ell_1}-\mu_{0,\ell_1},
\dotsc, \hat{\mu}_{\ell_p}-\mu_{0,\ell_p})\|< \eta$ implies
$\|\hat m_{i, k}-m_k(\tilde X_i)\|<\varepsilon$. Thus,
\begin{equation*}
  \max_{1\le i\le n} \tilde P(\|\hat m_i-m(\tilde X_i)\|\ge \varepsilon)
  \le \max_{1\le i\le n} \tilde P(\|(\hat\gamma'-\gamma', \hat\sigma_i-\sigma, \hat\mu_{\ell_1}-\mu_{0,\ell_1}, \dotsc,
  \hat{\mu}_{\ell_p}-\mu_{0,\ell_p})\|< \eta),
\end{equation*}
which converges to zero by
\Cref*{indep_shrinkage_sigma_consistency_assump,indep_shrinkage_mu_consistency_assump}.
This completes the verification of \Cref*{mu_consistency_assump}.

\Cref*{partition_assump} follows immediately from compactness of the set
$\mathcal{S}_1\times \dotsm \times \mathcal{S}_1$ and uniform continuity of
$m()$ on this set. \Cref*{mu_set_assump} follows from
\Cref*{indep_shrinkage_mu_consistency_assump} and
\Cref{constant_multiple_interior_set_lemma} below. This completes the proof of
\Cref*{indep_shrinkage_coverage_thm}.

\begin{lemma}\label{constant_multiple_interior_set_lemma}
  Suppose that, as $F$ ranges over all probability measures with respect to the
  Borel sigma algebra, $(\mu_{\ell_1}, \dotsc, \mu_{\ell_p})'$ is interior to the set of values of
  $\int (b^{\ell_1}, \dotsc, b^{\ell_p})'\, dF(b)$. Let $c\in\mathbb{R}$. Then,
  as $F$ ranges over all probability measures with respect to the Borel sigma
  algebra, $(c^{\ell_1}\mu_{\ell_1}, \dotsc, c^{\ell_p} \mu_{\ell_p})'$ is also
  in the interior of the set of values of
  $\int (b^{\ell_1}, \dotsc, b^{\ell_p})'\, dF(b)$.
\end{lemma}
\begin{proof}
  We need to show that, for any vector $r$ with $\|r\|$ small enough, there
  exists a probability measure $F$ such that
  $\int (b^{\ell_1}, \dotsc, b^{\ell_p})' \, dF(b)= (c^{\ell_1}\mu_{\ell_1}+r_1,
  \dotsc, c^{\ell_p} \mu_{\ell_p}+r_p)'$. Let
  $\tilde \mu_{\ell_k}= \mu_{\ell_k}+r_k/c^{\ell_k}$. For $\|r\|$ small enough,
  there exists a probability measure $\tilde F$ with
  $\int b^{\ell_k}\, dF(b)=\tilde \mu_{\ell_k}$ for each $k$. Let $F$ denote the
  probability measure of $c B$ when $B$ is a random variable distributed
  according to $\tilde F$. Then
  $\int b^{\ell_k}\, dF(b)= c^{\ell_k}\int b^{\ell_k} \, d\tilde F =
  c^{\ell_k}\tilde\mu_{\ell_k} = c^{\ell_k} \mu_{\ell_k} + r_k$ as required.
\end{proof}

\section{Details for simulations}\label{sec:appendix_empirics}

\Cref{sec:sim-details} gives details on the Monte Carlo designs
in \Cref*{sec:sim}. \Cref{sec:sim-heterosk} considers
an additional Monte Carlo exercise calibrated to the empirical application
in \Cref*{sec:empir-appl}.

\subsection{Details for panel data simulation designs}\label{sec:sim-details}

The simulation results reported
in \Cref*{sec:sim} consider the following six
distributions for $\theta_i$, each of which satisfies $\var(\theta_i)=\mu_2$:
\begin{enumerate}
\item Normal (kurtosis $\kappa=3$): $\theta_i \sim N(0,\mu_2)$.
\item Scaled chi-squared ($\kappa=15$):  $\theta_i \sim \sqrt{\mu_2/2} \cdot \chi^2(1)$.
\item 2-point ($\kappa = 1/(0.9 \cdot 0.1)-3 \approx 8.11$), with
  $\theta_{i}=0$ w.p. $0.9$ and $\theta_{i}=\mu_2/(0.9 \cdot 0.1)$ w.p. $0.1$.
\item 3-point ($\kappa = 2$):
  \begin{equation*}
\theta_i \sim \begin{cases}
-\sqrt{\mu_2/0.5} & \text{w.p. $0.25$}, \\
0 & \text{w.p. $0.5$}, \\
\sqrt{\mu_2/0.5} & \text{w.p. $0.25$}.
\end{cases}
\end{equation*}
\item Least favorable for robust EBCI\@: The (asymptotically as $n, T\to\infty$) least favorable
  distribution for the robust EBCI that exploits only second moments, i.e.,
  \begin{equation*}
\theta_i \sim
\begin{cases}
-\sqrt{\mu_2 / \min\lbrace \frac{m_2}{t_0(m_2,\alpha)},1 \rbrace} & \text{w.p. $\frac{1}{2}
  \min\{\frac{m_2}{t_0(m_2,\alpha)},1\}$}, \\
0 & \text{w.p. $1-\min\lbrace \frac{m_2}{t_0(m_2,\alpha)},1\rbrace$}, \\
\sqrt{\mu_2 / \min\lbrace \frac{m_2}{t_0(m_2,\alpha)},1 \rbrace} & \text{w.p. $\frac{1}{2}\min\lbrace \frac{m_2}{t_0(m_2,\alpha)},1\rbrace$},
\end{cases}
\end{equation*}
where $m_2=1/\mu_2$, and $t_0(m_2,\alpha)$ is the number defined in
\Cref*{theorem:bound-second-moment} with $\chi=\cva_\alpha(m_2)$. The kurtosis
$\kappa(\mu_2,\alpha) = 1/\min\lbrace
\frac{1/\mu_2}{t_0(1/\mu_2,\alpha)},1\rbrace$ depends on $\mu_2$ and $\alpha$.
\item Least favorable for parametric EBCI\@: The (asymptotically) least favorable
  distribution for the parametric EBCI\@. This is the same distribution as above,
  except that now $t_0(m_2,\alpha)$ is the number defined in
  \Cref*{theorem:bound-second-moment} with
  $\chi=z_{1-\alpha/2}/\sqrt{\mu_2/(1+\mu_2)}$.
\end{enumerate}

\subsection{Heteroskedastic design}\label{sec:sim-heterosk}

We now provide average coverage and length results for a heteroskedastic
simulation design. We base the design on the effect estimates and standard
errors obtained in the empirical application in \Cref*{sec:empir-appl}. Because
we do not have access to the underlying data set, we treat the standard errors
as known and impose exact conditional normality of the initial estimates. Let
$(\hat{\theta}_i,\hat{\sigma}_i)$, $i=1, \dotsc, n$, denote the $n=595$ baseline
shrinkage point estimates and associated standard errors from this application.
Note for reference that $E_n[\hat{\theta}_i] = 0.0602$, and
$E_n[(\hat{\theta}_i-\bar{\theta})^2] \cdot E_n[1/\hat{\sigma}_i^2] = 0.6698$,
where $E_{n}$ denotes the sample mean.

The simulation design imposes independence of $\theta_i$ and $\sigma_i$,
consistent with the moment independence assumption required by our baseline EBCI
procedure, see \Cref*{rem:conditional_coverage}. We calibrate the
design to match one of three values for the signal-to-noise ratio
$E[\varepsilon_i^2/\sigma_i^2] \in \lbrace 0.1, 0.5, 1 \rbrace$. Specifically, a
simulation sample $(Y_i, \theta_i, \sigma_i)$, $i=1, \dotsc, n$, is created as
follows:
\begin{enumerate}
\item Sample $\tilde{\theta}_i$, $i=1, \dotsc, n$, with replacement from the
  empirical distribution $\{\hat{\theta}_{j}\}_{j=1}^{n}$.
\item Sample $\sigma_i$, $i=1, \dotsc, n$, with replacement from the empirical
  distribution $\{\hat{\sigma}_{j}\}_{j=1}^{n}$.
\item Compute
  $\theta_i = \bar{\theta} + \sqrt{m/c} \cdot (\tilde{\theta}_i-\bar{\theta})$,
  $i=1, \dotsc, n$. Here $m$ is the desired population value of
  $E[\varepsilon_i^2/\sigma_i^2]$ and $c=0.6698$.
\item Draw $Y_i \overset{indep}{\sim} N(\theta_i, \sigma_i^2)$,
  $i=1, \dotsc, n$.
\end{enumerate}
The kurtosis of $\theta_i$ equals the sample kurtosis of $\hat{\theta}_i$, which
is 3.0773. We use precision weights $\omega_i = \sigma_i^{-2}$ when computing
the EBCIs, as in \Cref*{sec:empir-appl}.

\begin{table}[t]
  \centering
  \begin{threeparttable}
    \caption{Monte Carlo simulation results: heteroskedastic design.}\label{tab:sim_ch}
    \begin{tabular*}{0.9\linewidth}{@{\extracolsep{\fill}}@{}c@{}cccccc@{}}
      \multicolumn{1}{c}{ }      & \multicolumn{2}{c}{Robust, $\mu_2$ only} & \multicolumn{2}{c}{Robust, $\mu_2$ \& $\kappa$} & \multicolumn{2}{c}{Parametric} \\
      \cmidrule(rl){2-3} \cmidrule(rl){4-5}  \cmidrule(rl){6-7}
      $n$      & Oracle    & Baseline       & Oracle  & Baseline & Oracle & Baseline \\
      \midrule
      \multicolumn{7}{@{}l}{Panel A\@{}: Average coverage (\%), minimum across 3 DGPs}\\
      \cmidrule(r{50pt}){1-6}
      595 & 98.9 & 96.0 & 96.1 & 96.0 & 94.3 & 85.7 

      \\
      \multicolumn{7}{@{}l}{Panel B\@{}: Relative average length, average across 3 DGPs}\\
      \cmidrule(r{50pt}){1-6}
      595 & 1.56 & 1.51 & 1.00 & 1.48 & 0.89 & 0.86 

    \end{tabular*}
    \begin{tablenotes}
    \item \emph{Notes:} Nominal average confidence level $1-\alpha=95\%$. Top
      row: type of EBCI procedure. ``Oracle'': true $\mu_2$ and $\kappa$ (but
      not $\delta$) known. ``Baseline'': $\hat{\mu}_2$ and $\hat{\kappa}$
      estimates as in \Cref*{sec:baseline-implementation}. For each DGP,
      ``average coverage'' and ``average length'' refer to averages across
      observations $i=1, \dotsc, n$ and across 5,000 Monte Carlo repetitions.
      Average CI length is measured relative to the oracle robust EBCI that
      exploits $\mu_2$ and $\kappa$.
    \end{tablenotes}
  \end{threeparttable}
\end{table}

\Cref{tab:sim_ch} shows that our baseline implementation of the 95\% robust EBCI
achieves average coverage above the nominal confidence level, regardless of the
signal-to-noise ratio
$E[\varepsilon_i^2/\sigma_i^2] \in \lbrace 0.1, 0.5, 1 \rbrace$. This contrasts
with the feasible version of the parametric EBCI, which undercovers by 9.3
percentage points.

\section{Statistical power}\label{sec:power-details}

The efficiency calculations in \Cref*{fig:efficiency_unshrunk} of
\Cref*{sec:efficiency} show that our \ac{EBCI} is substantially shorter than the
conventional \ac{CI} based on the unshrunk estimate $Y_i$ if the signal-to-noise
ratio is small enough. Here, we perform analogous calculations using the
statistical power of tests based on a given \ac{CI} as the measure of
efficiency.

Consider testing $H_{0,i}:\theta_i=\theta_0$ for some null value $\theta_0$ by
rejecting when $\theta_0\notin CI_i$, where $CI_i$ is our robust \ac{EBCI}. As
with the efficiency calculations in \Cref*{sec:efficiency}, we consider
efficiency under the baseline model in \Cref*{eq:hierarch_theta}, and we
consider the asymptotic setting in which $\mu_{1,i}=X_i'\delta$, $\mu_2$,
$\sigma^2_i$ and $\kappa=3$ can be treated as known. We compute the average
power of this test (averaged over the baseline normal prior, conditional on
$X_i,\sigma_i$), and we compare it to the average power of the conventional
two-sided $z$-test based on the unshrunk estimate in the same setting. Since the
distribution of $\theta_i$ is atomless, the average power is given by the
rejection probability $P(\theta_0\notin CI_i\mid X_i, \sigma_i)$. Let
$d_{i}=(\mu_{1, i}-\theta_{0})/\sigma_{i}$ denote the standardized average
distance between the true parameter $\theta_{i}$ and the null $\theta_{0}$.
Under the baseline model in \Cref*{eq:hierarch_theta}, the average power of a
test based on the robust \ac{EBCI} given in \Cref*{eq:conditional_ci} with
$\kappa=3$ is thus given by
\begin{multline*}
  P\left( \theta_i\notin CI_i \mid X_i, \sigma_i \right) = P\left(
    \left|\frac{Y_i-\mu_{1,i}}{\sqrt{\sigma^2_i+\mu_2}}+\frac{d_{i}
      \sigma_{i}}{
        w_{EB, i}\sqrt{\sigma^2_i+\mu_2}}\right| >
    \frac{\cva_{\alpha}(\sigma^2_i/\mu_2,3)}{\sqrt{1+\mu_2/\sigma_{i}^{2}}} \Bigg| X_i,\sigma_i \right)  \\
  =r\left(
    \frac{d_{i}\sqrt{1-w_{EB, i}}}{w_{EB, i}},
    \cva_{\alpha}(1/w_{EB, i}-1, 3)\sqrt{1-w_{EB, i}} \right),
\end{multline*}
with $r$ given in \Cref*{eq:rb}, and we use the fact that
$Y_{i}-\mu_{1, i}\mid X_{i}, \sigma_{i} \sim
\mathcal{N}(0,\sigma_{i}^{2}+\mu_2)$ under \Cref*{eq:hierarch_theta}. The
two-sided $z$-test based on the unshrunk estimate $Y_i$ rejects when
$|Y_i-\theta_0|>z_{1-\alpha/2}\sigma_i$. By analogous reasoning, it follows that
the average power of this test is given by
\begin{equation*}
  P\left(\abs{Y_i-\theta_0}> z_{1-\alpha/2}\sigma_i \mid X_i, \sigma_i\right)
  =r\left( d_{i}\sqrt{1-w_{EB, i}}, z_{1-\alpha/2}\sqrt{1-w_{EB,i}} \right).
\end{equation*}
Both expressions depend only on $d_{i}$ and the shrinkage $w_{EB, i}$ (or,
equivalently, since $\mu_2/\sigma_i^2=w_{EB, i}/(1-w_{EB, i})$, the
signal-to-noise ratio $\mu_2/\sigma_i^2$).

\begin{figure}[p]
  \centering%
  \tikzset{font=\small}%
  \input{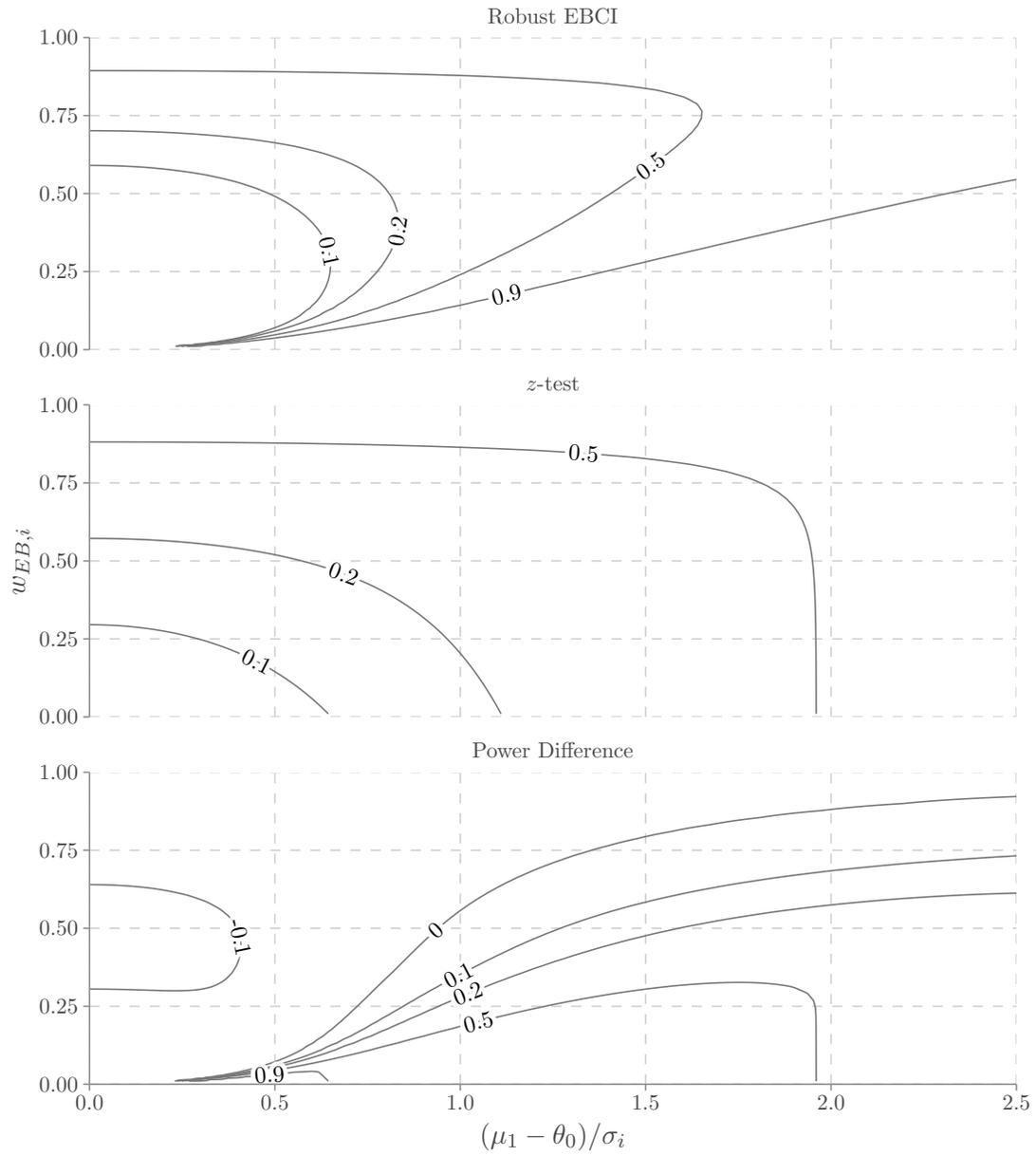}
  \caption{Average power of the robust \ac{EBCI} and the $z$-test based on the
    unshrunk estimate as a function of the normalized average distance to the
    null and of the shrinkage $w_{EB, i}$.}\label{fig:power_difference}
\end{figure}

\Cref{fig:power_difference} computes the power of the robust \ac{EBCI}-based
test and the $z$-test as a function of the normalized distance
$d_{i}=(\mu_{1,i}-\theta_0)/\sigma_i$ and the shrinkage $w_{EB, i}$ for
$\alpha=0.05$. The third panel shows the difference in power, with positive
values indicating greater power for the \ac{EBCI}-based test.

The graphs show that the \ac{EBCI}-based test is more powerful than the $z$-test
for a given shrinkage $w_{EB, i}$ (equivalently, given signal-to-noise ratio)
when the normalized distance is large enough, while being less powerful when it
is small enough. To get some intuition for this, note that the \ac{EBCI} differs
from the unshrunk \ac{CI} in two ways: it is shorter, and it uses shrinkage to
move the center of the \ac{CI} toward the regression line
$\mu_{1,i}=X_i'\delta$. Shortening the \ac{CI} makes the \ac{EBCI} more powerful
than the test based on the unshrunk \ac{CI}, but the effect of moving the center
of the \ac{CI} is ambiguous: it increases power when the regression line
$\mu_{1,i}$ is far from the null $\theta_0$, while decreasing power when
$\mu_{1,i}$ is close to $\theta_0$. On net, the graphs show that the
\ac{EBCI}-based test displays substantial gains in average power when the amount
of shrinkage is large, even for small to moderate distances to the null.

\section{Applications of general shrinkage}\label{sec:appendix_general}
Here we provide theoretical and numerical results for the soft thresholding
\ac{EBCI} and the Poisson \ac{EBCI}, discussed in
\Cref*{example:soft_thresholding,example:poisson} in
\Cref*{sec:general_shrinkage}.

\subsection{Soft thresholding}\label{sec:appendix_softthresh}

The soft thresholding \ac{EBCI} is obtained by calibrating the \ac{HPD} set in
the homoskedastic normal model with a baseline Laplace prior for $\theta_i$. The
\ac{HPD} set $\mathcal{S}(Y_{i}; \chi)$ in \Cref*{eq:hpd_st} takes the form of an
interval, and is available in closed form. In particular, it follows by direct
calculation that the posterior density for $\theta$ is given by
$p(\theta\mid
Y_{i})=e^{\overbar{c}(Y_{i})-\frac{1}{2\sigma^{2}}\theta^{2}+Y_{i}\theta/\sigma^{2}-\abs{\theta}\sqrt{2/\mu_{2}}}$,
where
$\overbar{c}(Y)=
\frac{1}{2}\log(2/\pi\sigma^{2})-\log(q(\sqrt{\sigma^{2}/\mu_{2}}-Y/\sigma\sqrt{2})+
q(\sqrt{\sigma^{2}/\mu_{2}}+Y/\sigma\sqrt{2}))$. Here
$q(x)=2e^{x^{2}}\Phi(-x \sqrt{2})$ is the scaled complementary error function.
Consequently, $\mathcal{S}(Y; \chi)$ equals the intersection of the
solution sets for two quadratic inequalities,
\begin{equation*}
  \mathcal{S}(Y; \chi)=\textstyle\left\{\theta\colon
    \frac{\theta^{2}}{2\sigma^{2}}-\left(\frac{Y}{\sigma^{2}}-\sqrt{\frac{2}{\mu_{2}}}\right)\theta\leq
    \chi+\overbar{c}(Y)
  \right\}\cap\left\{\theta\colon
    \frac{\theta^{2}}{2\sigma^{2}}-\left(\frac{Y}{\sigma^{2}}+\sqrt{\frac{2}{\mu_{2}}}\right)\theta\leq
    \chi+\overbar{c}(Y)
  \right\}.
\end{equation*}
Since the quadratic term is positive in both inequalities,
$\mathcal{S}(Y; \chi)$ is given by an intersection of two intervals, and is
therefore itself an interval. The non-coverage function
$\tilde{r}(\theta_i, \chi)$ in \Cref*{eq:conditional_coverage_general} is
computed via numerical quadrature. The linear program in \Cref*{eq:rho_general}
is solved by discretizing the support for $\theta_i$. In addition to computing a
robust soft thresholding \ac{EBCI}, we can similarly compute a parametric soft
thresholding \ac{EBCI}, with $\chi$ solving $E_F[r(\theta_i, \chi)]=\alpha$; here
$F$ is the Laplace distribution with second moment $\mu_2$.

\begin{figure}[t]
\centering
\begin{tabular}{cc}
\includegraphics[height=8.5cm]{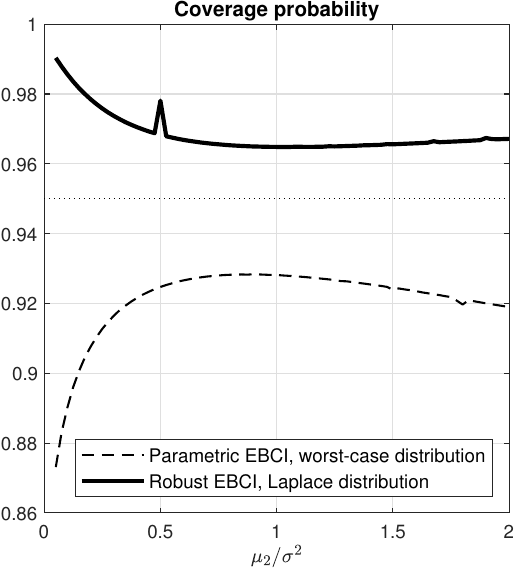} & \includegraphics[height=8.5cm]{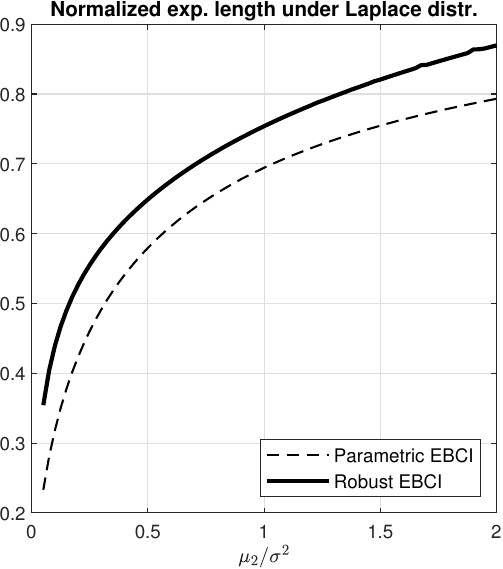}
\end{tabular}
\caption{Soft thresholding \acp{EBCI} in the normal means model, $\alpha=0.05$. The expected length is normalized by the length of the unshrunk \ac{CI}. The grid for $\theta_i$ for the linear program in \Cref*{eq:rho_general} is given by 500 points equally spaced on $[-10,10]$. Integrals over the $Y_i$ distribution are truncated at the endpoints $-10$ and 10.}\label{fig:softthresh}
\end{figure}

We now compute the coverage and expected length of the soft thresholding
\acp{EBCI}. We consider an asymptotic setting where $\mu_2=E[\theta^2]$ is
known, and this is the only constraint imposed when we compute the robust
\ac{EBCI}. \Cref{fig:softthresh} shows the coverage and expected length of the
parametric and robust \acp{EBCI} with $\alpha=0.05$. The worst-case coverage
(over all $\theta_i$-distributions with second moment $\mu_2$) of the nominal
95\% parametric \ac{EBCI} is below 88\% for small signal-to-noise ratios
$\mu_2/\sigma^2$. When $\theta_i$ is in fact Laplace-distributed, both the
parametric and robust soft thresholding \acp{EBCI} deliver substantial expected
length improvements relative to the unshrunk EBCI $Y_i \pm z_{1-\alpha/2}\sigma$.
For small values of $\mu_2/\sigma^2$, the length improvement exceeds that of the
linear \acp{EBCI} shown in \Cref*{fig:efficiency_unshrunk}.

\subsection{Poisson data}\label{sec:appendix_poisson}
Suppose now that $Y_i$ has a Poisson distribution with rate parameter
$\theta_i$, conditional on $\theta_i$. As a baseline prior for $\theta_i$, we use the conjugate
gamma distribution with shape parameter $k$ and scale parameter $\lambda$. Let
$\Gamma^{-1}(\alpha;k, \lambda)$ denote the $\alpha$-quantile of this
distribution. As candidate sets $\mathcal{S}(y;\chi)$, we use a modification of
the equal-tailed posterior credible set for $\theta_i$ under the baseline prior,
\begin{equation*}
  \mathcal{S}(y;\chi) = \left[\Gamma^{-1}\left(\frac{\alpha}{2};e^{-\chi}k
      +y, \frac{\lambda}{e^{-\chi}+\lambda}\right),
    \Gamma^{-1}\left(1-\frac{\alpha}{2};1+e^{-\chi}(k-1)+y, \frac{\lambda}{e^{-\chi}+\lambda}\right) \right],
\end{equation*}
where $1-\alpha$ is the nominal confidence level. For $\chi=0$, this corresponds
to the equal-tailed posterior credible interval under the baseline prior; we
call this the parametric \ac{EBCI}. As $\chi \to \infty$, the interval converges
to the ``unshrunk'' \citet{Garwood1936} confidence interval for the Poisson
parameter $\theta_i$, which has coverage at least $1-\alpha$ conditional on
$\theta_i$. We compute the value $\hat{\chi}\in(0,\infty)$ that leads to a
robust \ac{EBCI} numerically as in \Cref{sec:appendix_softthresh}, except that
we replace integrals over the distribution of $Y_i$ with (truncated) sums.

\begin{figure}[t]
\centering
\begin{tabular}{cc}
\includegraphics[height=8.5cm]{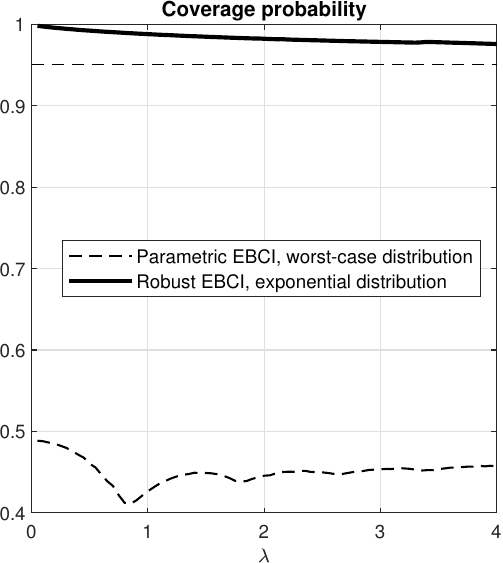} &
\includegraphics[height=8.5cm]{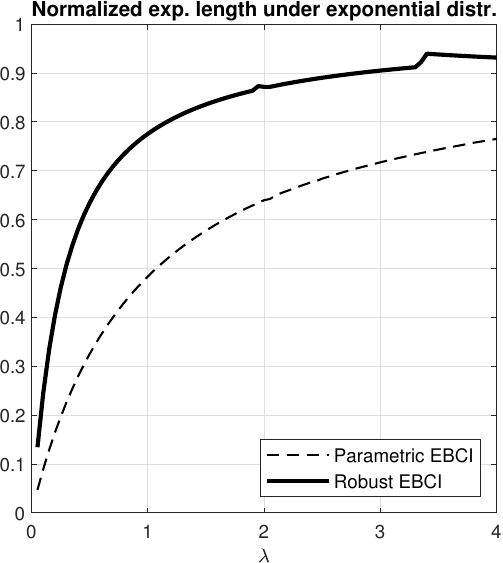}
\end{tabular}
\caption{Poisson \acp{EBCI}, $\alpha=0.05$. The expected length is normalized by
  that of the unshrunk \citet{Garwood1936} \ac{CI}. The grid for $\theta_i$ for
  the linear program in \Cref*{eq:rho_general} is given by 500 points equally
  spaced on $[10^{-6}, \Gamma^{-1}(0.999;1,\lambda)]$. The support for $Y_i$ is
  truncated above at 30.}\label{fig:poisson}
\end{figure}

\Cref{fig:poisson} displays the coverage and expected length for $k=1$, i.e.,
when the baseline $\theta_i$-distribution is exponential with mean $\lambda$. We
consider the asymptotic limit where the first two moments of $\theta_i$ are
known.\footnote{These moments are easily obtained from the first and second
  marginal moments of the data: $E[\theta]=E[Y]$ and $E[\theta^2]=E[Y^2]-E[Y]$.
  They equal $E[\theta]=k\lambda$ and $E[\theta^2]=k(k+1)\lambda^2$ under the
  baseline distribution.} We set $\alpha=0.05$. The worst-case coverage (over
all $\theta_i$-distributions with the same first and second moments as the
exponential distribution) of the nominal 95\% parametric \ac{EBCI} is
disastrously low for all values of $\lambda$ considered here. At the same time,
the robust \ac{EBCI} is over 50\% shorter on average than the unshrunk
\citet{Garwood1936} \ac{CI} when $\lambda \leq 0.3$, and more than 25\% shorter
when $\lambda \leq 0.85$.

\onehalfspacing
\phantomsection%
\addcontentsline{toc}{section}{References}
\bibliography{ebci_references}

\end{appendices}